\documentclass[english,aps,multicolumn]{revtex4}
\usepackage[T1]{fontenc}
\usepackage[latin1]{inputenc}
\usepackage{amsmath}
\usepackage{graphicx}
\usepackage{feynmf}
\usepackage{cancel}

\newcommand{\la}{\left\langle}
\newcommand{\ra}{\right\rangle}
\newcommand{\be}{\begin{equation}}
\newcommand{\ee}{\end{equation}}
\newcommand{\bea}{\begin{eqnarray}}
\newcommand{\eea}{\end{eqnarray}}
\newcommand{\ba}{\begin{array}}
\newcommand{\ea}{\end{array}}
\newcommand{\bi}{\begin{itemize}}
\newcommand{\ei}{\end{itemize}}
\newcommand{\ben}{\begin{enumerate}}
\newcommand{\een}{\end{enumerate}}

\makeatletter

\providecommand{\tabularnewline}{\\}

\usepackage{babel}
\makeatother
\begin{document}

\title{Introduction to Statistical Theory of Fluid Turbulence}

\author{Mahendra K. Verma}

\email{mkv@iitk.ac.in}

\homepage{https://sites.google.com/view/mahendra-verma/}

\affiliation{Department of Physics, IIT Kanpur, Kanpur 208016, India}

\begin{abstract}
This is a brief introduction to the statistical theory of hydrodynamic turbulence with an emphasis on its field-theoretic treatment. 
\end{abstract}
\maketitle

\tableofcontents

\section{Introduction}

Fluid and plasma flows exhibit complex random behaviour at high Reynolds
number; this phenomena is called turbulence. On the Earth, turbulence
is observed in atmosphere, channel and rivers flows etc. In the universe,
most of the astrophysical systems are turbulent. Some of the examples
are solar wind, convective zone in stars, galactic plasma, accretion
disk etc. 

Reynolds number, defined as $UL/\nu$ ($U$ is the large-scale velocity,
$L$ is the large length scale, and $\nu$ is the kinematic viscosity),
has to be large (typically 2000 or more) for turbulence to set in.
At large Reynolds number, there are many active modes which are nonlinearly
coupled. These modes show random behaviour along with rich structures
and long-range correlations. Presence of large number of modes and
long-range correlations makes turbulence a very difficult problem
that remains largely unsolved for more than hundred years. 

Fortunately, random motion and presence of large number of modes make
turbulence amenable to statistical analysis. Note that the energy
supplied at large-scales $(L)$ gets dissipated at small scales, say
$l_{d}$. Experiments and numerical simulations show that the velocity
difference $u(\mathbf{x+l})-u(\mathbf{x})$ has a universal probability
density function (pdf) for $l_{d}\ll l\ll L$. That is, the pdf is
independent of experimental conditions, forcing and dissipative mechanisms
etc.~\cite{Kolmogorov:DANS1941Degeneration,Kolmogorov:DANS1941Dissipation,Kolmogorov:DANS1941Structure}. Because of its universal behaviour, the above quantity has been
of major interest among physicists for many  years. Unfortunately,
we do not yet know how to derive the form of this pdf from the first
principle, but some of the moments have been computed analytically.
The range of scales $l$ satisfying $l_{d}\ll l\ll L$ is called inertial
range.

In 1941 Kolmogorov \cite{Kolmogorov:DANS1941Degeneration,Kolmogorov:DANS1941Dissipation,Kolmogorov:DANS1941Structure} computed an exact expression
for the third moment of velocity difference. He showed that under
vanishing viscosity, third moment for velocity difference for homogeneous,
isotropic, incompressible, and steady-state fluid turbulence is\[
\left\langle \left(u_{||}(\mathbf{x+l})-u_{||}(\mathbf{x})\right)^{3}\right\rangle =\frac{4}{5}\Pi l\]
where $||$ is the parallel component along $\mathbf{l}$, $\left\langle .\right\rangle $
stands for ensemble average, and $\Pi$ is the energy cascade rate,
which is also equal to the energy supply rate at large scale $L$
and the dissipation rate at the small scale $l_{d}$. Assuming fractal
structure for the velocity field, and $\Pi$ to be constant for all
$l$, it has been shown that the energy spectrum $E(k)$ is\[
E(k)=K_{Ko}\Pi^{2/3}k^{-5/3},\]
where $K_{Ko}$ is a universal constant, called Kolmogorov's constant,
and $L^{-1}\ll k\ll l_{d}^{-1}$. Numerical simulations and experiments
verify the above energy spectrum apart from a small deviation called
intermittency correction. 

Availability of powerful computers and sophisticated theoretical tools
have helped us understand several aspects of fluid turbulence. Some
of these theories have been motivated by Kolmogorov's theory for fluid
turbulence. Note that incompressible turbulence is better understood
than compressible turbulence. Therefore, our discussion is primarily
for incompressible plasma. \emph{In this paper we focus on the universal
statistical properties of fluid turbulence, which are valid in the
inertial range. In this paper we will review the statistical properties
of following quantities:}

\begin{enumerate}
\item Inertial-range energy spectrum for fluid turbulence.
\item Various energy fluxes in fluid turbulence.
\item Energy transfers between various wavenumber shells.
\item Fluctuations in the energy flux
\end{enumerate}

Many analytic calculations in fluid have been done using
field-theoretic techniques. Even though these methods are plagued with
some inconsistencies, many meaningful results have been obtained using
them. Here we will discuss items 1-3 in some detail.

As mentioned above, pdf of velocity difference in fluid turbulence
is still unsolved. We know from experiments and simulation that pdf
is close to Gaussian for small $\delta u$, but is nongaussian for
large $\delta u$. This phenomena is called intermittency. Note that
various moments called Structure functions are connected to pdf. It
can be shown that the structure functions are related to the {}``local
energy cascade rate'' $\Pi(k)$. Some phenomenological models, notably
by She and Leveque \cite{She:PRL1993} based on log-Poisson process, have
been developed to compute $\Pi(k)$; these models quite successfully
capture intermittency in both fluid and MHD turbulence. The predictions
of these models are in good agreement with numerical results.

Numerical simulations have provided many important data and clues
for understanding the dynamics of turbulence. They have motivated
new models, and have verified/rejected existing models. In that sense,
they have become another type of experiment, hence commonly termed
as numerical experiments. Modern computers have made reasonably high
resolution simulations possible  \cite{Gotoh:CPC2002,Yeung:PNAS2015,Ishihara:PRF2016,Verma:FD2018}).
Note that simulations are also used heavily for studying fluid flows
around aircrafts and vehicles, in atmospheres, engineering devices
like turbines etc. 

Fluid turbulence has a larger volume of literature. Here we will list
only some of them:  \citet{Kraichnan:JFM1959}, Monin and Yaglom~\cite{Monin:book:v1,Monin:book:v2}, Leslie \cite{Leslie:book}, McComb~\cite{McComb:book:Turbulence,McComb:book:RG,McComb:book:HIT,McComb:PRA1992Two_field},   \citet{Zhou:NASA1997}, \citet{Zhou:PR2010}, \citet{Yakhot:JSC1986}, and  \citet{Smith:ARFM1998} have
reviewed field-theoretic treatment of fluid turbulence. The 
books by  \citet{Frisch:book},  \citet{Lesieur:book:Turbulence}, \citet{Davidson:book:Turbulence}, and \citet{Verma:book:ET} cover
recent developments and phenomenological theories. Also, the review articles
by Orszag \cite{Orszag:CP1973}, Kraichnan and Montgomery \cite{Kraichnan:ROPP1980}, 
 Sreenivasan \cite{Sreenivasan:RMP1999}, \citet{Verma:PR2004}, and \citet{Alexakis:PR2018} are quite exhaustive.

The outline of the paper is as follows: Section \ref{sec:Definitions}
contains definition of various global and spectral quantities along
with their governing equations. In Section
\ref{sec:Mode-to-mode-Energy-Transfer} we discuss the formalism of
``mode-to-mode'' energy transfer rates in fluid
turbulence. Using this formalism, formulas for energy fluxes and
shell-to-shell energy transfer rates have been derived.  Section
\ref{sec:Turbulence-Models} contains the existing discussion
on Kolmogorov's phenomenology for fluid turbulence.
Sections V and VI contains very brief introduction on experiments
and simulations in fluid turbulence. 
 In Section
\ref{sec:Renormalization-Group} we introduce
Renormalization-group analysis of turbulence, with an emphasis on
McComb's procedure.  In Section \ref{sec:analytic-energy} we compute
various energy fluxes and shell-to-shell energy transfers in fluid
turbulence using field-theoretic techniques.  Section~\ref{sec:K41} contains a brief description of K41 theory, while Sec.~\ref{sec:beyond_K41} contains a  
intermittency models of fluid turbulence.  Section~\ref{sec:Struct_fn_2d} describes the properties of the structure functions for 2D turbulence.  Appendix
\ref{sec:Fourier-Series} contains the definitions of Fourier series
and transforms of fields in homogeneous turbulence. Appendix B
contains the Feynman diagrams for fluid turbulence; these diagrams are
used in the field-theoretic calculations. In the last two Appendix (C
and D) we briefly mention the mode-to-mode energy transfer formalism
for scalar and Rayleigh-B\'{e}nard convection.

\section{Governing equations \label{sec:Definitions}}

\subsection{Equations for Fluid Dynamics}

Navier-Stokes Equation

\begin{equation}
\frac{\partial\mathbf{u}}{\partial t}+(\mathbf{u}\cdot\nabla)\mathbf{u}=-\frac{1}{\rho}\nabla p_{th}+\nu\nabla^{2}\mathbf{u}+\frac{2\nu}{3}\nabla\nabla\cdot\mathbf{u},\label{eq:ux_compressible}\end{equation}
where $p_{th}$ is called thermodynamic pressure. The law of mass
conservation yields the following equation for density field
$\rho(\mathbf{x})$

\begin{equation}
\frac{\partial\rho}{\partial t}+\nabla\cdot\left(\rho\mathbf{u}\right)=0\label{eq:mass_conservation}\end{equation}
Pressure can be computed from $\rho$ using the equation of state
\begin{equation}
p=f\left(\rho\right)\label{eq:p_rho}\end{equation}
This completes the basic equations of fluids. Using these equations
we can determine the unknowns $(\mathbf{u,}\rho,p)$. Note that the
number of equations and unknowns are the same.

On nondimensionalization of the Navier-Stokes equation, the term $\nabla p$
becomes $\left(d\rho/dx'\right)/\rho$ $\times$ $(C_{s}/U)^{2}$,
where $C_{s}$ is the sound speed, $U$ is the typical velocity of
the flow, $x'$ is the position coordinate normalized with relative
to the length scale of the system \cite{Tritton:book}. $C_{s}\rightarrow\infty$
is the incompressible limit, which is widely studied because water,
the most commonly found fluid on earth, is almost incompressible ($\delta\rho/\rho<$0.01)
in most practical situations. The other limit $C_{s}\rightarrow0$
or $U\gg C_{s}$ (supersonic) is the fully compressible limit, and
it is described by Burgers equation that exhibits $ k^{-2} $ energy spectrum.  The energy
spectrum for both these extreme limits well known. When $U/C_{s}\ll1$
but nonzero, then we call the fluid to be nearly incompressible; Zank
and Matthaeus~\cite{Zank:PRL1990,Zank:PF1991} have given
theories for this limit. The energy and density spectra are not well
understood for arbitrary $U/C_{s}$.

For most part of this paper, we assume the fluid to be incompressible.
In most of the terrestrial experiments, the speed of water or air is
less than the sound speed. Hence, incompressibility is a good assumption
that simplifies the calculations significantly. The incompressibility
approximation can also be interpreted as the limit when volume of
a fluid parcel will not change along its path, that is, $d\rho/dt=0$.
From the continuity equation (\ref{eq:mass_conservation}), the incompressibility
condition reduces to

\begin{equation}
\nabla\cdot\mathbf{u}=0\label{eq:div_u_eq_0}\end{equation}
This is a constraint on the velocity field \textbf{u}. Note that incompressibility
does not imply constant density. However, for simplicity we take density
to be constant and equal to 1. Under this condition, 
Eqs. (\ref{eq:ux_compressible}, \ref{eq:mass_conservation}) reduce to
\begin{eqnarray}
\frac{\partial\mathbf{u}}{\partial t}+(\mathbf{u}\cdot\nabla)\mathbf{u} & = & -\nabla p+\nu\nabla^{2}\mathbf{u},\label{eq:ux}\\
\nabla\cdot\mathbf{u} & = & 0.\label{eq:div_u_eq_0}\end{eqnarray}

When we take divergence of the equation Eq. (\ref{eq:ux}), we obtain
Poisson's equation \begin{equation}
-\nabla^{2}p=\nabla\cdot\left[(\mathbf{u}\cdot\nabla)\mathbf{u}\right].
\end{equation}
Hence, given \textbf{u} fields at any given time, we can evaluate
$p$. Hence $p$ is a dependent variable in the incompressible limit. 

The Navier-Stokes equation is nonlinear, and that is the crux of the
problem. The viscous $(\nu\nabla^{2}\mathbf{u)}$ term dissipates
the input energy.
The ratio of the nonlinear vs. viscous dissipative term is called
Reynolds number $Re=UL/\nu$, where $U$ is the velocity scale, and
$L$ is the length scale. For turbulent flows, Reynolds number should
be high, typically more than 2000 or so~\cite{Lesieur:book:Turbulence}.

\subsection{Energy Equations and Conserved Quantities}

In this subsection we derive energy equations for compressible and
incompressible fluids. For compressible fluids we can construct equations
for energy using Eq. (\ref{eq:ux_compressible}). Following Landau
\cite{Landau:book:Fluid} we derive the following energy equation for
the kinetic energy \begin{equation}
\frac{\partial}{\partial t}\left(\frac{1}{2}\rho u^{2}+\rho\epsilon\right)=-\nabla\cdot\left[\left(\frac{1}{2}u^{2}+\epsilon\right)\rho\mathbf{u}\right]-\nabla\cdot (p\mathbf{u})+\Phi\label{eq:u2_compressible}\end{equation}
where $\epsilon$ is the internal energy function. The first term
in the RHS is the energy flux, and the second term is the work done
by the pressure, which enhances the energy of the system. The third
term,$\Phi$, a complex function of strain tensor, is the energy change
due to surface forces.

In the above equations we apply isoentropic and incompressibility
conditions. For the incompressible fluids we can choose $\rho=1$.
Landau \cite{Landau:book:Fluid} showed that under this condition $\epsilon$
is a constant. Hence, for incompressible fluid we treat $(u^{2})/2$
as total energy. For ideal incompressible fluid ($\nu=0)$ the energy
evolution equation is \bea
\frac{\partial}{\partial t}\frac{1}{2}\left(u^{2}\right)=-\nabla\cdot\left[\left(\frac{1}{2}u^{2}+p\right)\mathbf{u}\right] \eea
By applying Gauss law we find that \bea
\frac{\partial}{\partial t}\int\frac{1}{2}\left(u^{2}\right)d\mathbf{x}=-\oint\left[\left(\frac{1}{2}u^{2}+p\right)\mathbf{u}\right]\cdot d\mathbf{S} \eea
For the boundary condition $u_{n}=0$ or periodic boundary condition,
the total energy $\int$ $1/2(u^{2})$ is conserved.

There are some more important quantities in fluid turbulence. They
are listed in Table \ref{table:Globals}. Note that $\mathbf{\omega}$
is the vorticity field. %
\begin{table}

\caption{\label{table:Globals} Global Quantities in MHD}

\begin{tabular}{|c|c|c|c|}
\hline 
Quantity&
Symbol&
Definition&
Conserved in MHD?\tabularnewline
\hline
\hline 
Kinetic Energy&
$E$ &
$\int d\mathbf{x}u^{2}/2$ &
No\tabularnewline
\hline 
Kinetic Helicity&
$H_{K}$&
$\int d\mathbf{x}(\mathbf{u}\cdot\mathbf{\omega})/2$&
No\tabularnewline
\hline 
Enstrophy&
$\Omega$&
$\int d\mathbf{x}\omega^{2}/2$&
No\tabularnewline
\hline
\end{tabular}
\end{table}
By following the same procedure described above, we can show that in
addition to energy, $H_{K}$ is conserved in 3D fluids, while
 $\Omega$ is conserved in 2D fluids \cite{Leslie:book,Lesieur:book:Turbulence}. The
conserved quantities play very important role in turbulence.

Turbulent flow contains many interacting  ''modes'', and the
solution cannot be written in a simple way. A popular approach to
analyze the turbulent flows is to use statistical tools. We will describe
below the application of statistical methods to turbulence.

\subsection{Necessity for Statistical Theory of Turbulence}

In turbulent fluid the field variables are typically random both in
space and time. Hence the exact solutions given initial and boundary
conditions will not be very useful even when they were available (they
are not!). However statistical averages and probability distribution
functions are reproducible in experiments under steady state, and
they shed important light on the dynamics of turbulence. For this
reason many researchers study turbulence statistically. The idea is
to use the tools of statistical physics for understanding turbulence.
Unfortunately, only systems at equilibrium or near equilibrium have
been understood reasonably well, and a good understanding of nonequilibrium
systems (turbulence being one of them) is still lacking~\cite{Batchelor:book:Turbulence,Frisch:book}. 

The statistical description of turbulent flow starts by dividing the
field variables into mean and fluctuating parts. Then we compute
averages of various functions of fluctuating fields. There are three
types are averages: ensemble, temporal, and spatial averages. Ensemble
averages are computed by considering a large number of identical
systems and taking averages at corresponding instants over all these
systems.  Clearly, ensemble averaging demands heavily in experiments
and numerical simulations. So, we resort to temporal and/or spatial
averaging. Temporal averages are computed by measuring the quantity of
interest at a point over a long period and then averaging. Temporal
averages make sense for steady flows. Spatial averages are computed by
measuring the quantity of interest at various spatial points at a
given time, and then averaging.  Clearly, spatial averages are
meaningful for homogeneous systems.  Steady-state turbulent systems
are generally assumed to be ergodic, for which the temporal average is
equal to the ensemble average \cite{Frisch:book}.

Navier-Stokes equation, which is really Newton's equation, is 
invariant under Galilean transformation
\bea
x & = & x'+V_0 t' \\
t & = & t'
\eea
where $V_0$ is the velocity of the primed reference frame with relative
to the laboratory frame.  Clearly, we can eliminate mean velocity of the
flow by going to the frame whose velocity is the same as mean velocity
of the fluid.  Throughout this paper we will work in this reference
frame.

As discussed above, certain symmetries like homogeneity help us in
statistical description. Formally, homogeneity \emph{}indicates that
the average properties do not vary with absolute position in a particular
direction, but depends only on the separation between points. For
example, a homogeneous two-point correlation function is \bea
\left\langle u_{i}(\mathbf{x},t)u_{j}(\mathbf{x'},t)\right\rangle =C_{ij}(\mathbf{x-x'},t)=C_{ij}(\mathbf{r},t).\eea
Similarly, \emph{s}tationarity or \emph{}steady-state implies that
average properties depend on time difference, not on the absolute
time. That is, \be
\left\langle u_{i}(\mathbf{x},t)u_{j}(\mathbf{x},t')\right\rangle =C_{ij}(\mathbf{x},t-t').\ee
 Another important symmetry is isotropy. A system is said to be isotropic
if its average properties are invariant under rotation. For isotropic
systems\be
\left\langle u_{i}(\mathbf{x},t)u_{j}(\mathbf{x'},t)\right\rangle =C_{ij}(\left|\mathbf{x-x'}\right|,t)=C_{ij}(\left|\mathbf{r}\right|,t).\ee
Isotropy reduces the number of independent correlation functions.
Batchelor~\cite{Batchelor:book:Turbulence,Lesieur:book:Turbulence,Frisch:book} showed that the isotropic two-point
correlation could be written as\be
C_{ij}(\mathbf{r})=C^{(1)}(r)r_{i}r_{j}+C^{(2)}(r)\delta_{ij}\ee
where $C^{(1)}$ and $C^{(2)}$ are even functions of $r=|\mathbf{r}|.$
Hence we have reduced the independent correlation functions to two.
For incompressible flows, there is only one independent correlation
function \cite{Batchelor:book:Turbulence,Lesieur:book:Turbulence,Frisch:book} .

In turbulent fluid, fluctuations of all scales exist. Therefore, it
is quite convenient to use Fourier basis for the representation of
turbulent fluid velocity and magnetic field. Note that in recent times
another promising basis called wavelet is becoming popular. In this
paper we focus our attention on Fourier expansion, which is the topic
of the next subsection.

\subsection{Turbulence Equations in Spectral Space}

Turbulent fluid velocity $\mathbf{u}(\mathbf{x},t)$ is represented
in Fourier space as
\begin{eqnarray}
\mathbf{u}\left(\mathbf{x},t\right) & = & \int\frac{d\mathbf{k}}{(2\pi)^{d}}\mathbf{u}\left(\mathbf{k},t\right)\exp\left(i\mathbf{k\cdot x}\right)\\
\mathbf{u(}\mathbf{k},t) & = & \int d\mathbf{x}\mathbf{u}\left(\mathbf{x},t\right)\exp\left(-i\mathbf{k\cdot x}\right)\end{eqnarray}
where $d$ is the space dimensionality. 

In Fourier space, the equation for \emph{incompressible} fluid is
\cite{Lesieur:book:Turbulence,Verma:book:ET}
\begin{eqnarray}
\left(\frac{\partial}{\partial t}+\nu k^{2}\right)u_{i}(\mathbf{k},t) 
& = & -ik_{i}p\left(\mathbf{k},t\right)
-ik_{j}\int\frac{d\mathbf{p}}{(2\pi)^{d}}u_{j}(\mathbf{k-p},t)u_{i}(\mathbf{p},t)  \label{eq:uk} \end{eqnarray}
with the following constrains \begin{eqnarray}
\mathbf{k}\cdot\mathbf{u\left(\mathbf{k}\right)} & = & 0,\end{eqnarray}
The substitution of the incompressibility condition $\mathbf{k}\cdot\mathbf{u\left(\mathbf{k}\right)}=0$
into Eq. (\ref{eq:uk}) yields the following expression for the pressure
field \be
p\left(\mathbf{k}\right)=-\frac{k_{i}k_{j}}{k^{2}}\int\frac{d\mathbf{p}}{(2\pi)^{d}}\left[u_{j}(\mathbf{k-p},t)u_{i}(\mathbf{p},t)\right].\ee
Note that the density field has been taken to be a constant, and has
been set equal to 1. 

It is also customary to write the evolution equations symmetrically
in terms of $\mathbf{p}$ and \textbf{$\mathbf{k-p}$} variables.
The symmetrized equations are

\begin{eqnarray}
\left(\frac{\partial}{\partial t}+\nu k^{2}\right)u_{i}(\mathbf{k},t) & = & -\frac{i}{2}P_{ijm}^{+}(\mathbf{k})\int\frac{d\mathbf{p}}{(2\pi)^{d}}[u_{j}(\mathbf{p},t)u_{m}(\mathbf{k-p},t) ] \label{eq:uk_P}\end{eqnarray}
where \begin{eqnarray*}
P_{ijm}^{+}(\mathbf{k}) & = & k_{j}P_{im}(\mathbf{k})+k_{m}P_{ij}(\mathbf{k});\\
P_{im}(\mathbf{k}) & = & \delta_{im}-\frac{k_{i}k_{m}}{k^{2}};\end{eqnarray*}

Energy and other second-order quantities play important roles in fluid
turbulence. For a homogeneous system these quantities are defined
as \[
\left\langle u_{i}(\mathbf{k},t)u_{j}(\mathbf{k'},t)\right\rangle =C_{ij}(\mathbf{k},t)(2\pi)^{d}\delta(\mathbf{k+k'}).\]
The spectrum is also related to correlation function in real space\[
C_{ij}(\mathbf{r})=\int\frac{d\mathbf{k}}{(2\pi)^{d}}C_{ij}\left(\mathbf{k}\right)\exp\left(i\mathbf{k\cdot r}\right).\]

For isotropic situations
we can take $C_{ij}(\mathbf{k})$ to be an isotropic tensor, and it
can be written as \cite{Batchelor:book:Turbulence}\begin{equation}
C_{ij}(\mathbf{k})=P_{ij}(\mathbf{k})C(k).\label{eq:Cij_isotropic}\end{equation}
When turbulence is isotropic, then a quantity called 1D spectrum or
reduced spectrum $E(k)$ defined below is very useful.\begin{eqnarray*}
E=\frac{1}{2}\left\langle u^{2}\right\rangle  & =\frac{1}{2} & \int\frac{d\mathbf{k}}{(2\pi)^{d}}C_{ii}\left(\mathbf{k}\right)\\
\int E(k)dk & = & \frac{1}{2}\int dk\frac{S_{d}k^{d-1}}{(2\pi)^{d}}P_{ii}(\mathbf{k})C\left(\mathbf{k}\right)\\
 & = & \int dk\frac{S_{d}k^{d-1}(d-1)}{2(2\pi)^{d}}C\left(\mathbf{k}\right),\end{eqnarray*}
where $S_{d}=2\pi^{d/2}/\Gamma{(d/2)}$ is the area of $d-$dimensional
unit sphere. Therefore,\begin{equation}
E(k)=C(\mathbf{k})k^{d-1}\frac{S_{d}(d-1)}{2(2\pi)^{d}}.\label{eq:E(k)_eq_C(k)}\end{equation}
Note that the above formula is valid only for isotropic turbulence.
We have tabulated all the important spectra of fluid turbulence in Table
\ref{table:Spectra}.  %
\begin{table}

\caption{\label{table:Spectra} Various Spectra of Fluid Turbulence }

\begin{tabular}{|c|c|c|c|}
\hline 
Quantity&
Symbol&
Derived from&
Symbol for 1D\tabularnewline
\hline
\hline 
Kinetic energy spectrum&
$C\left(\mathbf{k}\right)$&
$\left\langle u_{i}(\mathbf{k})u_{j}(\mathbf{k'})\right\rangle $&
$E$ $\left(k\right)$\tabularnewline
\hline 
Enstrophy spectrum&
$\Omega\left(\mathbf{k}\right)$&
$\left\langle \omega_{i}(\mathbf{k})\omega_{j}(\mathbf{k'})\right\rangle $&
$\Omega\left(k\right)$
\tabularnewline
\hline
\end{tabular}
\end{table}

The global quantities defined in Table \ref{table:Globals} are related
to the spectra defined in Table \ref{table:Spectra} by Perceval's
theorem \cite{Batchelor:book:Turbulence}. Since the fields are homogeneous, Fourier
integrals are not well defined. In Appendix \ref{sec:Fourier-Series}
we show that energy spectra defined using correlation functions are
still meaningful because correlation functions vanish at large distances.
We consider energy per unit volume, which are finite for homogeneous
turbulence. As an example, the kinetic energy per unit volume is related
to energy spectrum in the following manner:

\[
\frac{1}{L^{d}}\int d\mathbf{x}\frac{1}{2}\left\langle u^{2}\right\rangle =\frac{1}{2}\int\frac{d\mathbf{k}}{(2\pi)^{d}}C_{ii}(\mathbf{k})=\int E(k)dk\]
Similar identities can be written for other fields. 

In three dimensions we have another important quantities called kinetic
helicities. In Fourier space kinetic helicity $H_{K}(\mathbf{k})$
is defined using\[
\left\langle u_{i}\left(\mathbf{k},t\right)\Omega_{j}\left(\mathbf{k'},t\right)\right\rangle =P_{ij}\left(\mathbf{k}\right)H_{K}(\mathbf{k})(2\pi)^{d}\delta(\mathbf{k+k'})\]
The total kinetic helicity $H_{K}$ can be written in terms of\begin{eqnarray*}
H_{K} & = & \frac{1}{2}\left\langle \mathbf{u}(\mathbf{x})\cdot\mathbf{\Omega}(\mathbf{x})\right\rangle \\
 & =\frac{1}{2} & \int\frac{d\mathbf{k}}{(2\pi)^{d}}\frac{d\mathbf{k'}}{(2\pi)^{d}}\left\langle \mathbf{u}(\mathbf{k})\cdot\mathbf{\Omega}(\mathbf{k'})\right\rangle \\
 & = & \int\frac{d\mathbf{k}}{(2\pi)^{d}}H_{K}(\mathbf{k})\\
 & = & \int dkH_{K}(k)\end{eqnarray*}
Therefore, one dimensional magnetic helicity $H_{M}$ is\[
H_{K}(k)=\frac{4\pi k^{2}}{(2\pi)^{3}}H_{K}(\mathbf{k}).\]
Using the definition $\mathbf{\Omega}(\mathbf{k})=i\mathbf{k}\times\mathbf{u}(\mathbf{k})$,
we obtain\[
\left\langle u_{i}\left(\mathbf{k},t\right)u_{j}\left(\mathbf{k'},t\right)\right\rangle =\left[P_{ij}\left(\mathbf{k}\right)C^{uu}(\mathbf{k})-i\epsilon_{ijl}k_{l}\frac{H_{K}(\mathbf{k})}{k^{2}}\right](2\pi)^{d}\delta(\mathbf{k+k'}).\]
Note that the magnetic helicity breaks mirror symmetry.

We can Fourier transform in time as well using \begin{eqnarray*}
\mathbf{u}\left(\mathbf{x},t\right) & = & \int d\hat{{k}}\mathbf{u}\left(\mathbf{k},\omega\right)\exp\left(i\mathbf{k\cdot x}-i\omega t\right)\\
\mathbf{u(}\mathbf{k},\omega) & = & \int d\mathbf{x}dt\mathbf{u}\left(\mathbf{x},t\right)\exp\left(-i\mathbf{k\cdot x}+i\omega t\right)\end{eqnarray*}
where $d\hat{{k}}=d\mathbf{k}d\omega/(2\pi)^{d+1}$. The resulting
fluid equations in $\hat{{k}}=(\mathbf{k},\omega)$ space are\begin{eqnarray}
\left(-i\omega+\nu k^{2}\right)u_{i}(\hat{{k}}) & = & -\frac{i}{2}P_{ijm}^{+}({\textbf{k}})\int_{\hat{{p}}+\hat{{q}}=\hat{{k}}}d\hat{{p}}\left[u_{j}(\hat{{p}})u_{m}(\hat{{q}})\right],\label{eq:MHDukw}\end{eqnarray}

After we have introduced the energy spectra and other second-order
correlation functions, we move on to discuss their evolution.

\subsection{Energy Equations }

The energy equation for general (compressible) Navier-Stokes is quite
complex. However, incompressible Navier-Stokes equation is
relatively simpler, and is discussed below.

From the evolution equations of fields, we can derive the following
spectral evolution equations for incompressible MHD\begin{eqnarray}
\left(\frac{\partial}{\partial t}+2\nu k^{2}\right)C\left(\mathbf{k},t\right) & = & \frac{2}{\left(d-1\right)\delta\left(\mathbf{k+k'}\right)}\int_{\mathbf{k'+p+q=0}}\frac{d\mathbf{p}}{(2\pi)^{2d}}[-\Im\left\langle \left(\mathbf{k'}\cdot\mathbf{u(q)}\right)\left(\mathbf{u(p)}\cdot\mathbf{u(k')}\right)\right\rangle 
\label{eq:Cuu(k)vst}
\end{eqnarray}
where $\Im$ stands for the imaginary part. Note that $\mathbf{k'+p+q=0}$
and $\mathbf{k'=-k}$. In Eq. (\ref{eq:Cuu(k)vst}) the first term
in the RHS provides the energy transfer from the velocity modes to
$\mathbf{u(k)}$ mode. Note that pressure couples with compressible
modes, hence it is absent in the incompressible equations. 

In a finite box, using $\left\langle \left|\mathbf{u}(\mathbf{k})\right|^{2}\right\rangle =C(\mathbf{k})/((d-1)L^{d})$,
and $\delta(\mathbf{k})(2\pi)^{d}=L^{d}$ (see Appendix A), we can
show that\begin{eqnarray*}
\left(\frac{\partial}{\partial t}+2\nu k^{2}\right)\frac{1}{2}\left\langle \left|\mathbf{u}(\mathbf{k})\right|^{2}\right\rangle  & = & \sum_{\mathbf{k}'+\mathbf{p}+\mathbf{q}=0}[-\Im\left\langle \left(\mathbf{k'}\cdot\mathbf{u(q)}\right)\left(\mathbf{u(p)}\cdot\mathbf{u(k')}\right)\right\rangle ],\end{eqnarray*}

Many important quantities, e.g. energy fluxes, can be derived from
the energy equations. We will discuss these quantities in the next
section.

\section{Mode-to-mode Energy Transfers and Fluxes in MHD Turbulence \label{sec:Mode-to-mode-Energy-Transfer}}

In turbulence energy exchange takes place between various Fourier
modes because of nonlinear interactions. Basic interactions in turbulence
involves a wavenumber triad $(\mathbf{k',p,q)}$ satisfying $\mathbf{k'+p+q=0}$.
Usually, energy gained by a mode in the triad is computed using the
\emph{combined energy transfer} from the other two modes \cite{Leslie:book}.
Dar et al. \cite{Dar:PD2001} devised a new scheme to compute
the energy transfer rate between two modes in a triad\emph{,} and
called it \emph{{}``mode-to-mode energy transfer}''~\cite{Verma:PR2004}. They computed
energy cascade rates and energy transfer rates between two wavenumber
shells using this scheme. We will review these ideas in this section.
Note that we are considering only the interactions of incompressible
modes.

\subsection{{}``Mode-to-Mode'' Energy Transfer in Fluid Turbulence \label{sub:'Mode-to-Mode'-fluid}}

In this subsection we discuss evolution of energy in turbulent fluid
\emph{in a periodic bo}x. 
We consider ideal case where viscous dissipation is zero $(\nu=0).$
The equations are \cite{Kraichnan:JFM1959,	Leslie:book}
\begin{eqnarray}
\frac{\partial}{\partial t}\frac{1}{2}\left|u\left(\mathbf{k'}\right)\right|^{2} & =\sum_{\mathbf{k'+p+q}=0} & -\frac{1}{2}\Im\left[\left(\mathbf{k'\cdot u(q)}\right)\left(\mathbf{u(k')}\cdot\mathbf{u(p)}\right)+\left(\mathbf{k'\cdot u(p)}\right)\left(\mathbf{u(k')}\cdot\mathbf{u(q)}\right)\right],\end{eqnarray}
where $\Im$ denotes the imaginary part. Note that the pressure does
not appear in the energy equation because of the incompressibility
condition.

Consider a case in which only three modes $\mathbf{u(k'),u(p),u(q)}$,
and their conjugates are nonzero. Then the above equation yields \begin{eqnarray}
\frac{\partial}{\partial t}\frac{1}{2}\left|u\left(\mathbf{k'}\right)\right|^{2} & = & \frac{1}{2}S^{uu}(\mathbf{k'|p,q)},\label{eq:Fluid_triad}\end{eqnarray}
where \begin{eqnarray}
S(\mathbf{k'|p,q)} & = & -\Im\left[\left(\mathbf{k'\cdot u(q)}\right)\left(\mathbf{u(k')}\cdot\mathbf{u(p)}\right)+\left(\mathbf{k'\cdot u(p)}\right)\left(\mathbf{u(k')}\cdot\mathbf{u(q)}\right)\right].\label{eq:Fluid_Sk,pq}\end{eqnarray}
Lesieur and other researchers physically interpreted $S(\mathbf{k'|p,q)}$
as the \emph{combined energy transfer rate} from modes $\mathbf{p}$
and $\mathbf{q}$ to mode $\mathbf{k'}$. The evolution equations
for $\left|u\left(\mathbf{p}\right)\right|^{2}$ and $\left|u\left(\mathbf{q}\right)\right|^{2}$
are similar to that for $\left|u\left(\mathbf{k'}\right)\right|^{2}$.
By adding the energy equations for all three modes, we obtain

\begin{eqnarray}
\frac{\partial}{\partial t}\left[\left|u\left(\mathbf{k'}\right)\right|^{2}+\left|u\left(\mathbf{p}\right)\right|^{2}+\left|u\left(\mathbf{q}\right)\right|^{2}\right]/2 & = & S^{uu}(\mathbf{k'|p,q)}+S^{uu}(\mathbf{p|q,k')}+S^{uu}(\mathbf{q|k',p)}\\
 & = & \Im[\left(\mathbf{q\cdot u(q)}\right)\left(\mathbf{u(k')}\cdot\mathbf{u(p)}\right)\\
 &  & +\left(\mathbf{p\cdot u(p)}\right)\left(\mathbf{u(k')}\cdot\mathbf{u(q)}\right)\\
 &  & +\left(\mathbf{k'\cdot u(k')}\right)\left(\mathbf{u(p)}\cdot\mathbf{u(q)}\right)]\end{eqnarray}
For incompressible fluid the right-hand-side is identically zero because
$\mathbf{k'\cdot u(k')}=0$. Hence the energy in each interacting
triad is conserved , i.e., \be
\left|u\left(\mathbf{k'}\right)\right|^{2}+\left|u\left(\mathbf{p}\right)\right|^{2}+\left|u\left(\mathbf{q}\right)\right|^{2}=const.\ee

The question is whether we can derive an expression for mode-to-mode
energy transfer rates from mode $\mathbf{p}$ to mode \emph{$\mathbf{k'}$,}
and from mode \emph{$\mathbf{q}$} to mode \emph{$\mathbf{k'}$} separately.
Dar et al. \cite{Dar:PD2001} showed that it is meaningful to talk about
energy transfer rate between two modes. They derived an expression
for the mode-to-mode energy transfer, and showed it to be unique apart
from an irrelevant arbitrary constant. They referred to this quantity
as {}``mode-to-mode energy transfer''. Even though they talk about
mode-to-mode transfer, they are still within the framework of triad
interaction, i.e., a triad is still the fundamental entity of interaction.

\subsubsection{Definition of Mode-to-Mode Transfer in a Triad}

Consider a triad ($\mathbf{k',p,q}$). Let the quantity $R^{uu}(\mathbf{k'|p|q})$
denote the energy transferred from mode \textbf{$\mathbf{p}$} to
mode \textbf{$\mathbf{k'}$} with mode \textbf{$\mathbf{q}$} playing
the role of a mediator. We wish to obtain an expression for $R$.

The $R$'s should satisfy the following relationships : 

\begin{enumerate}
\item The sum of energy transfer from mode \textbf{$\mathbf{p}$} to mode
\textbf{$\mathbf{k'}$ $(R^{uu}(\mathbf{k'|p|q})$),} and from mode
\textbf{$\mathbf{q}$} to mode \textbf{$\mathbf{k'}$} $(R^{uu}(\mathbf{k'|p|q}))$
should be equal to the total energy transferred to mode \textbf{$\mathbf{k'}$}
from modes \textbf{$\mathbf{p}$} and $\mathbf{q}$, i.e., $S^{uu}(\mathbf{k'|p,q})$
{[}see Eq.~(\ref{eq:Fluid_triad}){]}. That is, \begin{equation}
R^{uu}(\mathbf{k'|p|q})+R^{uu}(\mathbf{k'|q|p})=S^{uu}(\mathbf{k'|p,q}),\label{eq:Skpq}\end{equation}
\begin{equation}
R^{uu}(\mathbf{p|k'|q})+R^{uu}(\mathbf{p|q|k'})=S^{uu}(\mathbf{p|k',q}),\end{equation}
\begin{equation}
R^{uu}(\mathbf{q|k'|p})+R^{uu}(\mathbf{q|p|k'})=S^{uu}(\mathbf{q|k',p}).\end{equation}

\item By definition, the energy transferred from mode \textbf{$\mathbf{p}$}
to mode \textbf{$\mathbf{k}'$}, $R^{uu}(\mathbf{k'|p|q})$, will
be equal and opposite to the energy transferred from mode $\mathbf{k'}$
to mode $\mathbf{p}$, $R^{uu}(\mathbf{p|k'|q})$. Thus, \begin{equation}
R^{uu}(\mathbf{k'|p|q})+R^{uu}(\mathbf{p|k'|q})=0,\label{eq:Rkpq}\end{equation}
\begin{equation}
R^{uu}(\mathbf{k'|q|p})+R^{uu}(\mathbf{q|k'|p})=0,\label{eq:Rkqp}\end{equation}
\begin{equation}
R^{uu}(\mathbf{p|q|k'})+R^{uu}(\mathbf{q|p|k'})=0.\label{eq:Rpqk}\end{equation}

\end{enumerate}
These are six equations with six unknowns. However, the value of the
determinant formed from the Eqs.~(\ref{eq:Skpq}-\ref{eq:Rpqk})
is zero. Therefore we cannot find unique $R$'s given just these equations.
In the following discussion we will study the set of solutions of
the above equations.

\subsubsection{Solutions of equations of mode-to-mode transfer \label{sub:Solutions-mode-to-mode}}

Consider a function\begin{equation}
S^{uu}(\mathbf{k'|p|q})=-\Im\left(\left[\mathbf{k'}\cdot\mathbf{u}(\mathbf{q})\right]\left[\mathbf{u}(\mathbf{k}')\cdot\mathbf{u}(\mathbf{p})\right]\right)\label{eq:Suu_Dar}\end{equation}
Note that $S^{uu}(\mathbf{k'|p|q})$ is altogether different function
compared to $S(\mathbf{k'|p,q})$. In the expression for $S^{uu}(\mathbf{k'|p|q})$,
the field variables with the first and second arguments are dotted
together, while the field variable with the third argument is dotted
with the first argument. 

It is very easy to check that $S^{uu}(\mathbf{k'|p|q})$ satisfy the
Eqs. (\ref{eq:Skpq}-\ref{eq:Rpqk}). Note that $S^{uu}(\mathbf{k'|p|q})$
satisfy the Eqs. (\ref{eq:Rkpq}-\ref{eq:Rpqk}) because of incompressibility
condition. The above results implies that the set of $S^{uu}(||)$'s
is \textit{one instance} of the $R^{uu}(||)$'s. However, $S^{uu}(\mathbf{k'|p|q})$
is not a unique solution. If another solution $R^{uu}(\mathbf{k'|p|q})$
differs from $S(\mathbf{k'|p|q})$ by an arbitrary function $X_{\Delta}$,
i.e., $R^{uu}(\mathbf{k'|p|q})=S^{uu}(\mathbf{k'|p|q})+X_{\Delta}$,
then by inspection we can easily see that the solution of Eqs.~(\ref{eq:Skpq}-\ref{eq:Rpqk})
must be of the form \begin{equation}
R^{uu}(\mathbf{k'|p|q})=S^{uu}(\mathbf{k'|p|q})+X_{\Delta}\end{equation}
\begin{equation}
R^{uu}(\mathbf{k'|q|p})=S^{uu}(\mathbf{k'|q|p})-X_{\Delta}\end{equation}
\begin{equation}
R^{uu}(\mathbf{p|k'|q})=S^{uu}(\mathbf{p|k'|q})-X_{\Delta}\end{equation}
\begin{equation}
R^{uu}(\mathbf{p|q|k'})=S^{uu}(\mathbf{p|q|k'})+X_{\Delta}\end{equation}
\begin{equation}
R^{uu}(\mathbf{q|k'|p})=S^{uu}(\mathbf{q|k'|p})+X_{\Delta}\end{equation}
\begin{equation}
R^{uu}(\mathbf{q|p|k'})=S^{uu}(\mathbf{q|p|k'})-X_{\Delta}\end{equation}
 Hence, the solution differs from $S^{uu}(\mathbf{k'|p|q})$ by a
constant. %

\begin{figure}
\includegraphics{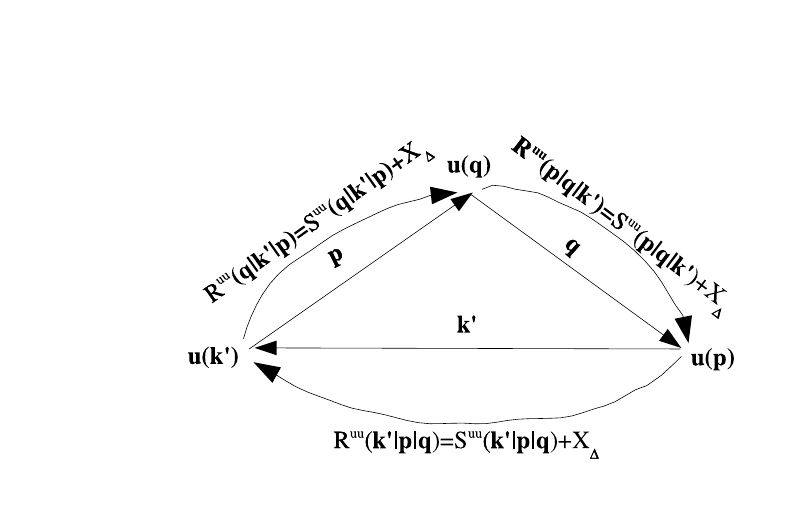}

\caption{\label{Fig:fluid-mode} Mode-to-mode energy transfer in fluid turbulence.
$S^{uu}({\textbf{k'|p|q}})$ represents energy transfer rate from
mode $\mathbf{u}(\mathbf{p})$ to mode $\mathbf{u}(\mathbf{k'})$
with the mediation of mode $\mathbf{u}(\mathbf{q})$. $X_{\Delta}$
is the arbitrary circulating transfer.}
\end{figure}

See Fig. \ref{Fig:fluid-mode} for illustration. A careful observation
of the figure indicates that the quantity $X_{\Delta}$ flows along
${\textbf{p}}\rightarrow{\mathbf{k'}}\rightarrow{\textbf{q}}\rightarrow{\textbf{p}}$,
circulating around the entire triad without changing the energy of
any of the modes. Therefore we will call it the \textit{Circulating
transfer}. Of the total energy transfer between two modes, $S^{uu}+X_{\Delta}$,
only $S^{uu}$ can bring about a change in modal energy. $X_{\Delta}$
transferred from mode \textbf{p} to mode \textbf{$\mathbf{k'}$} is
transferred back to mode \textbf{p} via mode \textbf{q}. Thus the
energy that is effectively transferred from mode \textbf{p} to mode
\textbf{$\mathbf{k'}$} is just $S^{uu}(\mathbf{k'|p|q})$. Therefore
$S^{uu}(\mathbf{k'|p|q})$ can be termed as the \textit{effective
mode-to-mode energy transfer} from mode \textbf{p} to mode $\mathbf{k'}$. 

Note that $X_{\Delta}$ cannot be calculated even by simulation or experiment,
because we can experimentally compute only the energy transfer rate
to a mode, which is a sum of two mode-to-mode energy transfers. The
mode-to-mode energy transfer rate is really an abstract quantity,
somewhat similar to {}``gauges'' in electrodynamics.  Recently, using symmetry arguments,~\cite{Verma:book:ET} showed that $X_{\Delta} = 0$. 

The terms $u_{j}\partial_{j}u_{i}$ and $u_{i}u_{j}\partial_{j}u_{i}$
are nonlinear terms in the Navier-Stokes equation and the energy equation
respectively. When we look at the formula (\ref{eq:Suu_Dar}) carefully,
we find that the $u_{j}(\mathbf{q})$ term is contracted with $k_{j}$
in the formula. Hence, $u_{j}$ field is the mediator in the energy
exchange between first $(u_{i})$ and third field $(u_{i})$ of $u_{i}u_{j}\partial_{j}u_{i}$.

In this following discussion we will compute the energy fluxes and
the shell-to-shell energy transfer rates using $S^{uu}(\mathbf{k'|p|q})$.

\subsection{Shell-to-Shell Energy Transfer in Fluid Turbulence Using Mode-to-mode
Formalism \label{sub:Shell-to-Shell-fluid}}

In turbulence energy transfer takes place from one region of the wavenumber
space to another region. Domaradzki and Rogallo \cite{Domaradzki:PF1990}
have discussed the energy transfer between two shells using the combined
energy transfer $S^{uu}({\textbf{k'|p,q}})$. They interpret the quantity
\begin{equation}
T_{nm}^{uu}=\frac{1}{2}\sum_{\mathbf{k'}\in n}\sum_{\mathbf{p}\in m}S^{uu}(\mathbf{k'|p,q}).\label{eq:shell_old_defn}\end{equation}
 as the rate of energy transfer from shell \textit{$m$} to shell
\textit{$n$} . Note that \textbf{$\mathbf{k'}$}-sum is over shell
$n$, $\mathbf{p}$-sum over shell $m$, and $\mathbf{k'+p+q}=0$.
However, Domaradzki and Rogallo \cite{Domaradzki:PF1990} themselves points
out that it may not be entirely correct to interpret the formula (\ref{eq:shell_old_defn})
as the shell-to-shell energy transfer. The reason for this is as follows.

\begin{figure}
\includegraphics[scale=0.7]{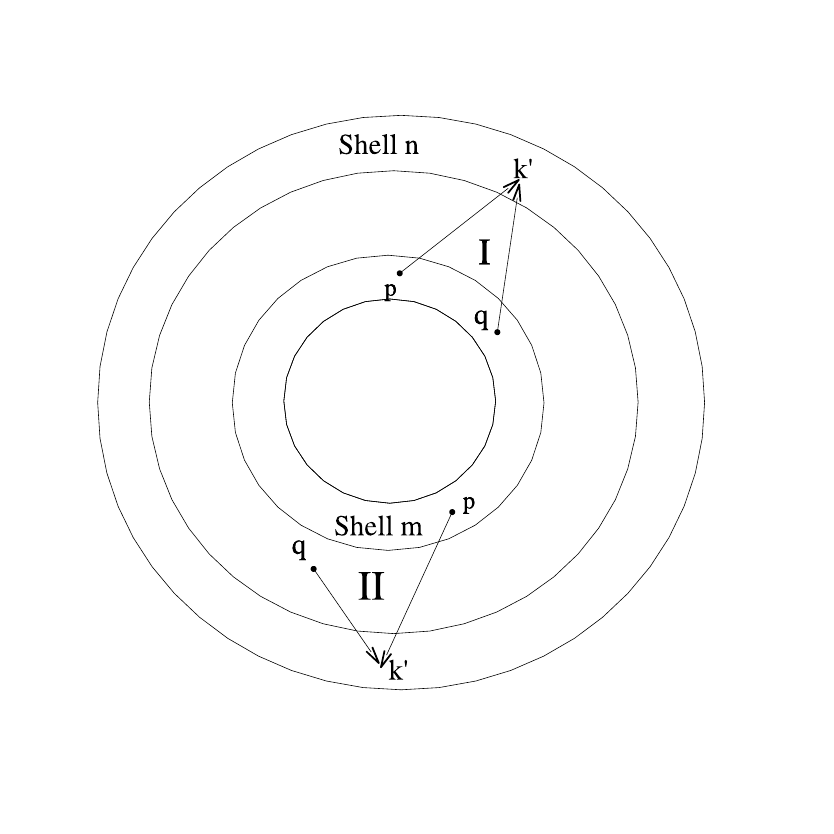}
\caption{\label{Fig:shell-to-shell} Shell-to-shell energy transfer from wavenumber-shell
$m$ to wavenumber-shell $n$. The triads involved in this process
fall in two categories: Type I, where both $\mathbf{p}$ and $\mathbf{q}$
are inside shell $m$, and Type II, where only $\mathbf{p}$ is inside
shell $m$. }
\end{figure}

In the energy transfer between two shells \textit{m} and \textit{n},
two types of wavenumber triads are involved, as shown in Fig. \ref{Fig:shell-to-shell}.
The real energy transfer from the shell \textit{$m$} to the shell
\textit{$n$} takes place through both $\mathbf{k'}$-$\mathbf{p}$
and $\mathbf{k'}$-$\mathbf{q}$ legs of triad I, but only through
$\mathbf{k'}$-$\mathbf{p}$ leg of triad II. But in Eq. (\ref{eq:shell_old_defn})
summation erroneously includes $\mathbf{k'}$-$\mathbf{q}$ leg of
triad II also along with the three legs given above. Hence  Domaradzki
and Ragallo's formalism \cite{Domaradzki:PF1990} do not yield totally correct
shell-to-shell energy transfers, as was pointed out by Domaradzki
and Rogallo themselves. We will show below how Dar et al.'s formalism
\cite{Dar:PD2001} overcomes this difficulty.

By definition of the the mode-to-mode transfer function $R^{uu}(\mathbf{k'|p|q})$,
the energy transfer from shell \textit{$m$} to shell \textit{$n$}
can be defined as \begin{equation}
T_{nm}^{uu}=\sum_{\mathbf{k'}\in n}\sum_{\mathbf{p}\in m}R^{uu}(\mathbf{k'|p|q})\label{eq:shell_new_defn}\end{equation}
 where the $\mathbf{k'}$-sum is over the shell $n$, and $\mathbf{p}$-sum
is over the shell $m$. The quantity $R^{uu}$ can be written as a
sum of an effective transfer $S^{uu}(\mathbf{k'|p|q})$ and a circulating
transfer $X_{\Delta}$. As discussed in the last section, the circulating
transfer $X_{\Delta}$ does not contribute to the energy change of
modes. From Figs. \ref{Fig:fluid-mode} and \ref{Fig:shell-to-shell}
we can see that $X_{\Delta}$ flows from the shell \textit{$m$} to
the shell \textit{$n$} and then flows back to \textit{$m$} indirectly
through the mode $\mathbf{q}$. Therefore the \textit{effective} energy
transfer from the shell \textit{m} to the shell \textit{n} is just
$S^{uu}(\mathbf{k'|p|q})$ summed over all the \textbf{$\mathbf{k'}$-}modes
in the shell \textit{$n$} and all the \textbf{$\mathbf{p}$-}modes
in the shell $m$, i.e., \begin{equation}
T_{nm}^{uu}=\sum_{\mathbf{k'}\in n}\sum_{\mathbf{p}\in m}S^{uu}(\mathbf{k'|p|q}).\label{eq:shell_eff_defn}\end{equation}
 Clearly, the energy transfer through $\mathbf{k'-q}$ of the triad
II of Fig. \ref{Fig:shell-to-shell} is not present in $T_{nm}^{uu}$
in Dar et al.'s formalism because $\mathbf{q}\notin m$. Hence, the
formalism of the mode-to-mode energy transfer rates provides us a
correct and convenient method to compute the shell-to-shell energy
transfer rates in fluid turbulence.

\subsection{Energy Cascade Rates in Fluid Turbulence Using Mode-to-mode Formalism
\label{sub:Energy-Cascade-Rates-fluid}}

The kinetic energy cascade rate (or flux) $\Pi$ in fluid turbulence
is defined as the rate of loss of kinetic energy by the modes inside
a sphere to the modes outside the sphere. Let $k_{0}$ be the radius
of the sphere under consideration. Kraichnan \cite{Kraichnan:JFM1959}, Leslie
\cite{Leslie:book} and others have computed the energy flux in fluid
turbulence using $S^{uu}(\mathbf{k'|p,q})$\begin{equation}
\Pi(k_{0})=-\sum_{|\mathbf{k}|<k_{0}}\sum_{|\mathbf{p}|>k_{0}}\frac{1}{2}S^{uu}(\mathbf{k'|p,q}).\label{eq:flux_Krai59}\end{equation}
 Although the energy cascade rate in fluid turbulence can be found
by the above formula, the mode-to-mode approach of Dar et al. \cite{Dar:PD2001}
provides a more natural way of looking at the energy flux. Since $R^{uu}(\mathbf{k'|p|q})$
represents energy transfer from $\mathbf{p}$ to $\mathbf{k'}$ with
\textbf{$\mathbf{q}$} as a mediator, we may alternatively write the
energy flux as \begin{equation}
\Pi(k_{0})=\sum_{|\mathbf{k'}|>k_{0}}\sum_{|\mathbf{p}|<k_{0}}R^{uu}(\mathbf{k'|p|q}).\label{eq:fluid_flux_Dar_R}\end{equation}
 However, $R^{uu}(\mathbf{k'|p|q})=S^{uu}(\mathbf{k'|p|q})+X_{\Delta}$,
and the circulating transfer $X_{\Delta}$ makes no contribution to
the energy flux from the sphere because the energy lost from the
sphere through $X_{\Delta}$ returns to the sphere. Hence,\begin{equation}
\Pi(k_{0})=\sum_{|\mathbf{k}'|>k_{0}}\sum_{|\mathbf{p}|<k_{0}}S^{uu}(\mathbf{k'|p|q}).\label{eq:fluid_flux_Dar}\end{equation}
 Both the formulas given above, Eqs. (\ref{eq:flux_Krai59}) and (\ref{eq:fluid_flux_Dar}),
are equivalent as shown by Dar et al. \cite{Dar:thesis}.

Frisch \cite{Frisch:book} has derived a formula for energy flux as\be
\Pi(k_{0})=\left\langle \mathbf{u_{k_{0}}^{<}}\cdot\left(\mathbf{u_{k_{0}}^{<}}\cdot\nabla\mathbf{u_{k_{0}}^{>}}\right)\right\rangle +\left\langle \mathbf{u_{k_{0}}^{<}}\cdot\left(\mathbf{u_{k_{0}}^{>}}\cdot\nabla\mathbf{u_{k_{0}}^{>}}\right)\right\rangle .\ee
It is easy to see that the above formula is consistent with mode-to-mode
formalism. As discussed in the Subsection \ref{sub:Solutions-mode-to-mode},
the second field of both the terms are mediators in the energy transfer.
Hence in mode-to-mode formalism, the above formula will translate
to \[
\Pi(k_{0})=\sum_{k>k_{0}}\sum_{p<k_{0}}-\Im\left[\left(\mathbf{k'}\cdot\mathbf{u^{<}}(\mathbf{q})\right)\left(\mathbf{u^{<}}(\mathbf{p})\cdot\mathbf{u^{>}}(\mathbf{k'})\right)+\left(\mathbf{k}'\cdot\mathbf{u^{>}}(\mathbf{q})\right)\left(\mathbf{u^{<}}(\mathbf{p})\cdot\mathbf{u^{>}}(\mathbf{k'})\right)\right],\]
which is same as mode-to-mode formula (\ref{eq:fluid_flux_Dar}) of
Dar et al. \cite{Dar:PD2001}.

The above quantities are computed numerically or theoretically.

\subsection{Digression to Infinite Box \label{sub:Digression-to-Infinite-box:flux}}

In the above discussion we assumed that the fluid is contained in
a finite volume. In simulations, box size is typically taken to $2\pi$.
However, most analytic calculations assume infinite box. It is quite
easy to transform the equations given above to those for infinite
box using the method described in Appendix. Here, the evolution of
energy spectrum is given by (see Section \ref{sec:Definitions})\begin{eqnarray}
\left(\frac{\partial}{\partial t}+2\nu k^{2}\right)C\left(\mathbf{k},t\right) & = & \frac{2}{\left(d-1\right)\delta\left(\mathbf{k+k'}\right)}\int_{\mathbf{k'+p+q=0}}\frac{d\mathbf{p}}{(2\pi)^{2d}}
 \left[S^{uu}(\mathbf{k'}|\mathbf{p}|\mathbf{q})\right]\label{eq:Cuu(k)-evolve}\end{eqnarray}
The shell-to-shell energy transfer rate $T_{nm}$ from the $m$-th
shell to the $n$-th shell is\begin{equation}
T_{nm}=\frac{1}{(2\pi)^{d}\delta(\mathbf{k'+p+q)}}\int_{k'\in n}\frac{d\mathbf{k'}}{(2\pi)^{d}}\int_{p\in m}\frac{d\mathbf{p}}{(2\pi)^{d}}\left\langle S^{uu}(\mathbf{k'|p|q})\right\rangle ,\label{eq:T_MHD_FT}\end{equation}
In terms of Fourier transform, the energy cascade rate from a sphere
of radius $k_{0}$is \begin{equation}
\Pi(k_0)=\frac{1}{(2\pi)^{d}\delta(\mathbf{k'+p+q)}}\int_{k>k_{0}}\frac{d\mathbf{k'}}{(2\pi)^{d}}\int_{p<p_{0}}\frac{d\mathbf{p}}{(2\pi)^{d}}\left\langle S^{uu}(\mathbf{k'|p|q})\right\rangle .\label{eq:Pi_MHD_FT}\end{equation}

For isotropic flows, after some manipulation and using Eq. (\ref{eq:E(k)_eq_C(k)}),
we obtain \cite{Leslie:book}\begin{equation}
\left(\frac{\partial}{\partial t}+2\nu k^{2}\right)E(k,t)=T(k,t),\label{eq:dtE(k,t)_eq_T(k,t)}\end{equation}
where $T(k,t)$, called \emph{transfer function}, can be written in
terms of $S^{YX}(\mathbf{k'}|\mathbf{p}|\mathbf{q})$. The above formulas
will be used in analytic calculations.

The mode-to-mode formalism discussed here is quite general, and it
can be applied to scalar turbulence \cite{Verma:IJMPB2001}, MHD turbulence,
Rayleigh-B\'{e}nard convection, enstrophy, Electron MHD etc. Some of these
issues are discussed in Appendices C and D. One key assumption however is
incompressibility.  In the next section we will discuss various turbulence
phenomenologies and models of fluid turbulence.

\section{Turbulence Phenomenological Models}
 \label{sec:Turbulence-Models}

In the last two sections we introduced Navier-Stokes equation, and
spectral quantities like the energy spectra and fluxes. These quantities
have been analyzed using (a) phenomenological (b) numerical (c) analytical
(d) experimental methods. In the present section we will present the
most important phenomenological model called Kolmogorov's phenomenology
of turbulence.

\subsection{Kolmogorov's 1941 Theory for Fluid Turbulence \label{sub:Kolmogorov's-1941-Theory}}

For homogeneous, isotropic, incompressible, and steady fluid turbulence
with vanishing viscosity (large $Re$), Kolmogorov \cite{Kolmogorov:DANS1941Degeneration,Kolmogorov:DANS1941Dissipation,Kolmogorov:DANS1941Structure}
derived an exact relation that 

\begin{equation}
\left\langle \left(\bigtriangleup u\right)_{\parallel}^{3}\right\rangle =-\frac{4}{5}\epsilon l\label{eq:K41}\end{equation}
where $(\bigtriangleup u)_{||}$ is component of $\mathbf{u}(\mathbf{x}+\mathbf{l})-\mathbf{u}(\mathbf{x})$
along $\mathbf{l}$, $\epsilon$ is the dissipation rate, and $l$
lies between forcing scale $(L)$ and dissipative scales $(l_{d})$,
i.e., $l_{d}\ll l\ll L$. This intermediate range of scales is called
inertial range. Note that the above relationship is universal, which
holds independent of forcing and dissipative mechanisms, properties
of fluid (viscosity), and initial conditions. Therefore it finds applications
in wide spectrum of phenomena, e. g., atmosphere, ocean, channels,
pipes, and astrophysical objects like stars, accretion disks etc. 

More popular than Eq. (\ref{eq:K41}) is its equivalent statement
on energy spectrum. If we assume $\bigtriangleup u$ to be fractal,
and $\epsilon$ to be independent of scale, then\be
\left\langle \left(\bigtriangleup u\right)^{2}\right\rangle \propto\epsilon^{2/3}l^{2/3}\ee
Fourier transform of the above equation yields\begin{equation}
E(k)=K_{Ko}\epsilon^{2/3}k^{-5/3}\label{eq:Kolm_fluid}\end{equation}
where $K_{Ko}$ is a universal constant, commonly known as Kolmogorov's
constant. Eq. (\ref{eq:Kolm_fluid}) has been supported by numerous experiments
and numerical simulations. Kolmogorov's constant $K_{Ko}$ has been
found to lie between 1.4-1.6 or so. It is quite amazing that complex
interactions among fluid eddies in various different situations can
be quite well approximated by Eq. (\ref{eq:Kolm_fluid}). 

Kolmogorov's derivation of Eq. (\ref{eq:K41}) is quite involved.
However, Eqs. (\ref{eq:K41}, \ref{eq:Kolm_fluid}) can be derived
using scaling arguments (dimensional analysis) under the assumption
that 

\begin{enumerate}
\item The energy spectrum in the inertial range does not depend on the large-scaling
forcing processes and the small-scale dissipative processes, hence
it must be a power law in the local wavenumber.
\item The energy transfer in fluid turbulence is local in the wavenumber
space. The energy supplied to the fluid at the forcing scale cascades
to smaller scales, and so on. Under steady-state the energy cascade
rate is constant in the wavenumber space, i. e., $\Pi(k)=constant=\epsilon$. 
\end{enumerate}

In the framework of Kolmogorov's theory, several interesting deductions
can be made.

\begin{enumerate}
\item Kolmogorov's theory assumes homogeneity and isotropy. In real flows,
large-scales (forcing) as well as dissipative scales do not satisfy
these properties. However, experiments and numerical simulations show
that in the inertial range ($l_{d}\ll l\ll L$), the fluid flows are
typically homogeneous and isotropic.
\item The velocity fluctuations at any scale $l$ goes as\be
u_{l}\approx\epsilon^{1/3}l^{1/3}.\ee
Therefore, the effective time-scale for the interaction among eddies
of size $l$ is \be
\tau_{l}\approx\frac{l}{u_{l}}\approx\epsilon^{-1/3}l^{2/3}.\ee

\item An extrapolation of Kolmogorov's scaling to the forcing and the dissipative
scales yields\be
\epsilon\approx\frac{u_{L}^{3}}{L}\approx\frac{u_{l_{d}}^{3}}{l_{d}}.\ee
Taking $\nu\approx u_{l_{d}}l_{d}$, one gets \be
l_{d}\approx\left(\frac{\nu^{3}}{\epsilon}\right)^{1/4}.\ee
Note that the dissipation scale, also known as Kolmogorov's scale,
depends on the large-scale quantity $\epsilon$ apart from kinematic
viscosity.
\item From the definition of Reynolds number\be
Re=\frac{U_{L}L}{\nu}\approx\frac{U_{L}L}{u_{l_{d}}l_{d}}\approx\left(\frac{L}{l_{d}}\right)^{4/3}\ee
Therefore, \be
\frac{L}{l_{d}}\approx Re^{3/4}.\ee
Onset of turbulence depends on geometry, initial conditions, noise
etc. Still, in most experiments turbulences sets in after $Re$ of
2000 or more. Therefore, in three dimensions, number of active modes
$(L/l_{d})^{3}$ is larger than 26 million. These large number of
modes make the problem quite complex and intractable.
\item Space dimension does not appear in the scaling arguments. Hence, one
may expect Kolmogorov's scaling to hold in all dimensions. It is however
found that the above scaling law is applicable in three dimension
only. In two dimension (2D), conservation of enstrophy changes the
behaviour significantly (see next two sections).
The solution for one-dimensional incompressible Navier-Stokes is $\mathbf{u}(\mathbf{x},t)=const$,
which is a trivial solution.
\item Mode-to-mode energy transfer term $S(k|p|q)$ measures the strength
of nonlinear interaction. Kolmogorov's theory implicitly assumes that
energy cascades from larger to smaller scales. It is called local
energy transfer in Fourier space. These issues will be discussed in
Section \ref{sec:analytic-energy}.
\item Careful experiments show that the spectral index is close to 1.71
instead of 1.67. This correction of $\approx0.04$ is universal and
is due to the small-scale structures. This phenomena is known as intermittency,
and will be discussed in Section \ref{sec:Intermittency}.
\item Kolmogorov's model for turbulence works only for incompressible flow.
It is connected to the fact that incompressible flow has local energy
transfer in wavenumber space. Note that Burgers equation, which represents
compressible flow $(U\gg C_{s})$, has $k^{-2}$ energy spectrum,
very different from Kolmogorov's spectrum.
\end{enumerate}
Kolmogorov's theory of turbulence had a major impact on turbulence
research because of its universality. Properties of scalar, MHD, Burgers,
Electron MHD, wave turbulence have been studied using similar arguments.

As discussed in earlier sections, apart from energy spectra, there
are many other quantities of interest in turbulence. Some of them
are kinetic helicity, enstrophy etc. The statistical properties of
these quantities are quite interesting, and they are addressed using
Absolute Equilibrium State discussed below.

\subsection{Absolute Equilibrium States \label{sub:Absolute-Equilibrium-States}}

In fluid turbulence when viscosity is identically zero (inviscid limit),
kinetic energy is conserved in the incompressible limit. Now consider
independent Fourier modes (transverse to wavenumbers) as state variables
$y_{a}(t)$. Lee~\cite{Lee:QAM1952} and Kraichnan~\cite{Kraichnan:JFM1973} have shown that these variables
move in a constant energy surface, and the motion is area preserving
like in Liouville's theorem. Now we look for equilibrium probability-distribution
function $P(\{ y_{a}\})$ for these state variables. Once we assume
ergodicity, the ideal incompressible fluid turbulence can be mapped
to equilibrium statistical mechanics~\cite{Lee:QAM1952,Kraichnan:JFM1973}.   

By applying the usual arguments of equilibrium statistical mechanics
we can deduce that at equilibrium, the probability distribution function
will be \[
P(y_{1},...,y_{m})=\frac{1}{Z}\exp{\left(-\frac{1}{2}\sigma\sum_{a=1}^{m}y_{a}^{2}\right)},\]
where $\sigma$ is a positive constant. The parameter $\sigma$ corresponds
to inverse temperature in the Boltzmann distribution. Clearly \[
\left\langle y_{a}^{2}\right\rangle =\int\Pi_{i}dy_{i}y_{a}^{2}P(\{ y_{i}\})=\frac{1}{\sigma},\]
 independent of $a$. Hence energy spectrum $C(\mathbf{k})$ is constant,
and 1-d spectrum will be proportional to $k^{d-1}$ \cite{Lesieur:book:Turbulence}.
This is very different from Kolmogorov's spectrum for large $Re$
turbulence. Hence, the physics of turbulence at $\nu=0$ (inviscid)
differs greatly from the physics at $\nu\rightarrow0$. This is not
surprising because (a) turbulence is a nonequilibrium process, and
(b) Navier-Stokes equation is singular in $\nu$.

Even though nature of inviscid flow is very different from turbulent
flow, Kraichnan and Chen \cite{Kraichnan:PD1989} suggested that the tendency
of the energy cascade in turbulent flow could be anticipated from
the absolute equilibrium states. Using absolute equilibrium theory,
Kraichnan \cite{Kraichnan:JFM1971_2D3D} showed that in two dimensions, enstrophy
cascades forward, but energy cascades backward (see also Lesieur \cite{Lesieur:book:Turbulence}).
The above prediction holds good for real fluids.

\section{Experimental Results on Turbulence\label{sec:Solar-Wind}}

Analytical results are very rare in turbulence research because of
complex nature of turbulence. Therefore, experiments and numerical
simulations play very important role in turbulence research. In fluid
turbulence, engineers have been able to obtain necessary information
from experiments (e.g., wind tunnels), and successfully design complex
machines like aeroplanes, spacecrafts etc. This aspect of fluid turbulence
is not being covered here. For details on experiments, refer to 
books on fluid turbulence, e.g.,  \cite{Monin:book:v1,Monin:book:v2,Davidson:book:Turbulence}.

\section{Numerical Investigation of Fluid Turbulence \label{sec:Numerical-Investigation-MHD}}

Like experiments, numerical simulations help us test existing models
and theories, and inspire new one. In addition, numerical simulations
can be performed for conditions which may be impossible in real experiments,
and all the field components can be probed everywhere, and at all
times. Recent exponential growth in computing power has fueled major
growth in this area of research. Of course, numerical simulations
have limitations as well. Even the best computers of today cannot
resolve all the scales in a turbulent flow. We will investigate these
issues in this section.

There are many numerical methods to simulate turbulence on a computer.
Engineers have devised many clever schemes to simulate flows in
complex geometries; however, their attention is typically at large
scales.  Physicists normally focus on intermediate and small scales in
a simple geometry because these scales obey universal laws. Since
nonlinear equations are generally quite sensitive, one needs to
compute both the spatial and temporal derivatives as accurately as
possible. It has been shown that spatial derivative could be computed
{}``exactly'' using Fourier transforms given enough resolutions
\cite{Canuto:book:SpectralFluid,Boyd:book:Spectral}.  Therefore, physicists typically choose spectral
method to simulate turbulence. Note however that several researchers
have used higher order finite-difference scheme and have obtained
comparable results.

\subsection{Numerical Solution of fluid Equations using Pseudo-Spectral Method}

In this subsection we will briefly sketch the spectral method for
3D flows. For details refer to \cite{Canuto:book:SpectralFluid,Boyd:book:Spectral}. The
fluid equations in Fourier space is written as

\begin{eqnarray*}
\frac{\partial\mathbf{u}\left(\mathbf{k},t\right)}{\partial t} & = & -i\mathbf{k}p\left(\mathbf{k},t\right)-FT\left[\mathbf{u}\left(\mathbf{k},t\right)\cdot\nabla\mathbf{\mathbf{u}\left(\mathbf{k},t\right)}\right]-\nu k^{2}\mathbf{u}\left(\mathbf{k},t\right)+\mathbf{f}\end{eqnarray*}
where $FT$ stands for Fourier transform, and $\mathbf{f}(\mathbf{k},t)$
is the forcing function. The flow is assumed to be incompressible,
i. e., $\mathbf{k}\cdot\mathbf{u}\left(\mathbf{k},t\right)=0$. We
assume periodic boundary condition with real-space box size as $(2\pi)\times(2\pi)\times(2\pi)$,
and Fourier-space box size as $(nx,ny,nz)$. The allowed wavenumbers
are $\mathbf{k}=(k_{x},k_{y},k_{z})$ with $k_{x}=(-n_{x}/2:n_{x}/2),k_{y}=(-n_{y}/2:n_{y}/2),k_{z}=(-n_{z}/2:n_{z}/2)$.
The reality condition implies that $\mathbf{z^{\pm}}\left(\mathbf{-k}\right)=\mathbf{z^{\pm*}}\left(\mathbf{k}\right)$,
therefore, we need to consider only half of the modes \cite{Canuto:book:SpectralFluid}.
Typically we take $(-n_{x}/2:n_{x}/2,-n_{y}/2:n_{y}/2,0:n_{z}/2)$,
hence, we have $N=n_{x}*n_{y}*(n_{z}/2+1)$ coupled ordinary differential
equations. The objective is to solve for the field variables at a
later time given initial conditions. The following important issues
are involved in this method:

\begin{enumerate}
\item The Navier-Stokes equation is converted to nondimensionalized form,
and then solved numerically. The parameter $\nu$ is inverse Reynold's
number. Hence, for turbulent flows, $\nu$ is chosen to be quite small
(typically $10^{-3}$ or $10^{-4}$). In Section \ref{sub:Kolmogorov's-1941-Theory}
we deduced using Kolmogorov's phenomenology that the number of active
modes are\be
N\sim\nu^{-9/4}.\ee
 If we choose a moderate Reynolds number $\nu^{-1}=10^{4}$, $N$
will be $10^{9}$, which is a very large number even for the most
powerful supercomputers. To overcome this difficulty, researchers
apply some tricks; the most popular among them are introduction of
hyperviscosity and hyperresistivity, and large-eddy simulations. Hyperviscous
(hyperresistive) terms are of the form $(\nu_{j})k^{2j}\mathbf{\mathbf{u}\left(\mathbf{k}\right)}$
with $j\ge2$; these terms become active only at large wavenumbers,
and are expected not to affect the inertial range physics, which is
of interest to us. Because of this property, the usage of hyperviscosity
and hyperresistivity has become very popular in turbulence simulations.
Large-eddy simulations are discussed in various books (e.g.,
see Pope \cite{Pope:book}).

\item The computation of the nonlinear terms is the most expensive part
of turbulence simulation. A naive calculation involving convolution
will take $O(N^{2})$ floating point operations. It is instead efficiently
computed using Fast Fourier Transform (FFT) as follows: \\
(a) Compute $\mathbf{\mathbf{u}\left(x\right)}$ from $\mathbf{\mathbf{u}\left(\mathbf{k}\right)}$
using Inverse FFT. \\
(b) Compute $u_{i}(\mathbf{x})u_{j}(\mathbf{x})$ in real space by
multiplying the fields at each space points.\\
(c) Compute $FFT[u_{i}(\mathbf{x})u_{j}(\mathbf{x})]$ using FFT.\\
(d) Compute $ik_{j}FFT[u_{i}(\mathbf{x})u_{j}(\mathbf{x})]$ by multiplying
by $k_{j}$ and summing over all $j$. This vector is $-FFT\left[\mathbf{u}\left(\mathbf{k},t\right)\cdot\nabla\mathbf{\mathbf{u}\left(\mathbf{k},t\right)}\right]$.\\
Since FFT takes $O(N\log N)$, the above method is quite efficient.
The multiplication is done in real space, therefore this method is
called pseudo-spectral method instead of just spectral method.
\item Products $u_{i}(\mathbf{x})u_{j}(\mathbf{x})$ produce modes with
wavenumbers larger than $k_{max}$. On FFT, these modes get aliased
with $k<k_{max}$ and will provide incorrect value for the convolution.
To overcome this difficulty, last 1/3 modes of fields $z_{i}^{\pm}(\mathbf{k})$
are set to zero (zero padding), and then FFTs are performed. This
scheme is called $2/3$ rule. For details refer to Canuto et al. \cite{Canuto:book:SpectralFluid}.
\item Pressure is computed by taking the dot product of Navier-Stokes equation
with $\mathbf{k}$. Using incompressibility condition one obtains\[
p\left(\mathbf{k},t\right)=\frac{i\mathbf{k}}{k^{2}}\cdot FT\left[\mathbf{u}\left(\mathbf{x},t\right)\cdot\nabla\mathbf{\mathbf{u}\left(\mathbf{x},t\right)}\right].\]
To compute $p(\mathbf{k})$ we use already computed nonlinear term. 
\item Once the right-hand side of the Navier-Stokes equation could be computed,
we could time advance the equation using one of the standard techniques.
The viscous terms are advanced using an implicit method called Crank-Nicholson's
scheme. However, the nonlinear terms are advanced using Adam-Bashforth
or Runge-Kutta scheme. One uses either second or third order scheme.
Choice of $dt$ is determined by CFL criteria $(dt<(\bigtriangleup x)/U_{rms})$.
By repeated application of time-advancing, we can reach the desired
final time.
\item When forcing $\mathbf{f}=0$, the total energy gets dissipated due
to viscosity. This is called decaying simulation. On the contrary,
forced simulation have nonzero forcing $(\mathbf{f}\ne0)$, which
feed energy into the system, and the system typically reaches a steady-state
in several eddy turnover time. Forcing in turbulent systems are typically
at large-scale eddies (shaking, stirring etc.). Therefore, in forced
turbulence $\mathbf{f}$ is typically applied at small wavenumbers,
which could feed kinetic energy and kinetic helicity.
\end{enumerate}
Spectral method has several disadvantages as well. This method can
not be easily applied to nonperiodic flows. That is the reason why
engineers hardly use spectral method. Note however that even in aperiodic
flows with complex boundaries, the flows at small length-scale can
be quite homogeneous, and can be simulated using spectral method.
Spectral simulations are very popular among physicists who try to
probe universal features of small-scale turbulent flows. 

The numerical results on turbulent energy spectrum and fluxes are
described in many turbulence literature, e. g., Pope~\cite{Pope:book}.
In Section 8 we describe a numerical result on shell-to-shell energy
transfer in $512^{3}$ simulation. In recent times a technique called
large-eddy simulation (LES) has become very popular. LES enables us
to perform turbulence simulations on smaller grids. In this paper
we do not cover this topic.

In the next three sections we will describe the field-theoretic calculation
of renormalized viscosity.

\section{Renormalization Group Analysis of Fluid Turbulence \label{sec:Renormalization-Group}}

In Section \ref{sec:Turbulence-Models} we discussed various existing
turbulence models. Here we will describe some field-theoretic calculations.

Field theory is well developed, and has been applied to many areas
of physics, e.g., Quantum Electrodynamics, Condensed Matter Physics
etc. In this theory, the equations are expanded perturbatively in
terms of nonlinear term, which are considered small. In fluid turbulence
the nonlinear term is not small; the ratio of nonlinear to linear
(viscous) term is Reynolds numbers, which is large in turbulence regime.
This problem appears in many areas of physics including Quantum Chromodynamics
(QCD), Strongly Correlated Systems, Quantum Gravity etc., and is largely
unsolved.  To overcome the above difficulty, some clever schemes
have been adopted such as Direct Interaction
Approximation, Renormalization Groups (RG), Eddy-damped quasi-normal
Markovian approximations, etc. We discuss
some of them below. A simple-minded calculation of Green's function
shows divergence at small wavenumbers (infrared divergence). One way
to solve this problem is by introducing an infrared cutoff for the
integral. The reader is referred to  \citet{Kraichnan:JFM1959} and \citet{Leslie:book} for details.
RG technique, to be described below, is a systematic procedure to
cure this problem.

\subsection{Renormalization Groups in Turbulence \label{sub:Renormalization-Groups-in-Turb}}

Renormalization Group Theory (RG) is a technique which is applied to
complex problems involving many length scales. Many researchers have
applied RG to  turbulence. Over the years, several
different RG applications for turbulence has been discovered. Broadly
speaking, they fall in three different categories:

\subsubsection*{Yakhot-Orszag (YO) Perturbative approach }

Yakhot and Orszag's \cite{Yakhot:JSC1986} work, motivated by  \citet{Forster:PRA1977} and \citet{Fournier:PRA1978}, is the
first comprehensive application of RG to turbulence. It is based on
Wilson's shell-elimination procedure~\cite{Wilson:PR1974}. Also refer to  Smith and Woodruff
\cite{Smith:ARFM1998} for details. Here the renormalized parameter is function
of forcing noise spectrum $D(k)=D_{0}k^{-y}$. It is shown that the
local Reynolds number $\bar{\lambda}$ is\[
\bar{\lambda}=\frac{\lambda_{0}^{2}D_{0}}{\nu^{3}(\Lambda)\Lambda^{\epsilon}},\]
where $\lambda_{0}$ is the expansion parameter, $\Lambda$ is the
cutoff wavenumber, and $\epsilon=4+y-d$ \cite{Yakhot:JSC1986}. It is found
that $\nu(\Lambda)$ increases as $\Lambda$ decreases, therefore,
$\bar{\lambda}$ remains small (may not be less that one though) compared
to $Re$ as the wavenumber shells are eliminated. Hence, the {}``effective''
expansion parameter is small even when the Reynolds number may be
large.

The RG analysis of Yakhot and Orszag \cite{Yakhot:JSC1986} yielded Kolmogorov's
constant $K_{Ko}=1.617$, turbulent Prandtl number for high-Reynolds-number
heat transfer $P_{t}=0.7179$, Batchelor constant $Ba=1.161$ etc.
These numbers are quite close to the experimental results. Hence,
Yakhot and Orszag's method appears to be highly successful. However
there are several criticisms to the YO scheme. Kolmogorov's spectrum
results in the YO scheme for $\epsilon=4$, far away from $\epsilon=0$,
hence epsilon-expansion is questionable. YO proposed that higher order
nonlinearities are {}``irrelevant'' in the RG sense for $\epsilon=0$,
and are marginal when $\epsilon=4$.  \citet{Eyink:PF1994} objected
to this claim and demonstrated that the higher order nonlinearities
are marginal regardless of $\epsilon$.  \citet{Kraichnan:PF1987YO}
compared YO's procedure with Kraichnan's Direct Interaction Approximation
\cite{Kraichnan:JFM1959} and raised certain objections regarding distant-interaction
in YO scheme. For details, refer to  \citet{Zhou:NASA1997}
and  \citet{Smith:ARFM1998}.

\subsubsection*{Self-consistent approach of McComb and Zhou}

This is one of the nonperturbative method, which is often used in
Quantum Field theory. In this method, a self-consistent equation of
the full propagator is written in terms of itself and the proper vertex
part. The equation may contain many (possibly infinite) terms, but
it is truncated at some order. Then the equation is solved iteratively.
 \citet{McComb:book:Turbulence} and \citet{Zhou:PRE1993RG} have
applied this scheme to fluid turbulence, and have calculated renormalized
viscosity and Kolmogorov's constant successfully. Direct Interaction
Approximation of Kraichnan is quite similar to self-consistent theory
(eg. \citet{Smith:ARFM1998}).

The difficulty with this method is that it is not rigorous. In McComb
and Zhou's procedures, the vertex correction is not taken into account.
Verma \cite{Verma:PP1999,Verma:PRE2001,Verma:Pramana2003Nonhelical,Verma:Pramana2003Helical} has applied the self-consistent
theory to MHD turbulence.

\subsubsection*{Callan-Symanzik Equation for Turbulence}

\citet{DeDominicis:PRA1979}  and  \citet{Teodorovich:JETP1989}
obtained the RG equation using functional integral. Teodorovich
obtained $K_{Ko}=2.447$, which is in not in good agreement with the
experimental data, though it is not too far away. It has been shown
that Wilson's shell-renormalization and RG through Callan-Symanzik
equation are equivalent procedure. However, careful comparison of RG
schemes in turbulence is not completely worked out.

In the following discussion we will discuss McComb's RG scheme is
some detail. The other schemes have been discussed in great lengths
in several books and review articles. After renormalization, in Section
\ref{sec:analytic-energy} we will discuss the computation of energy
fluxes. These calculations are done using self-consistent field theory,
a scheme very similar to DIA. At the end we will describe Eddy-damped
quasi-normal Markovian approximation, which is very similar to the
energy flux calculation.

\subsection{Physical Meaning of Renormalization in Turbulence}

The field theorists have been using renormalization techniques since
1940s. However, the physical meaning of renormalization became clear
after path-breaking work of Wilson \cite{Wilson:PR1974}. Here renormalization
is a variation of parameters as we go from one length scale to the
next. Following Wilson, renormalized viscosity and resistivity can
also be interpreted as scale-dependent parameters. We coarse-grain
the physical space and look for an effective theory at a larger scale.
In this method, we sum up all the interactions at smaller scales,
and as a outcome we obtain terms that can be treated as a correction
to viscosity and resistivity. The corrected viscosity and resistivity
are called {}``effective'' or renormalized dissipative parameters.
This procedure of coarse graining is also called shell elimination
in wavenumber space. We carry on with this averaging process till
we reach inertial range. In the inertial range the {}``effective''
or renormalized parameters follow a universal powerlaw, e. g., renormalized
viscosity $\nu(l)\propto l^{4/3}$. This is the renormalization procedure
in turbulence. Note that the renormalized parameters are independent
of microscopic viscosity or resistivity. 

\begin{figure}
\includegraphics[%
  scale=1]{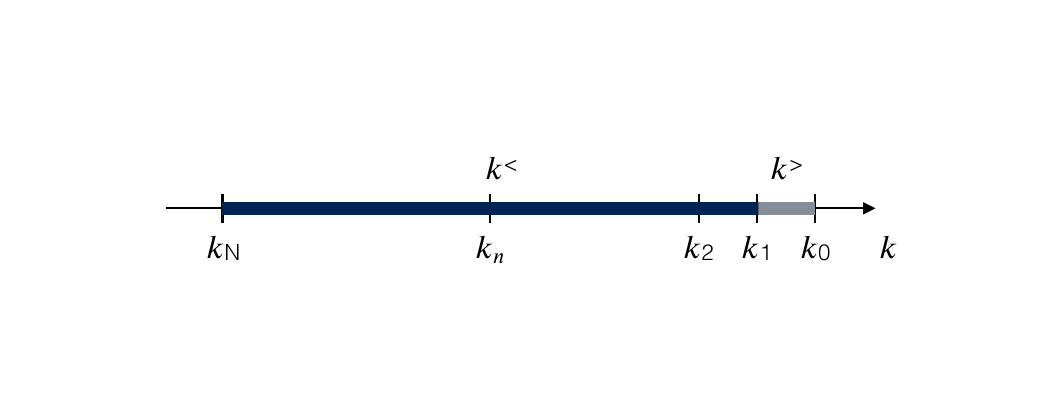}
\caption{\label{Fig:k0kN} The wavenumber shells to be averaged during renormalization
procedure. }
\end{figure}

In viscosity renormalization the large wavenumber shells are eliminated,
and the interaction involving these shells are summed. Hence, we move
from larger wavenumbers to smaller wavenumbers. However, it is also
possible to go from smaller wavenumbers to larger wavenumber by summing
the smaller wavenumber shells, e.g, for shear flows. This process
is not coarse-graining, but it is a perfectly valid RG procedure,
and is useful when the small wavenumber modes (large length scales)
are linear. This scheme is followed in Quantum Electrodynamics (QED),
where the electromagnetic field is negligible at a large distance
(small wavenumbers) from a charge particle, while the field becomes
nonzero at short distances (large wavenumber). In QED, the charge
of a particle gets renormalized when we come closer to the charge
particle, i. e., from smaller wavenumbers to larger wavenumbers. See
Fig. \ref{Fig:k0kN} for an illustration of wavenumber shells to be
averaged.%
In the following subsection we will calculate renormalized viscosity
using RG procedure.

\subsection{Renormalization of viscosity using self-consistent procedure \label{sub:Renormalization-of-viscosity}}

In this subsection we compute renormalized viscosity using self-consistent
procedure. This work was done by McComb and his group workers~\cite{McComb:book:Turbulence,McComb:book:HIT}. The
renormalization of viscosity is performed \emph{from large wavenumber
to smaller wavenumbers. }

McComb and his group workers took the following form of Kolmogorov's
spectrum for kinetic energy \begin{eqnarray}
E(k) & = & K_{Ko}\Pi^{2/3}k^{-5/3},\label{eq:Euk}\end{eqnarray}
where $K_{Ko}$ is Kolmogorov's constant, and $\Pi$
is the total energy flux. The incompressible fluid equations in the
Fourier space are \begin{eqnarray}
\left(-i\omega+\nu k^{2}\right)u_{i}\left(\hat{{k}}\right) & = & -\frac{i}{2}P_{ijm}^{+}(\mathbf{k})\int_{\hat{{p}}+\hat{{q}}=\hat{{k}}}d\hat{{p}}\left[u_{j}(\hat{{p}})u_{m}(\hat{{q}})\right],\label{eq:udot}\end{eqnarray}
where\begin{eqnarray}
P_{ijm}^{+}(\mathbf{k}) & = & k_{j}P_{im}(\mathbf{k})+k_{m}P_{ij}(\mathbf{k}),\label{eq:Pp}\end{eqnarray}
Here $\nu$ is viscosity, and $d$ is the space dimensionality.

\begin{figure}
\includegraphics[scale=0.6]{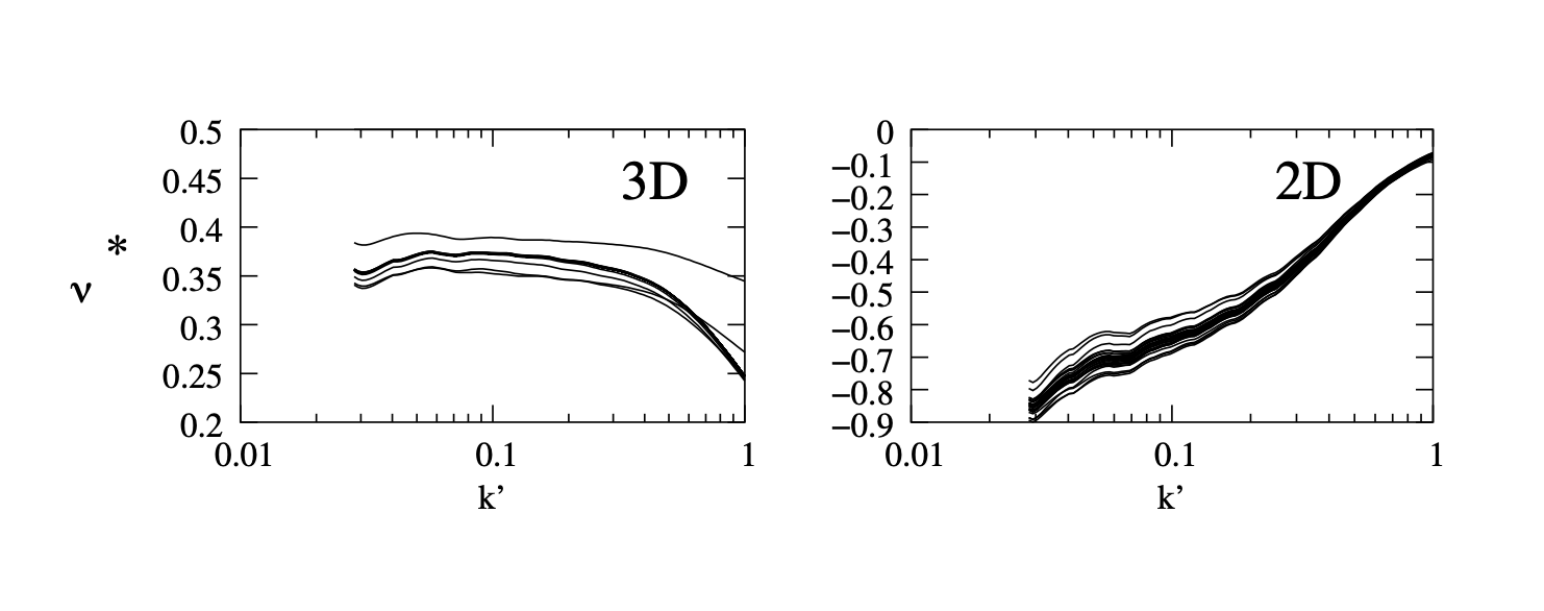}
\caption{\label{Fig:Fluid-RG} Plot of $\nu^{*}(k')$ vs. $k'$ for 2D and
3D fluid turbulence. In 2D, $\nu^{*}$ is negative.}
\end{figure}

In this RG procedure the wavenumber range $(k_{N},k_{0})$ is divided
logarithmically into $N$ shells. The $n$th shell is $(k_{n},k_{n-1})$
where $k_{n}=h^{n}k_{0}\,\,(h<1)$. In the following discussion, the
elimination of the first shell $(k_{1},k_{0})$ is carried out, and
modified NS equation is obtained. Then one proceeds iteratively to
eliminate higher shells and get a general expression for the modified
fluid equation. The renormalization group procedure is as follows:

\begin{enumerate}
\item The the spectral space is divided in two parts: 1. the shell $(k_{1},k_{0})=k^{>}$,
which is to be eliminated; 2. $(k_{N},k_{1})=k^{<}$, set of modes
to be retained. Note that $\nu_{(0)}$ denote the viscosity and resistivity
before the elimination of the first shell.
\item Rewrite Eqs.~(\ref{eq:udot}) for $k^{<}$ and $k^{>}$.
The equations for $u_{i}^{<}(\hat{k})$ and $b_{i}^{<}(\hat{k})$
modes are 
\begin{eqnarray}
\left(-i\omega+\Sigma_{(0)}(k)\right)u_{i}^{<}(\hat{k}) & = & -\frac{i}{2}P_{ijm}^{+}({\textbf{k}})\int d\hat{p}([u_{j}^{<}(\hat{p})u_{m}^{<}(\hat{k}-\hat{p})]\nonumber \\
 &  & +2[u_{j}^{<}(\hat{p})u_{m}^{>}(\hat{k}-\hat{p})]+[u_{j}^{>}(\hat{p})u_{m}^{>}(\hat{k}-\hat{p})])
\label{eq:ukless}
\end{eqnarray}
 The $\Sigma$s appearing in the equations are usually called the
{}``self-energy'' in Quantum field theory language. In the first
iteration, $\Sigma_{(0)}=\nu_{(0)}k^{2}$. The equation for $u_{i}^{>}(\hat{k})$
modes can be obtained by interchanging $<$ and $>$ in the above
equations.
\item The terms given in the second and third brackets in the Right-hand
side of Eqs.~(\ref{eq:ukless}) are calculated
perturbatively. Since we are interested in the statistical properties
of $\mathbf{u}$ fluctuations, we perform the usual ensemble average
of the system \cite{Yakhot:JSC1986}. It is assumed that $\mathbf{u}^{>}(\hat{k})$
has Gaussian distributions with zero mean, while $\mathbf{u}(\hat{k})$
is unaffected by the averaging process. Hence,\begin{eqnarray}
\left\langle u_{i}^{>}(\hat{k})\right\rangle  & = & 0\label{eqn:avgbegin}\\
\left\langle u_{i}^{<}(\hat{k})\right\rangle  & = & u_{i}^{<}(\hat{k})
\label{eqn:avgend}
\end{eqnarray}
and\begin{eqnarray}
\left\langle u_{i}^{>}(\hat{p})u_{j}^{>}(\hat{q})\right\rangle  & = & P_{ij}(\mathbf{p})C(\hat{p})\delta(\hat{p}+\hat{q})\label{eq:nonhelical-uu}\end{eqnarray}
The triple order correlations $\left\langle u_{i}^{>}(\hat{k})u_{j}^{>}(\hat{p})u_{m}^{>}(\hat{q})\right\rangle $
are zero due to Gaussian nature of the fluctuations. Here, $X$ stands
for $u$ or $b$. In addition, we also neglect the contribution from
the triple nonlinearity $\left\langle u^{<}(\hat{k})u_{j}^{<}(\hat{p})u_{m}^{<}(\hat{q})\right\rangle $,
as done in many of the turbulence RG calculations \cite{Yakhot:JSC1986,McComb:book:Turbulence}.
The effects of triple nonlinearity can be included following the scheme
of Zhou and Vahala \cite{Zhou:PRA1988}.
\item To the first order, the second bracketed terms of 
Eqs.~(\ref{eq:ukless}) vanish, but the nonvanishing third bracketed terms
yield corrections to $\Sigma$s. Refer to Appendix C for details.
Eqs. (\ref{eq:ukless}) can now be approximated
by \begin{eqnarray}
\left(-i\omega+\Sigma_{(0)}+\delta\Sigma_{(0)}\right)u_{i}^{<}(\hat{k}) & = & -\frac{i}{2}P_{ijm}^{+}({\textbf{k}})\int d\hat{p}[u_{j}^{<}(\hat{p})u_{m}^{<}(\hat{k}-\hat{p})]\end{eqnarray}
 with \begin{eqnarray}
\delta\Sigma_{(0)}^{uu}(k) & = & \frac{1}{(d-1)}\int_{\hat{p}+\hat{q}=\hat{k}}^{\Delta}d\hat{p}[S(k,p,q)G(\hat{p})C(\hat{q})]\label{eq:Sigmauu}\end{eqnarray}
where\be
S(k,p,q)=kp((d-3)z+2z^{3}+(d-1)xy).\ee
The integral $\Delta$ is to be done over the first shell.
\item The frequency dependence of the correlation function is taken as:
$C(k,\omega)=2C(k)\Re(G(k,\omega))$. In other words, the relaxation
time-scale of correlation function is assumed to be the same as that
of corresponding Green's function. Since we are interested in the
large time-scale behaviour of turbulence, we take the limit $\omega$
going to zero. Under these assumptions, the frequency integration
of the above equations yield 
\begin{eqnarray}
\delta\nu_{(0)}(k) & = & \frac{1}{(d-1)k^{2}}\int_{\textbf{p+q=k}}^{\Delta}\frac{d\mathbf{p}}{(2\pi)^{d}}\frac{S(k,p,q)C(q)}{\nu_{(0)}(p)p^{2}+\nu_{(0)}(q)q^{2}}
\label{eq:delta-nu}
\end{eqnarray}
Note that $\nu(k)=\Sigma^{uu}(k)/k^{2}$ . There are some important
points to remember in the above step. The frequency integral in the
above is done using contour integral. It can be shown that the integrals
are nonzero only when both the components appearing the denominator
are of the same sign. For example, first term of Eq. (\ref{eq:delta-nu})
is nonzero only when both $\nu_{(0)}(p)$ and $\nu_{(0)}(q)$ are
of the same sign.
\item Let us denote $\nu_{(1)}(k)$ as the renormalized viscosity after
the first step of wavenumber elimination. Hence, \begin{eqnarray}
\nu_{(1)}(k) & = & \nu_{(0)}(k)+\delta\nu_{(0)}(k);\label{eq:eta_1}\end{eqnarray}
We keep eliminating the shells one after the other by the above procedure.
After $n+1$ iterations we obtain \begin{eqnarray}
\nu_{(n+1)}(k)=\nu_{(n)}(k)+\delta\nu_{(n)}(k)\label{nu_n}\end{eqnarray}
where the equation for $\delta\nu_{(n)}(k)$ is the same as the Eqs.~(\ref{eq:delta-nu})
except that $\nu_{(0)}(k)$ appearing in the equation is to be replaced
by $\nu_{(n)}(k)$. Clearly $\nu_{(n+1)}(k)$ is the renormalized
viscosity and resistivity after the elimination of the $(n+1)$th
shell.
\item We need to compute $\delta\nu_{(n)}$ for various $n$. These computations,
however, require $\nu_{(n)}$. In our scheme we solve these equations
iteratively. In Eqs.~(\ref{eq:delta-nu}, \ref{eq:delta-eta}) we
substitute $C(k)$ by one dimensional energy spectrum $E(k)$\[
C(k)=\frac{2(2\pi)^{d}}{S_{d}(d-1)}k^{-(d-1)}E(k)\]
 where $S_{d}$ is the surface area of $d$-dimensional spheres. We
assume that $E(k)$ follows Eqs.~(\ref{eq:Euk}) . Regarding $\nu_{(n)}$,
we attempt the following form of solution \[
\nu_{(n)}(k_{n}k')=(K_{Ko})^{1/2}\Pi^{1/3}k_{n}^{-4/3}\nu_{(n)}^{*}(k')\]
 with $k=k_{n+1}k'\,(k'<1)$. We expect $\nu_{(n)}^{*}(k')$ to be
a universal functions for large $n$. The substitution of $C(k),\nu_{(n)}(k)$
yields the following equations: \begin{eqnarray}
\delta\nu_{(n)}^{*}(k') & = & \frac{1}{(d-1)}\int_{\textbf{p'+q'=k'}}d\mathbf{q}'\frac{2}{(d-1)S_{d}}\frac{E^{u}(q')}{q'^{d-1}}[S(k',p',q')\frac{1}{\nu_{(n)}^{*}(hp')p'^{2}+\nu_{(n)}^{*}(hq')q'^{2}}]\label{eq:delta_nu*}\\
\nu_{(n+1)}^{*}(k') & = & h^{4/3}\nu_{(n)}^{*}(hk')+h^{-4/3}\delta\nu_{(n)}^{*}(k')\end{eqnarray}
 where the integrals in the above equations are performed iteratively
over a region $1\leq p',q'\leq1/h$ with the constraint that $\mathbf{p}'+\mathbf{q}'=\mathbf{k}'$.
Fournier and Frisch \cite{Fournier:PRA1978} showed the above volume integral
in $d$ dimension to be \begin{equation}
\int_{\mathbf{p}'+\mathbf{q}'=\mathbf{k}'}d\mathbf{p'}=S_{d-1}\int dp'dq'\left(\frac{p'q'}{k'}\right)^{d-2}\left(\sin\alpha\right)^{d-3},\label{eq:volume-integral}\end{equation}
where $\alpha$ is the angle between vectors $\mathbf{p}'$ and $\mathbf{q}'$.
\item Now the above equations are solved self-consistently with $h=0.7$.
This value is about middle of the range (0.55-0.75) estimated to be
the reasonable values of $h$ by Zhou \emph{et al.} \cite{Zhou:NASA1997}.
One starts with constant value of $\nu_{(0)}^{*}$, and compute the
integrals using Gauss quadrature technique. Once $\delta\nu_{(0)}^{*}$
has been computed, $\nu_{(1)}^{*}$ is computed. This process is iterated
till $\nu_{(m+1)}^{*}(k')\approx\nu_{(m)}^{*}(k')$, that is, till
they converge. The result of our RG analysis is given below.
\end{enumerate}
McComb and coworkers \cite{McComb:book:Turbulence,McComb:PRA1992Two_field,Zhou:NASA1997} successfully
applied the above self-consistent renormalization group theory to
2D and 3D fluid turbulence. They found that $\nu^{*}(k')$ converges
quite quickly. For 3D the value of $\nu^{*}(k'\rightarrow0)$ is approximately
0.38. See Fig. \ref{Fig:Fluid-RG} for an illustration.

For 2D turbulence $\nu^{*}(k')$ is negative as shown in Fig. \ref{Fig:Fluid-RG}.
The function $\nu^{*}$ is not very well behaved as $k'\rightarrow0$.
Still, negative renormalized viscosity is consistent with negative
eddy viscosity obtained using Test Field Model \cite{Kraichnan:JFM1971_2D3D} and
EDQNM calculations \cite{Pouquet:JFM1978}. We estimate $\nu^{*}\approx-0.60$. 
\begin{figure}
\includegraphics[scale=0.6]{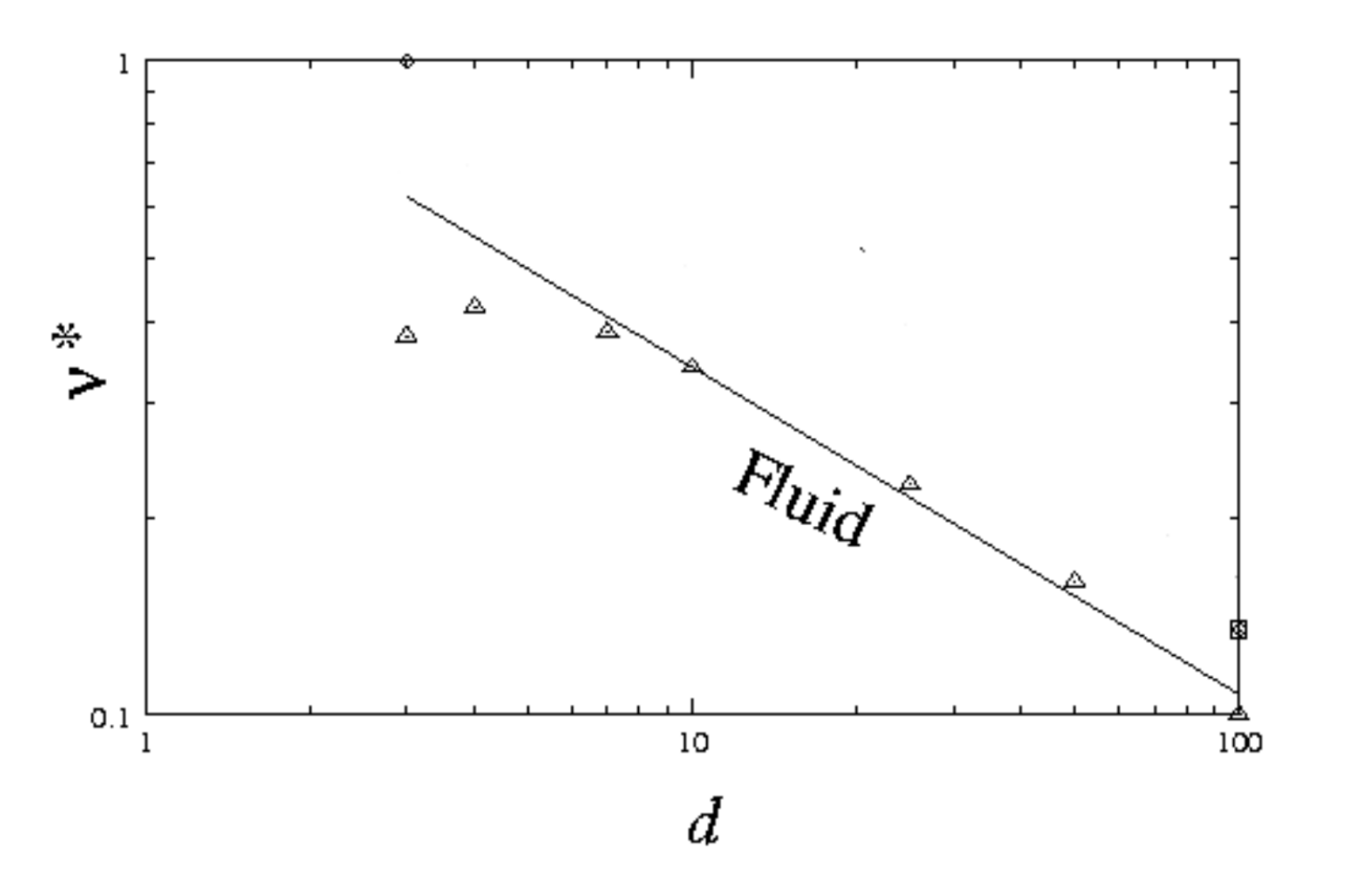}
\caption{\label{Fig:nuvsd} The plot of asymptotic $\nu^{*}$vs. $d$ . For
large $d$, the plot fits quite well with predicted $d^{-1/2}$ curve. Adopted from Verma \cite{Verma:PR2004}.}
\end{figure}

 For large $d$ , $\nu^{*}=\eta^{*}$, and it decreases as $d^{-1/2}$
 (see Fig. \ref{Fig:nuvsd}). $\nu^{*}$ for pure fluid turbulence also decreases as $d^{-1/2}$,
as shown in the same figure. This is evident from Eqs. (\ref{eq:delta_nu*})
\cite{Fournier:PRA1978}. For large $d$ \begin{eqnarray}
\int dp'dq'\left(\frac{p'q'}{k'}\right)^{d-2}\left(\sin\alpha\right)^{d-3}... & \sim & d^{-1/2},\\
\frac{S_{d-1}}{\left(d-1\right)^{2}S_{d}} & \sim & \frac{1}{d^{2}}\left(\frac{d}{2\pi}\right)^{1/2},\nonumber \\
S,-S_{6},-S_{8},S_{9}(k',p',q') & = & kpd(z+xy),\label{eq:SSSS}\end{eqnarray}
which leads to\[
\nu^{*}\delta\nu^{*}\propto\frac{1}{d^{2}}\left(\frac{d}{2\pi}\right)^{1/2}d^{-1/2}d\]
hence $\nu^{*}\propto d^{-1/2}$.

In conclusion the above RG procedure shows that \begin{eqnarray}
E(k) & = & K_{Ko}\Pi^{2/3}k^{-5/3},\label{eq:fluid-Ek}\\
\nu(k=k_{n}k') & = & K_{Ko}^{1/2}\Pi^{1/3}k_{n}^{-4/3}\nu^{*}(k'),\label{eq:fluid-nu}\end{eqnarray}
is a consistent solution of renormalization group equation. Here,
$K_{Ko}$ is Kolmogorov's constant, $\Pi$ is the energy flux, and
$\nu^{*}(k')$ is a universal function that is a constant as $k'\rightarrow0$.

\subsubsection{Helical turbulence}

Helical turbulence is defined for space dimension $d=3$. We can extend
the above the RG analysis to helical turbulence (Zhou \cite{Zhou:PRA1988,Avinash:Pramana2006,Verma:book:ET}).
All the steps are the same except Eqs. (\ref{eq:nonhelical-uu}) are
replaced by\begin{eqnarray}
\left\langle u_{i}^{>}(\hat{p})u_{j}^{>}(\hat{q})\right\rangle  & = & \left[P_{ij}({\textbf{p)}}C^{uu}(\hat{p})-i\epsilon_{ijl}\frac{p_{l}}{p^{2}}H_{K}(\hat{p})\right](2\pi)^{4}\delta(\hat{p}+\hat{q})\end{eqnarray}
 Because of helicities, the equation for change in renormalized self-energy
(\ref{eq:delta-nu}) gets altered to\begin{eqnarray*}
\delta\nu_{(0)}(k) & = & \frac{1}{(d-1)k^{2}}\int_{\textbf{p+q=k}}^{\Delta}\frac{d{\textbf{p}}}{(2\pi)^{d}}[\frac{S(k,p,q)C^{uu}(q)+S'(k,p,q)H_{K}(q)}{\nu_{(0)}(p)p^{2}+\nu_{(0)}(q)q^{2}}\end{eqnarray*}
where $S'_{i}$ defined below can be shown to be zero.\begin{eqnarray*}
S'(k,p,q) & = & P_{bjm}^{+}(k)P_{mab}^{+}(p)\epsilon_{jal}q_{l}=0,\end{eqnarray*}
The argument for vanishing of $S'$ is follows. Since $\delta\nu$
is a proper scalar, and $H_{K}$ is a pseudo scalar, $S'(k,p,q)$
will be also be a pseudo scalar. In addition, $S'(k,p,q)$ are also
linear in $k,p$ and $q$. This implies that $S'_{i}(k,p,q)$ must
be proportional to $\mathbf{q}\cdot(\mathbf{k}\times\mathbf{p})$,
which will be zero because $\mathbf{k}=\mathbf{p}+\mathbf{q}$. Hence
$S'(k,p,q)$ turn out to be zero. \emph{Hence, helicities do not alter
the already calculated $\delta(\nu)_{(n)}(k)$ in the earlier section}. 

In fluid turbulence, there are some other interesting variations of
field-theoretic calculations by DeDominicis and Martin \cite{DeDominicis:PRA1979},
Bhattacharjee \cite{Bhattacharjee:PF1991}, Carati \cite{Carati:PRA1990} and others.

In the next section we will compute energy fluxes for fluid turbulence
using field-theoretic techniques.

\section{Field-theoretic Calculation of Energy Fluxes and Shell-to-shell Energy
Transfer \label{sec:analytic-energy}}

In this section we present calculation of energy flux using
field-theoretic method. We assume the turbulence to be homogeneous
and isotropic. Even though the real-world turbulence do not satisfy
these properties, many conclusions drawn using these assumption provide
us with important insights into the energy transfer mechanisms at
small scales. The field-theoretic procedure requires Fourier space
integrations of functions involving products of energy spectrum and
the Greens functions. Since there is a general agreement on Kolmogorov-like
spectrum for fluid turbulence, $E(k)\propto k^{-5/3}$ is taken for
the energy spectrum. For the Greens function, we substitute the {}``renormalized''
or {}``dressed'' Greens function computed in the previous section
\cite{Verma:PRE2001} (see Section \ref{sub:Renormalization-of-viscosity}). 

Most of these works are based on perturbative expansion to first order.  We assume that the Fourier modes are quasi-Gaussian that yield the following to first order in perturbation:
\bea
\Pi_u(k_{0}) & = & \sum_{p \le k_{0}} \sum_{k>k_{0}} \Im \left[ \la {\bf  \{  k \cdot u(q) \} \{ u(p) \cdot u^*(k) \} }  \ra \right] \nonumber \\
& \sim &  \sum_{p \le k_{0}} \sum_{k>k_{0}} \int_0^t  dt' G({\bf k}, t-t')  \la {\bf  u(p}, t) \cdot {\bf  u(-p},t') \ra
\la {\bf  u(q}, t) \cdot {\bf  u(-q},t') \ra \nonumber \\
& = & \Pi_u = \mathrm{const},
\label{eq:K41:flux}
\eea
where $G({\bf k}, t-t')$ is the Green's function, and $t,t'$ are two different times.  The quasi-Gaussian nature of the modes yields $\la u u u \ra \sim \la u u \ra \la u u\ra$.  The above integral converges and  yields a constant.  Some of the details of calculations are given below.

\subsection{Calculation of Energy Flux}

As described in Section \ref{sec:Mode-to-mode-Energy-Transfer}
the energy flux from a wavenumber sphere of radius $k_{0}$ to the
outside of the sphere of the same radius is \begin{eqnarray}
\Pi(k_{0}) & = & \frac{1}{(2\pi)^{d}\delta(\mathbf{k'+p+q})}\int_{k'>k_{0}}\frac{d\mathbf{k'}}{(2\pi)^{d}}\int_{p<k_{0}}\frac{d\mathbf{p}}{(2\pi)^{d}}\left\langle S^{uu}(\mathbf{k'|p|q})\right\rangle \label{eq:flux-MHD}\end{eqnarray}
 We assume that the kinetic energy is forced at small wavenumbers.

We analytically calculate the above energy fluxes in the inertial
range to leading order in perturbation series. It was assumed that
$\mathbf{u}(\mathbf{k})$ is quasi-Gaussian as in EDQNM approximation.
Under this approximation, the triple correlation $\left\langle XXX\right\rangle $
is zero to zeroth order, but nonzero to first oder. To first order
$\left\langle XXX\right\rangle $ is written in terms of $\left\langle XXXX\right\rangle $,
which is replaced by its Gaussian value, a sum of products of second-order
moment. Consequently, the ensemble average of $S$, $\left\langle S\right\rangle $,
is zero to the zeroth order, but is nonzero to the first order. The
first order terms for $\left\langle S(k|p|q)\right\rangle $ in terms
of Feynman diagrams are given in Appendix C. They are given below
in terms of Green's functions and correlation functions:

\begin{eqnarray}
\left\langle S(k|p|q)\right\rangle  & = & \int_{-\infty}^{t}dt'(2\pi)^{d}[T_{1}(k,p,q)G(k,t-t')C(p,t,t')C(q,t,t')\nonumber \\
 &  & \hspace{1cm}+T_{2}(k,p,q)G(p,t-t')C(k,t,t')C(q,t,t')\nonumber \\
 &  & \hspace{1cm}+T_{3}(k,p,q)G(q,t-t')C(k,t,t')C(p,t,t')]\delta(\mathbf{k'}+\mathbf{p}+\mathbf{q})\label{eqn:Suu-nonhelical}\end{eqnarray}
 where $T_{i}(k,p,q)$ are functions of wavevectors $k,p$, and $q$
given in Appendix B.

The Greens functions can be written in terms of {}``effective''
or {}``renormalized'' viscosity $\nu(k)$ and resistivity $\eta(k)$
computed in Section \ref{sec:Renormalization-Group} \[
G(k,t-t')=\theta{(t-t')}\exp\left(-\nu(k)k^{2}(t-t')\right)\]
 The relaxation time for $C(k,t,t')$ is assumed to be the same as
that of $G(k,t,t')$. Therefore the time dependence of the unequal-time
correlation functions will be \[
C^{uu}(k,t,t')=\theta{(t-t')}\exp\left(-\nu(k)k^{2}(t-t')\right)C^{uu,bb}(k,t,t)\]
 The above forms of Green's and correlation functions are substituted
in the expression of $\left\langle S\right\rangle $, and the $t'$
integral is performed. Now Eq. (\ref{eq:flux-MHD}) yields the following
flux formula for $\Pi(k_{0})$: \begin{eqnarray}
\Pi(k_{0}) & = & \int_{k>k_{0}}\frac{d\mathbf{k}}{(2\pi)^{d}}\int_{p<k_{0}}\frac{d\mathbf{k}}{(2\pi)^{d}}\frac{1}{\nu(k)k^{2}+\nu(p)p^{2}+\nu(q)q^{2}}\times[T_{1}(k,p,q)C^{uu}(p)C^{uu}(q)\nonumber \\
 &  & \hspace{3cm}+T_{2}(k,p,q)C^{uu}(k)C^{uu}(q)+T_{3}(k,p,q)C^{uu}(k)C^{uu}(p)].\label{eqn:Pi_mhd}\end{eqnarray}
 The expressions for the other fluxes can be obtained similarly.

The equal-time correlation function $C(k,t,t)$ at the steady-state
can be written in terms of one dimensional energy spectrum as \[
C(k,t,t)=\frac{2(2\pi)^{d}}{S_{d}(d-1)}k^{-(d-1)}E(k),\]
 where $S_{d}$ is the surface area of $d$-dimensional unit spheres.
We are interested in the fluxes in the inertial range. Therefore,
we substitute Kolmogorov's spectrum {[}Eqs.(\ref{eq:Euk}){]}
for the energy spectrum. The effective viscosity is proportional to
$k^{-4/3}$, i.e., \be
\nu(k)=(K^{u})^{1/2}\Pi^{1/3}k^{-4/3}\nu^{*},\ee
 and the parameter $\nu^{*}$ was calculated in Section \ref{sec:Renormalization-Group}.

We nondimensionalize Eq.~(\ref{eqn:Pi_mhd}) by substituting \cite{Leslie:book}\begin{equation}
k=\frac{k_{0}}{u};\,\,\,\,\, p=\frac{k_{0}}{u}v;\,\,\,\,\, q=\frac{k_{0}}{u}w.\end{equation}
 Application of Eq. (\ref{eq:volume-integral}) yields\begin{equation}
\Pi=(K_{Ko})^{3/2}\Pi\left[\frac{4S_{d-1}}{(d-1)^{2}S_{d}}\int_{0}^{1}dv\ln{(1/v)}\int_{1-v}^{1+v}dw(vw)^{d-2}(\sin\alpha)^{d-3}F(v,w)\right],\label{eq:Pik0}\end{equation}
 where the integral $F(v,w)$ are \begin{eqnarray}
F & = & \frac{1}{\nu^{*}(1+v^{2/3}+w^{2/3})}[t_{1}(v,w)(vw)^{-d-\frac{2}{3}}+t_{2}(v,w)w^{-d-\frac{2}{3}}+t_{3}(v,w)v^{-d-\frac{2}{3}}],\end{eqnarray}
 Here $t_{i}(v,w)=T_{i}(k,kv,kw)/k^{2}$. Note that the energy fluxes
are constant, consistent with the Kolmogorov's picture. We compute
the bracketed term (denoted by $I$) numerically using Gaussian-quadrature
method, and found it to be convergent. Using $I$ the constant
$K_{Ko}$ can be calculated as \begin{equation}
K_{Ko}=(I)^{-2/3}.\label{eq:KoI}\end{equation}

For 3D turbulence, the value of constant $K_{Ko}$ computed using
Eqs. (\ref{eq:Pik0}, \ref{eq:KoI}) is 1.58. This number is very
good agreement with numerical and experimental estimate of Kolmogorov's
constant. For 2D turbulence, we substitute $\nu^{*}=-0.60$ in the
above equations. The computation yields $K_{Ko}^{2D}\approx6.3$.

For large $d$\begin{eqnarray}
\int dp'dq'\left(\frac{p'q'}{k'}\right)^{d-2}\left(\sin\alpha\right)^{d-3}... & \sim & d^{-1/2},\\
\frac{S_{d-1}}{\left(d-1\right)^{2}S_{d}} & \sim & \frac{1}{d^{2}}\left(\frac{d}{2\pi}\right)^{1/2},\nonumber \\
\nu^{*}=\eta^{*} & \sim & d^{-1/2}\\
t_{1}=-t_{2} & = & kpd(z+xy),\label{eq:t-large-d}\end{eqnarray}
and $t_{3}=0$. Using Eq. (\ref{eq:t-large-d}) and by matching the
dimensions, it can be shown that $K\propto d^{-1/3}$. This result
is due to Fournier \emph{et al.} \cite{Fournier:PRA1978}.

All the above conclusions are for large Reynolds number or $\nu\rightarrow0$
limit. The behaviour of Navier-Stokes equation for viscosity $\nu=0$
(inviscid) is very different, and has been analyzed using absolute
equilibrium theory (see Section \ref{sub:Absolute-Equilibrium-States}).
It can be shown using this theory that under steady state, energy
is equipartitioned among all the modes, hence $C(k)=const$
\cite{Orszag:CP1973}. Using this result we can compute mode-to-mode energy
transfer rates $\left\langle S^{uu}(k|p|q)\right\rangle $ to first
order in perturbation theory (Eq. {[}\ref{eqn:Suu-nonhelical}{]}),
which yields\[
\left\langle S^{uu}(k|p|q)\right\rangle \propto\int\frac{\left(T_{1}(k,p,q)+T_{5}(k,p,q)+T_{9}(k,p,q)\right)Const}{\nu(k)k^{2}+\nu(p)p^{2}+\nu(q)q^{2}}=0\]
because $T_{1}(k,p,q)+T_{5}(k,p,q)+T_{9}(k,p,q)=0$. Hence, under
steady-state, their is no energy transfer among Fourier modes in inviscid
Navier-Stokes. In other words {}``principle of detailed balance''
holds here. Note that the above result holds for all space dimensions.
Contrast this result with the turbulence situation when energy preferentially
gets transferred from smaller wavenumber to larger wavenumber. This
example contrasts equilibrium and nonequilibrium systems.

After completing the discussion on energy fluxes for fluid turbulence,
we now move on to theoretical computation of shell-to-shell energy
transfer.

\subsection{Field-theoretic Calculation of Shell-to-shell Energy Transfer \label{sub:Field-theoretic-Calculation-of-shell}}

Energy transfers between wavenumber shells provide us with important
insights into the dynamics of turbulence. Kolmogorov's fluid turbulence
model is based on local energy transfer between wavenumber shells.
There are several quantitative theories in fluid turbulence about
the amount of energy transfer between neighbouring wavenumber shells.
For examples, Kraichnan \cite{Kraichnan:JFM1971_2D3D} showed that $35$\% of the
energy transfer comes from wavenumber triads where the smallest wave-number
is greater than one-half of the middle wavenumber. 

\begin{figure}
	\includegraphics[scale=0.5]{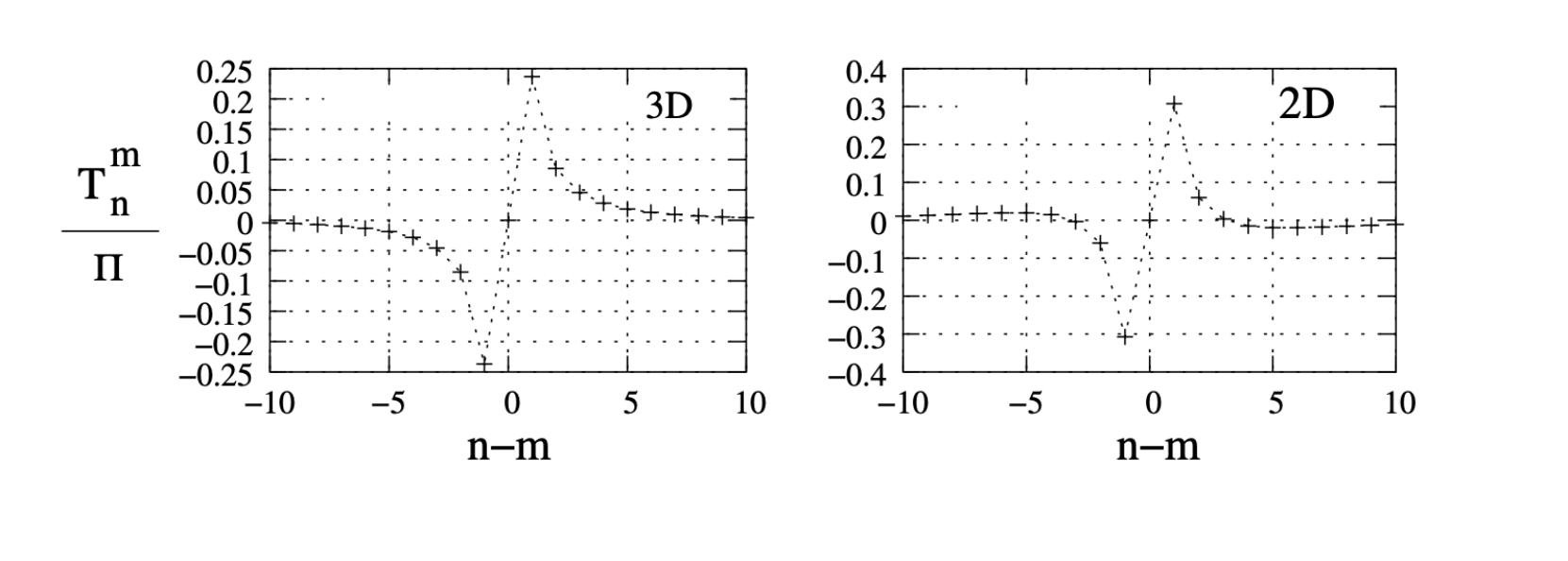}
\caption{\label{Fig:Fluid-T} Plots of shell-to-shell energy transfer rates
$T_{nm}/\Pi$ vs. $n-m$ for 3D and 2D fluid turbulence. }
\end{figure}

In this subsection we will compute the shell-to-shell energy transfer
in turbulence using field-theoretic method \cite{Verma:PP2005}. The procedure
is identical to the one described for energy fluxes. Recall that the
energy transfer rates from the $m$-th shell to the $n$-th shell
is\be
T_{nm}=\sum_{\mathbf{k'}\in n}\sum_{\mathbf{p}\in m}S^{uu}(\mathbf{k'|p|q}).\ee
The $\mathbf{p}$-sum is over $m$-th shell, and the $\mathbf{k'}$-sum
is over $n$-th shell (Section \ref{sec:Mode-to-mode-Energy-Transfer}).
The terms of $S$'s are the same as in flux calculation, however,
the limits of the integrals are different. The shells are binned logarithmically
with $n$-th shell being $(k_{0}s^{n-1},k_{0}s^{n})$. We nondimensionalize
the equations using the transformation \cite{Leslie:book}\begin{equation}
k=\frac{a}{u};\hspace{1cm}p=\frac{a}{u}v;\hspace{1cm}q=\frac{a}{u}w,\end{equation}
 where $a=k_{0}s^{n-1}$. The resulting equation is\begin{equation}
\frac{T_{nm}}{\Pi}=K_{Ko}^{3/2}\frac{4S_{d-1}}{(d-1)^{2}S_{d}}\int_{s^{-1}}^{1}\frac{du}{u}\int_{us^{m-n}}^{us^{m-n+1}}dv\int_{|1-v|}^{1+v}dw\left(vw\right)^{d-2}\left(\sin{\alpha}\right)^{d-3}F(v,w),\label{eqn:shell_final}\end{equation}
where$F(v,w)$ was computed in the previous section. The renormalized
parameters $\nu^{*}$, and Kolmogorov's constant $K_{Ko}$ required
to compute $T_{nm}/\Pi$ are taken from the previous calculations.
From Eq. (\ref{eqn:shell_final}) we can draw the following inferences:

\begin{enumerate}
\item The shell-to-shell energy transfer rate is a function of $n-m$, that
is, $\Phi_{nm}=\Phi_{(n-i)(m-i)}$. Hence, the turbulent energy transfer
rates in the inertial range are all self-similar. Of course, this
is true only in the inertial range.
\item $T_{nn}/\Pi=0$. 
\end{enumerate}
We compute the integral of Eq. (\ref{eqn:shell_final}) and substitute
the value of $K_{Ko}$, which yields $T_{nm}$. The plots of $T_{nm}$for
2D and 3D fluids are shown in Fig. \ref{Fig:Fluid-T}. In 3D the energy
transfer is forward and local. In 2D however the energy transfer is
forward for the nearest neighbours, but it is backward for fourth
neighbour onward; these backward transfers are one of the major factors
in the inverse cascade of energy \cite{Verma:PP2005}. The sum of all these
transfers is negative energy flux, consistent with the inverse cascade
result of Kraichnan \cite{Kraichnan:JFM1971_2D3D}. For details refer to Verma et
al. \cite{Verma:PP2005}.

\begin{figure}
\includegraphics[scale=0.5]{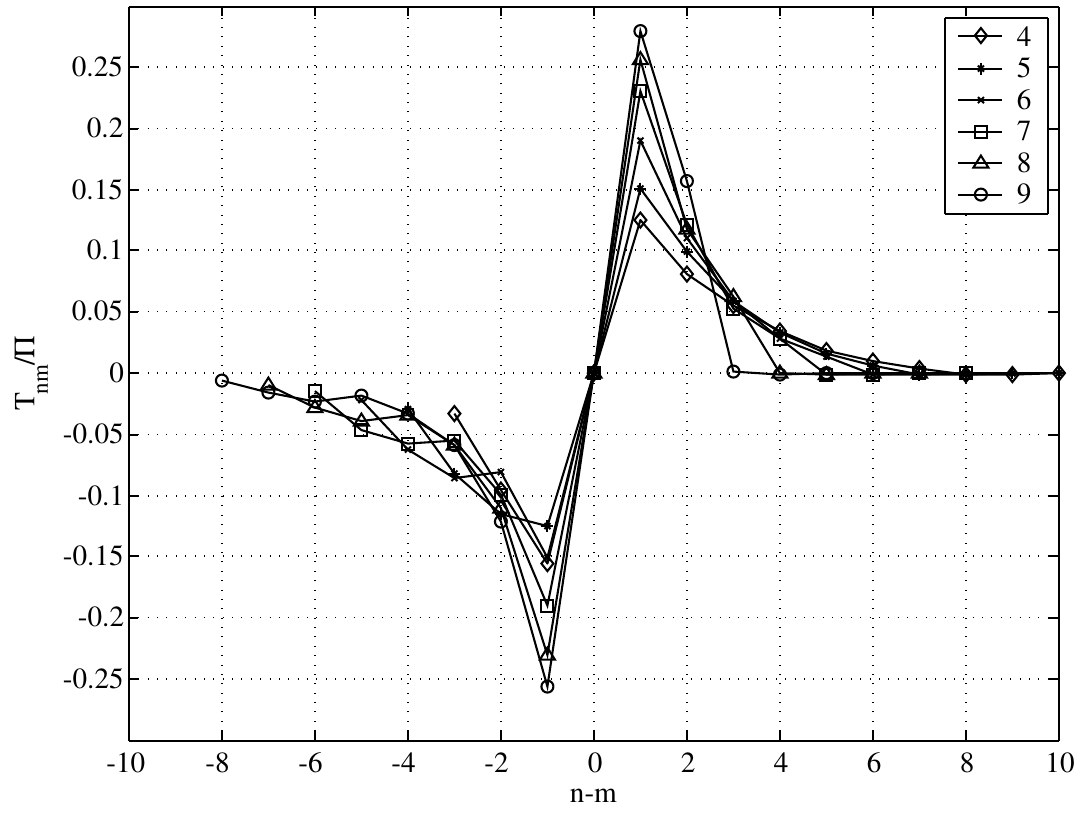}
\caption{ \label{fig:shell-num} Plot of normalized 
shell-to-shell energy transfer $T_{nm}/\Pi$ vs. $n-m$ for
$d=3$obtained from numerical simulations on $512^{3}$DNS. The $n$th
shell is $(k_{0}s^{n}:k_{0}s^{n+1})$ with $s=2^{1/4}.$The energy
transfer is maximum for $n=m\pm1$, hence the energy transfer is local
and self-similar. The energy transfer is also forward. Taken from
Verma et al.~\cite{Verma:PP2005}. }
\end{figure}

Verma et al.~\cite{Verma:PP2005} computed the shell-to-shell energy transfer
in 3D fluid turbulence using numerical simulations. Their result is
shown in Fig.~\ref{fig:shell-num}.
Comparison of Fig. \ref{Fig:Fluid-T} and Fig. \ref{fig:shell-num}
shows that theoretical and numerical computation of shell-to-shell
energy transfer are consistent with each other.

Incompressible fluid turbulence is nonlocal in real space due to incompressibility
condition. Field-theoretic calculation also reveals that mode-to-mode
transfer $S(k|p|q)$ is large when $p\ll k$, but small for $k\sim p\sim q$,
hence Navier-Stokes equation is nonlocal in Fourier space too. However,
in 3D shell-to-shell energy transfer rate $T_{nm}$ is forward and
most significant to the next-neighbouring shell. Hence, shell-to-shell
energy transfer rate is local even though the interactions appear
to be nonlocal in both real and Fourier space. Refer to Zhou \cite{Zhou:PF1993},
Domaradzki and Rogallo \cite{Domaradzki:PF1990}, Verma et al. \cite{Verma:PP2005},
and Verma \cite{Verma:Pramana2005proceeding}.

With this we conclude our discussion on shell-to-shell energy transfer
in hydrodynamic turbulence.

\subsection{EDQNM Calculation of Fluid Turbulence \label{sub:EDQNM-Calculation}}

Eddy-damped quasi-normal Markovian (EDQNM) calculation of turbulence
is very similar to the field-theoretic calculation of energy evolution.
This scheme was first invented by Orszag \cite{Orszag:CP1973} for Fluid
turbulence. 

The Navier-Stokes equation is symbolically written as \[
\left(\frac{d}{dt}+\zeta k^{2}\right)X(\mathbf{k},t)=\sum_{\mathbf{p+q=k}}X(\mathbf{p},t)X(\mathbf{q},t),\]
where $X$ stands for the field \textbf{u}, $X(\mathbf{p},t)X(\mathbf{q},t)$
represents all the nonlinear terms, and $\zeta$ is the dissipation
coefficient $(\nu$). The evolution of second and third moment would
be

\begin{eqnarray}
\left(\frac{d}{dt}+2\zeta k^{2}\right)\left\langle X(\mathbf{k},t)X(\mathbf{-k},t)\right\rangle  & = & \sum_{\mathbf{p+q=k}}\left\langle X(\mathbf{-k},t)X(\mathbf{p},t)X(\mathbf{q},t)\right\rangle \label{eq:EDQNM-C}\\
\left(\frac{d}{dt}+\zeta(k^{2}+p^{2}+q^{2})\right)\left\langle X(\mathbf{-k},t)X(\mathbf{p},t)X(\mathbf{q},t)\right\rangle  & = & \sum_{\mathbf{p+q+r+s=0}}\left\langle X(\mathbf{q},t)X(\mathbf{p},t)X(\mathbf{r},t)X(\mathbf{s},t)\right\rangle \nonumber \end{eqnarray}
If $X$ were Gaussian, third-order moment would vanish. However, quasi-normal
approximation gives nonzero triple correlation; here we replace $\left\langle XXXX\right\rangle $
by its Gaussian value, which is a sum of products of second-order
moments. Hence, \begin{eqnarray*}
\left\langle X(\mathbf{-k},t)X(\mathbf{p},t)X(\mathbf{q},t)\right\rangle  & = & \int_{0}^{t}d\tau\exp{-\zeta(k^{2}+p^{2}+q^{2})(t-\tau)}\sum_{\mathbf{p+q=k}}\\
 &  & [\left\langle X(\mathbf{q},\tau)X(-\mathbf{q},\tau)\right\rangle \left\langle X(\mathbf{p},\tau)X(\mathbf{p},\tau)\right\rangle +...],\end{eqnarray*}
where $...$ refers to other products of second-order moments. The
substitution of the above in Eq. (\ref{eq:EDQNM-C}) yields a closed
form equation for second-order correlation functions. Orszag \cite{Orszag:CP1973}
discovered that the solution of the above equation was plagued by
problems like negative energy. To cure this problem, a suitable linear
relaxation operator of the triple correlation (denoted by $\mu$)
was introduced (Eddy-damped approximation). In addition, it was assumed
that the characteristic evolution time of $\left\langle XX\right\rangle \left\langle XX\right\rangle $
is larger than $\left(\mu_{kpq}+\nu(k^{2}+p^{2}+q^{2})\right)^{-1}$(Markovian
approximation). As a result the following form of energy evolution
equation is obtained\begin{equation}
\left(\frac{d}{dt}+2\zeta k^{2}\right)\left\langle X(\mathbf{k},t)X(\mathbf{-k},t)\right\rangle =\int d\mathbf{p}\theta_{kpq}(t)\sum_{\mathbf{p+q=k}}[\left\langle X(\mathbf{q},t)X(-\mathbf{q},t\right\rangle \left\langle X(\mathbf{p},t)X(-\mathbf{p},t\right\rangle +...],\label{eq:EDQNM-final-eqn}\end{equation}
where \[
\theta_{kpq}(t)=\left(1-\exp{-(\mu_{k}+\mu_{p}+\mu_{q})t}\right)/\left(\mu_{k}+\mu_{p}+\mu_{q}\right)\]
with \begin{equation}
\mu_{k}=\left(\nu+\eta\right)k^{2}+C_{s}\left(\int_{0}^{k}dq\left(E^{u}(q)\right)q^{2}\right)^{1/2}.\label{eq:mu-EDQNM}\end{equation}
The first and second terms represent viscous and nonlinear eddy-distortion
rates respectively. Note that homogeneity and isotropy are assumed
in EDQNM analysis too.

The right-hand side of Eq. (\ref{eq:EDQNM-final-eqn}) is very similar
to the perturbative expansion of $S^{uu}(k|p|q)$ (under $t\rightarrow\infty$).
The term $\mu_{k}$ of Eq. (\ref{eq:mu-EDQNM}) is nothing but the
renormalized dissipative parameters. Thus, field-theoretic techniques
for turbulence is quite similar to EDQNM calculation. There is a bit
of difference however. In field-theory, we typically compute asymptotic
energy fluxes in the inertial range. On the contrary, energy is numerically
evolved in EDQNM calculations.

\section{Energy flux in real space; Kolmogorov's four-fifth law (K41)}
\label{sec:K41}

\begin{figure}[htbp]
	\begin{center}
		\includegraphics[scale = 0.8]{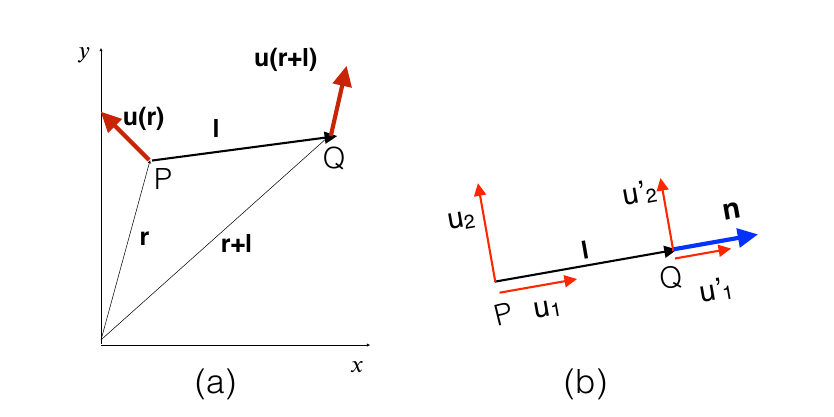}
	\end{center}
	\caption{(a)  The velocity fields at the two points P and Q  are ${\bf u(r)}$ and ${\bf u(r+l)}$ respectively.  (b) The components of velocity fields along ${\bf l}$ are $u_1$ and $u_1'$, while the perpendicular components  are $u_2$, $u_3$, $u_2'$, and $u_3'$.  Here ${\bf n = l}/l$ is the unit vector along ${\bf l}$.  }
	\label{fig:K41:delta_u_r}
\end{figure}

In this section, we briefly describe the kinetic  energy flux in real space.  The kinetic energy flows from one scale to next and to next, $l \rightarrow l/2 \rightarrow l/4 ...$, till the dissipation scales where the energy gets dissipated.   Our focus is on the  formulation  by  \citet{Kolmogorov:DANS1941Structure,Kolmogorov:DANS1941Dissipation}, who connected the third-order structure   function to the energy flux.  For the quantification of correlation and structure functions, we consider two real space points, ${\bf r}$ and ${\bf r+l}$, where the velocity fields are  ${\bf u(r)}$ and ${\bf u(r+l)}$ respectively.  It is convenient to denote ${\bf r+l}$, $u_i({\bf r})$,  $u_i({\bf r+l})$, $\partial/\partial x_i$,  $\partial/\partial x_i'$ using ${\bf r'}$, $u_i$, $u_i'$, $\partial_i$, $\partial_i'$  respectively.  See Fig.~\ref{fig:K41:delta_u_r} for an illustration.

Kolmogorov considered statistically homogeneous, isotropic, and steady turbulence, with forcing employed at  large  scales.  In such flows, the second-order correlation function for the velocity field is~ \cite{Kolmogorov:DANS1941Structure, Batchelor:book:Turbulence} 
\bea 
\la u_i({\bf r}, t) u_j({\bf r+l}, t) \ra = \la u_i u'_j \ra =  C_{ij}(l).
\eea
where $u_i, u_j$ are the velocity components,  $t$ is time, and $\la . \ra$ represents ensemble average.  Note that $C_{ij}(l)$ is a function of $l$ or $|{\bf l}|$, and it is independent of ${\bf r}$, $t$, and orientation of  ${\bf l}$.  Also, the above correlation is for equal time.  Similarly, the third-order longitudinal structure function  $ S_3(l) $ is defined as
\bea
S_3(l)  = \la [{\bf (u'-u)} \cdot \hat{l}]^3 \ra,
\label{eq:K41:S3_l_def}
\eea
where $\hat{l}$ is the unit vector along  vector ${\bf l}$. 

Starting from Navier-Stokes equation, under the assumption of homogeneity  and isotropy,  \citet{Kolmogorov:DANS1941Dissipation} derived  the following evolution equation for $\la u_i u'_j \ra$:
\bea
\frac{\partial}{\partial t} \frac{1}{2}  \la u_i u_i' \ra & = &
\frac{1}{2} \la  u_i' \frac{\partial}{\partial t} u_i \ra + \frac{1}{2}  \la  u_i \frac{\partial}{\partial t} u_i' \ra 
\nonumber \\
& = &\frac{1}{2}  \left[ - \partial_j \la u_i'  (u_j u_i) \ra - \partial_j'  \la u_i (u_j' u_i') \ra
-\cancel{\partial_i \la p u_i' \ra}  -\cancel{\partial_i' \la p' u_i \ra }   \right.
\nonumber \\
&& \left. +  \la u_i' F_{\mathrm{LS},i} \ra +   \la u_i F_{\mathrm{LS},i}' \ra 
+\nu \la u_i' \nabla^2 u_i \ra +\nu \la u_i \nabla'^2 u_i' \ra \right] \nonumber \\
&= & \frac{1}{2}  \left[   \partial_j' \la u_i'  (u_j u_i) \ra - \partial_j'  \la u_i (u_j' u_i') \ra 
+2  \la u_i' F_{\mathrm{LS},i} \ra + 2 \nu \nabla'^2 \la u_i u_i' \ra \right]   \nonumber \\
& = & \frac{1}{4} \nabla_l \cdot \la |{\bf u'-u}|^2 ({\bf u'-u}) \ra  +   \la F_{\mathrm{LS},i}  u_i'  \ra + \nu \nabla'^2 \la u_i u_i' \ra \nonumber \\
& = & T_u({\bf l}) + \mathcal{F}_\mathrm{LS}({\bf l}) - D_u({\bf l}).
\label{eq:K41:d_uiuj_dt}
\eea     
The above derivation makes use of tensorial and symmetry properties (homogeneity and isotropy) of the correlation functions.  For example, $ \la  p u'_i \ra  = \la p' u_i  \ra= 0$ due to isotropy. For  details, refer to  \citet{Kolmogorov:DANS1941Dissipation, Landau:book:Fluid, Frisch:book, Brachet:book_chapter}.  

In Eq.~(\ref{eq:K41:d_uiuj_dt}), $T_u({\bf l})$ corresponds to the spectral energy transfer term  $T_u({\bf k})$; and  $\mathcal{F}_u({\bf l})$ and $ D_u({\bf l})$ are the respective correlations of the energy injection rate and the dissipation rate.  Further, Kolmogorov \citet{Kolmogorov:DANS1941Dissipation} assumed the following:
\begin{enumerate}
	\item  $\partial \la u_i u_i' \ra /\partial t =0$ due to the steady nature of the flow.
	
	\item $\nu \rightarrow 0$, hence the dissipation wavenumber ($k_d$) is at infinity.
	
	\item The  flow is forced at large scales, hence $\mathcal{F}_\mathrm{LS}(l) \approx \epsilon_u$, and it is approximately the same at all scales. 
\end{enumerate}
In addition, we focus on the inertial range,  $1/k_d \ll l \ll L$, where the viscous dissipation  $D_u({\bf l}) = 2 \nu \nabla'^2 \la u_i u_i' \ra \rightarrow 0$. Therefore,  Eq.~(\ref{eq:K41:d_uiuj_dt}) yields
\bea
\mathcal{F}_\mathrm{LS}(l) \approx \epsilon_u \approx - T_u(l) = - \frac{1}{4} \nabla_l \cdot \la |{\bf u'-u}|^2 ({\bf u'-u}) \ra. \label{eq:RS:steady_Kolm_without_visc}
\eea
We denote
\bea
{\bf Q(l)} & = &  \la |{\bf u'-u}|^2 ({\bf u'-u}) \ra, \\
\bar{S}_3(l) & = &  \la |{\bf u'-u}|^2 \{ ({\bf u'-u})\cdot \hat{l} \} \ra .
\eea   
Since ${\bf Q(l)} $ is a isotropic vector, we deduce that
\bea
{\bf Q(l)} = \bar{S}_3(l) \hat{l},
\eea
whose substitution in Eq.~(\ref{eq:RS:steady_Kolm_without_visc}) yields the following equation (in spherical coordinate system for ${\bf l}$):
\bea
-\frac{1}{4} \frac{1}{l^2} \frac{d}{dl} \left[ l^2 \bar{S}_3(l) \right] = \epsilon_u.
\label{eq:RS:S3_bar_epsilon}
\eea
The solution of the above equation is
\bea
\bar{S}_3(l) = -\frac{4}{3} \epsilon_u l,
\label{eq:RS:S3_bar_4by3_rule}
\eea
which is the four-third law. 

It is easy to show that~ \cite{Kolmogorov:DANS1941Dissipation,Frisch:book} 
\bea
\bar{S}_3(l) = \frac{1}{3} \left[ l \frac{d}{dl} S_3(k) + 4 S_3(l) \right] 
= \frac{1}{3l^3}  \frac{d}{dl} \left[ l^4 S_3(k) \right].
\label{eq:RS:S3_S3bar_reln}
\eea
Using Eqs.~(\ref{eq:RS:S3_bar_4by3_rule}, \ref{eq:RS:S3_S3bar_reln}), we  immediately deduce the third-order longitudinal structure function as
\bea
S_3(l)  = \la [{\bf (u'-u) } \cdot \hat{l}]^3 \ra  = -\frac{4}{5} \epsilon_u l.
\label{eq:K41:S3_l}
\eea
This is the Kolmogorov's four-fifth law, which has been verified in various experiments and simulations~\cite{Frisch:book, Davidson:book:Turbulence}.  Using Eqs.~(\ref{eq:RS:S3_bar_4by3_rule}, \ref{eq:K41:S3_l}), we deduce the third-order transverse structure function as
\bea
\la |{\bf u_\perp'-u_\perp}|^2 ({\bf u'-u}) \ra = -\left(4/3-4/5 \right) \epsilon_u l  = -\frac{8}{15} \epsilon_u l,
\eea
where ${\bf u}_\perp$ is the velocity component perpendicular to ${\bf l}$, and $ |{\bf u_\perp'-u_\perp}|^2 =  |{\bf u'-u}|^2 -  |({\bf u'-u})\cdot \hat{l}|^2$.    \citet{Eyink:Nonlinearity2002} derived the above relations earlier. 

The prefactors for the longitudinal and transverse structure functions, $4/5$ and $8/15$, are of the same order. This is consistent with the observations of  \citet{Dhruva:PRE1997} who employed atmospheric turbulence data to deduce that the longitudinal and transverse structure functions are close to each other.   \citet{Shen:PF2002} arrived at a similar conclusion based on their studies on sheared and unshared wind-tunnel turbulence.

The aforementioned four-third and four-fifth laws have been derived under the assumption that the external force is applied at the large scales leading to  $\mathcal{F}_\mathrm{LS}(l) \approx \epsilon_u$, a constant.  However, we can generalise the above laws to {\em isotropic} forcing  employed in the inertial range.  Suppose that  in the inertial range, the correlation function related to energy injection rate  is $\mathcal{F}_u(l)$, then Eq.~(\ref{eq:RS:steady_Kolm_without_visc}) translates to
\bea
\mathcal{F}_u(l)  = - T_u(l) = - \frac{1}{4} \nabla_l \cdot \la |{\bf u'-u}|^2 ({\bf u'-u}) \ra.
\label{eq:RS:steady_Kolm_F_l}
\eea  
whose solution is
\bea
\bar{S}_3(l) = -\frac{4}{ l^2}  \int^l  dl' \mathcal{F}_u(l') l'^2.
\label{eq:RS:S3_bar_with_F_l}
\eea
For   $\mathcal{F}_u(l) = \epsilon_u$, we recover the four-third law.

\begin{figure}[htbp]
	\begin{center}
		\includegraphics[scale = 0.5]{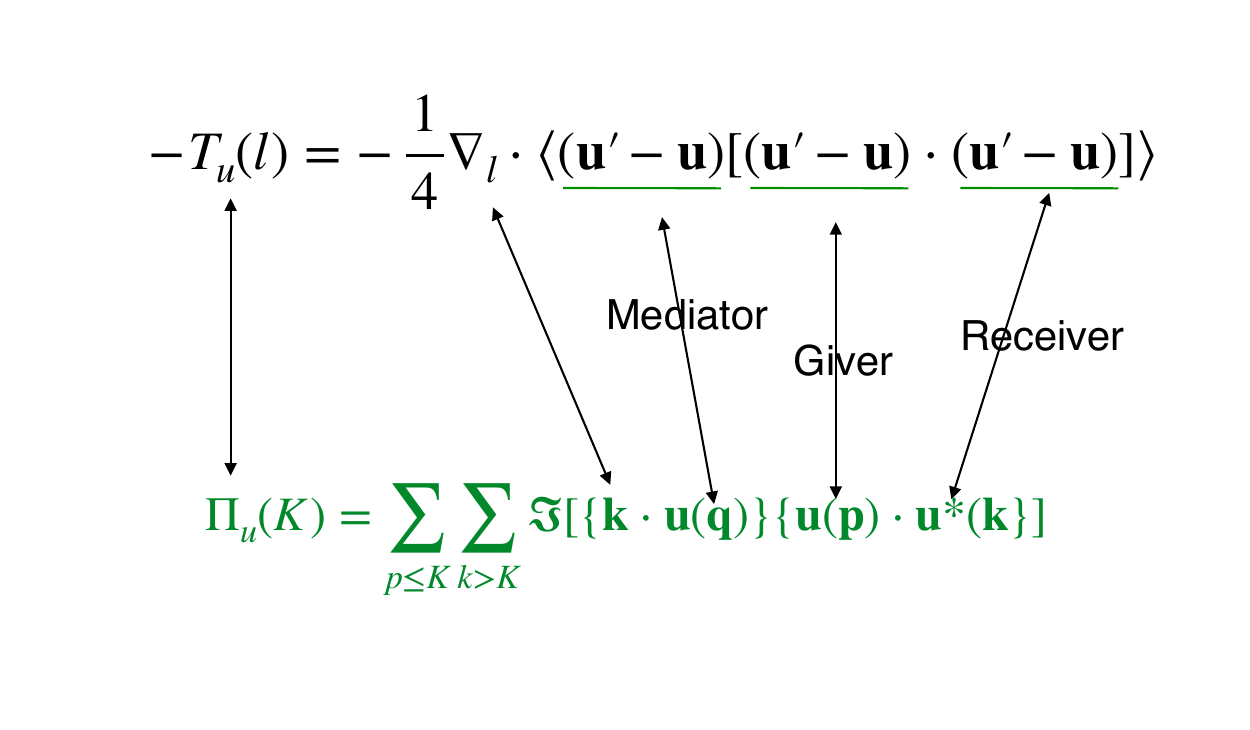}
	\end{center}
	\caption{A figure illustrating connections between  $-T_u(l)$ and the energy flux $\Pi_u(K)$.  We contrast the receiver, giver, and mediator fields in both the formulas.}
	\label{fig:K41:connection}
\end{figure}

The above real-space description is closely related to the corresponding Fourier space description.   We show below that Eq.~(\ref{eq:RS:steady_Kolm_without_visc}) is related to its Fourier space equivalent.   See Fig.~\ref{fig:K41:connection} for an illustration.   In  $-T_u(l)$, the product $|{\bf u'-u}|^2$ includes the giver and receiver fields, while the third $\delta {\bf u}$ is the mediator field.  Since the real space $\delta {\bf u}$  is a linear superposition of many Fourier modes,  $\Pi_u(k)$, which is  the complement of $T_u(l)$, involves  sums over ${\bf p}$ and ${\bf k}$ modes.  Also note that $K \leftrightarrow 1/l$ is the radius of the wavenumber sphere.    

Both, the four-fifth law and  the variable energy flux equation, $d\Pi_u/dk = - \mathcal{F}_u(k) $, are  exact relations in statistical sense.  This is because their derivation is  based on conservation of energy.  An interesting departure however is isotropy.  The Fourier-space definition of flux  has been employed to anisotropic system, for example to quasi-static MHD turbulence after angular averaging.  However, Kolmogorov's four-fifth law assumes isotropy, though there have been attempts to generalise it to anisotropic turbulence~ \cite{Danaila:PD2012b, Ching:book}.     

The skewness $S$ for a turbulent flow is defined using
\bea
\left\langle \left( \frac{\partial u_1}{\partial x_1} \right)^2 \right\rangle^{3/2}/
\left\langle \left( \frac{\partial u_1}{\partial x_1} \right)^3 \right\rangle = \frac{1}{S}.
\eea
\cite{Kolmogorov:DANS1941Dissipation} assumed that
\bea
\la   (\Delta u)_\parallel^2   \ra 
= \la \frac{1}{S} S_3(l) \ra^{2/3}  \sim  \Pi^{2/3} l^{2/3}.  
\eea
Using the above it can be deduced that 
\bea
\la {\bf u(r+l) \cdot u(r)} \ra \sim C - \Pi_u^{2/3} l^{2/3},
\eea
where $C$ is a constant.  The  Fourier transform of the above correlation yields Kolmogorov's spectrum, apart from a constant~ \cite{Batchelor:book:Turbulence}.  

A trivial generalisation  of the above relations to higher order structure functions yields
\bea
S_q(l) =  \la (\Delta u)_\parallel^q \ra = C_q \Pi^{q/3} l^{q/3} \sim l^{\zeta_q}
\eea
leading to  $\zeta_q = q/3$, and $C_q$  as constants.  The exponents $\zeta_q$'s are called  {\em intermittency exponents}.  However, contrary to the above predictions, the experiments and numerical simulations reveal that $\zeta_q < q/3$, specially for large $q$'s; this is the topic of the next section.

\section{Beyond K41; Fluctuations in energy flux}
\label{sec:beyond_K41}

In Kolmogorov's theory of hydrodynamic turbulence, the kinetic energy cascades to intermediate scales and then to smaller scales.   In Fourier space, conservation of energy  for a shell  leads to $\Pi_u(k) = \mathrm{const}$ in the inertial range.  However, this theory does not   provide  information about the distribution of energy  among the daughter eddies.    An uneven distribution of the energy flux among the daughter eddies  leads to fluctuations in  the flux, whose quantification is the topic of the present section.

\citet{Landau:book:Fluid} pointed out that  the viscous dissipation ($\nu \omega^2$) in a turbulent flow is singular.  That is, there are tiny regions of strong viscous dissipation in a sea where the average dissipation is weak.   The above phenomenon,  observed in many experiments and numerical simulations, is due to an uneven distribution of energy flux among the daughter eddies.  This distribution has been studied using various models, which will be presented  In the following discussion.  We refer the reader to  \citet{Frisch:book},  \citet{Dubrulle:JFM2019}, and  \citet{Stolovitzky:RMP1994} for details.

\subsection{Fractal model}

\citet{Frisch:JFM1978} and  \citet{Frisch:book} constructed a fractal-based model, popularly known as {\em $\beta$ model},  for the energy cascade.  They  assumed  fluid structures to be a fractal  with a fractal dimension of $D$. Here,  the fraction of ``active" space in a turbulent cascade decreases as a power law:
\bea
p_l = \beta^n =  \left( \frac{l}{L} \right)^{(3-D)},
\eea
where $l =  2^{-n} L$ is the length scale of the eddies at the $n$th level.  Hence, the energy flux at length scale $l$ would be
\bea
\Pi_l = \frac{u_l^3}{l} p_l = \frac{u_l^3}{l} \left( \frac{l}{L} \right)^{(3-D)}.
\eea
Constancy of energy flux in the inertial range implies that $\Pi_l = \Pi$, i.e.,
\bea
\Pi_l = \frac{U_L^3}{L} .
\eea
This leads to
\bea
u_l = U_L  \left( \frac{l}{L} \right)^{1/3- (3-D)/3}.
\eea
Therefore, in this model, the $q$th-order structure function is
\bea
S_q(l) = \la (\Delta u)_\parallel^q \ra \sim \la u_l^q \ra \sim u_l^q p_l
\sim U_L^q  \left( \frac{l}{L} \right)^{q[1/3- (3-D)/3]}  \left( \frac{l}{L} \right)^{(3-D)}.
\eea
The above expression yields the intermittency exponent as
\bea
\zeta_q = \frac{q}{3} + (3-D)  \left( 1-\frac{q}{3} \right).
\label{eq:K41:beta_zeta}
\eea

\begin{figure}[htbp]
	\begin{center}
		\includegraphics[scale = 0.5]{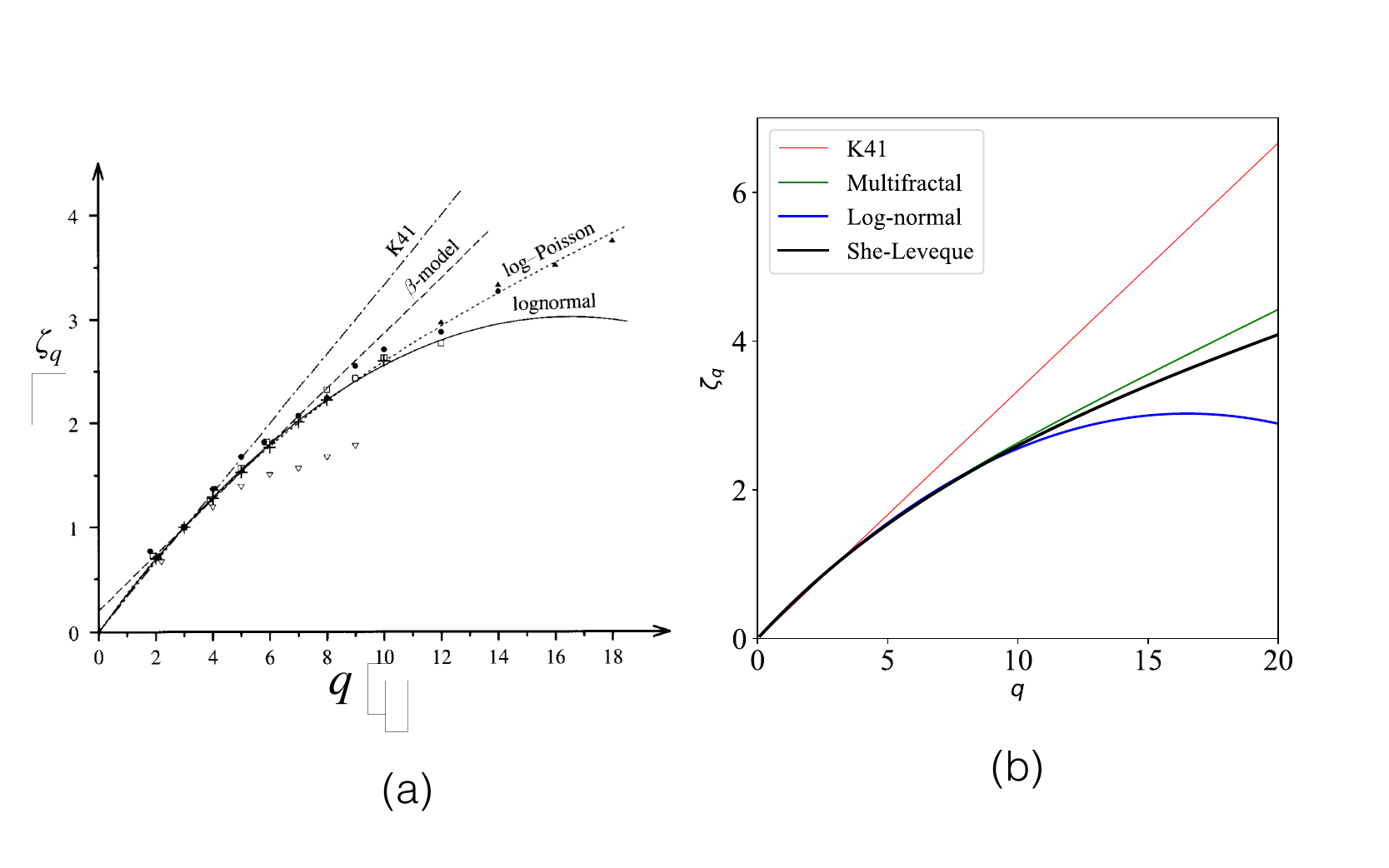}
	\end{center}
	\caption{Plots of $\zeta_q$ vs. $q$ for various models and those computed using experimental data. The experimental results~\cite{Anselmet:JFM1984} are shown as triangles and circles.   The plot of $\beta$-model is constructed using $D=2.8$,  that of multifractal model using $p=0.7$, and  that of log-normal model using $\mu=0.2$.  Figure (a) is adopted  from a figure of \citet{Frisch:book}.   }
	\label{fig:K41:zeta_q}
\end{figure}

Note that $\zeta_q=q/3$ for $D=3$.  This situation corresponds to homogeneous dissipation among all the eddies  at any scale.  Imagine a cube being divided into 8 equal-sized cubes, which are successively cut into 8 cubes each.  For $D=3$, each cube at level 1 will receive $\Pi_u/8$ units of energy flux, and then they will pass on $\Pi_u/64$ units of energy flux to daughter cubes, and  so on.   Such distribution of $\Pi_u$ corresponds to $p_l=1$.    Now contrast the above with structures with $D < 3$.  Here, the energy flux is divided as in fractals, e.g., three-dimensional Sierpinski Gasket  or Menger sponge~ \cite{Addison:book:Fractals}.  These structures, as well the energy cascade in them, are self-similar.  

In Fig.~\ref{fig:K41:zeta_q}(a) we illustrate a plot of $\zeta_q$ vs.~$q$ for $D=2.8$.  The model predictions match with experimental results of  \citet{Anselmet:JFM1984} only till $q\approx 8$~ \cite{Frisch:book}.  The above discrepancy is corrected in multifractal models of turbulence.

\subsection{Multifractal model}
The $\beta$-model of turbulence discussed in the previous subsection assumes the turbulent structures to be a homogeneous fractal.  Experimental observations however reveal that the structures are inhomogeneous, that is, the fractal dimension of the turbulent structures varies with position.  Hence, researchers have proposed multifractal model of turbulence~ \cite{Frisch:CP1985,Frisch:book,Meneveau:PRL1987,Meneveau:NPB1987,Addison:book:Fractals,Grassberger:PD1983}.  In this subsection we illustrate the central idea of multifractality using   \citet{Meneveau:PRL1987}'s model. 
\begin{figure}[htbp]
	\begin{center}
		\includegraphics[scale = 0.4]{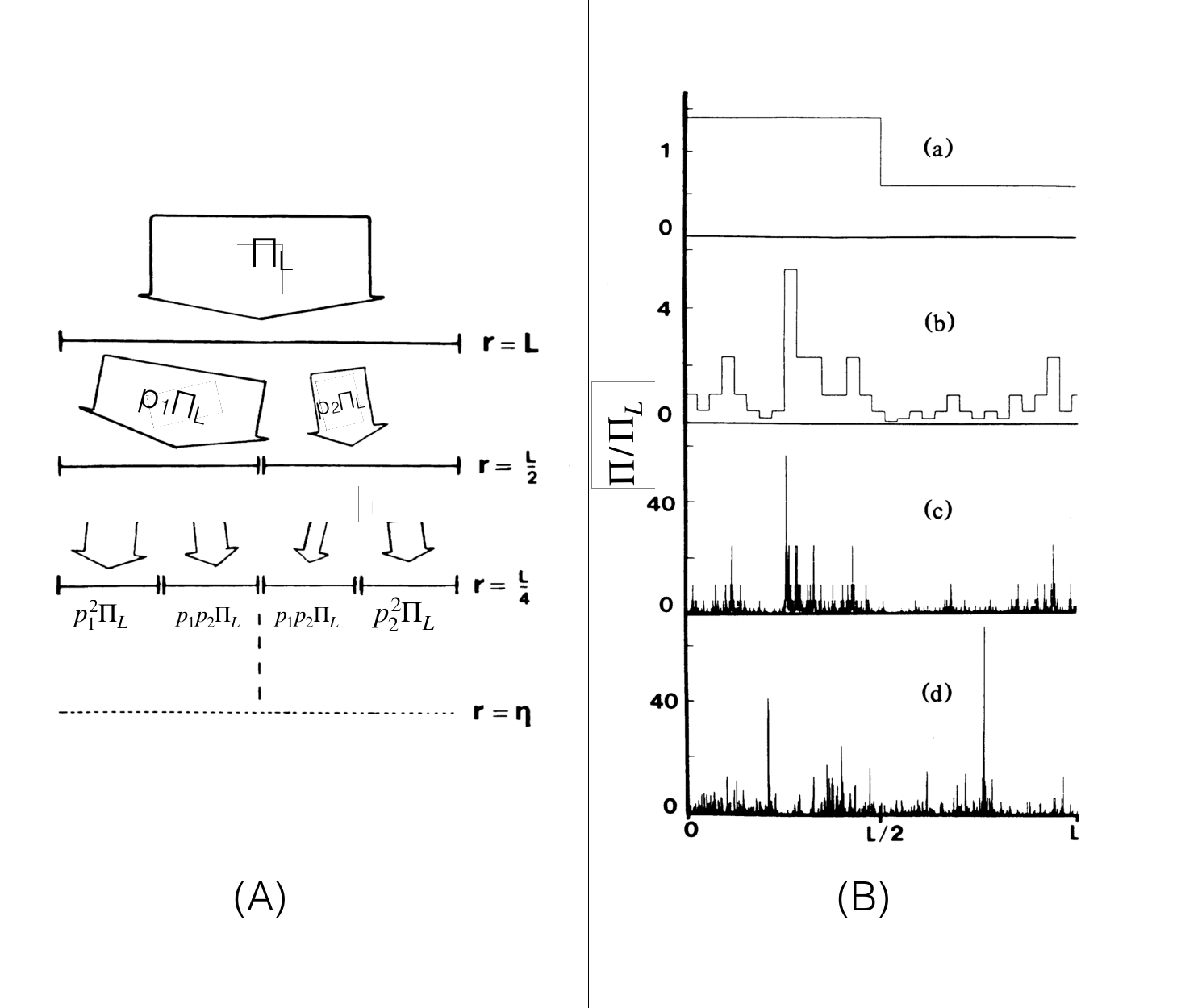}
	\end{center}
	\caption{ (A) A schematic diagram illustrating distribution of energy flux during a one-dimensional cascade process.  In the multifractal model of  \citet{Meneveau:PRL1987}, the two daughter eddies receive fraction  $p_1$ and $p_2 = 1-p_1$ of $\Pi_l$.  This cascade process is continued successively.  (B) The  distribution of $\Pi_l$ at the  (a) first stage, (b) fifth stage, (c) twelfth stage.  (d)  A distribution of $\Pi_l$  in an experimental signal resembles distribution of (c).  From   \citet{Meneveau:PRL1987}.    }
	\label{fig:K41:meneveau}
\end{figure}

Therefore, the fractal dimension for the three-dimensional structures would be
\bea
D_{q} = \bar{D}_q + 2.
\label{eq:K41:Dq_Meneveau2}
\eea
Note that $D_q $ is function of $q$ due to multifractal nature of the structures.  This feature is related to space variability of the fractal dimension~ \cite{Frisch:book,Addison:book:Fractals,Grassberger:PD1983}.   In contrast, for the $\beta$-model, $D_q=D$ for all $q$'s due the fractal  nature of the flow.   The intermittency exponent $\zeta_q$ for the above multifractal model can be computed by substituting  $D_q$ of Eq.~(\ref{eq:K41:Dq_Meneveau2}) in Eq.~(\ref{eq:K41:beta_zeta}) that yields~ \cite{Meneveau:NPB1987}
\bea
\zeta_q = \frac{q}{3} + (3-2-\bar{D}_{q/3})  \left( 1-\frac{q}{3} \right) = \left( \frac{q}{3} -1 \right) \bar{D}_{q/3} + 1.
\label{eq:K41:multifractal_zeta_q}
\eea
We illustrate the above $\zeta_q$ in  Fig.~\ref{fig:K41:zeta_q}(b).  We remark that the deviation of $\zeta_q$ from $q/3$ is  due to the spatial intermittency of  $\Pi_l$, hence it is called {\em intermittency correction}.   \citet{Meneveau:PRL1987,Meneveau:NPB1987} also computed $\zeta_q$ using  data from several experiments (grid turbulence, wake of a circular cylinder, atmospheric turbulence).  The   model $\zeta_q$  with $p=0.7$ provides best fit to the experimental results.

\subsection{Lognormal model}

The  multifractal model of  \citet{Meneveau:PRL1987} is  made probabilistic by   a modified prescription, according to which the incident energy flux for an arbitrary eddy is~ \cite{Kolmogorov:JFM1963,Gurvich:PF1967,Frisch:book}
\bea
\Pi_l = \eta_1 \eta_2 ... \eta_n \Pi = \Pi \prod_i \eta_i ,
\label{eq:K41:Pir}
\eea
where $\eta_i$ is  a random variable.  Since $\Pi_l$ is a product of random variables, it follows a lognormal distribution.  
\bea
\log \Pi_l =   \log \Pi + \sum_i  \log \eta_i
\eea
follows a gaussian distribution, or
\bea
P(\Pi_l) = \frac{1}{\sigma \Pi_l \sqrt{2 \pi}} \exp\left( -\frac{(\log \Pi_l - M)^2}{2\sigma^2} \right),
\label{eq:K41:log_normal_distr}
\eea
where $M$ and $\sigma$ are respectively the mean and standard deviation of $\Pi_l$~ \cite{Kolmogorov:JFM1963,Gurvich:PF1967,Frisch:book}.  

Using the above probability distribution we deduce that
\bea
\frac{\la \Pi_l^q \ra}{\la \Pi_l \ra^q} =  \exp \left( M q+ \frac{1}{2} \sigma^2 q^2 \right).
\label{eq:K41:Pi_l^q}
\eea
Let us postulate that the above ratio is
\bea
\frac{\la \Pi_l^q \ra}{\la \Pi_l \ra^q} =  A_q \left( \frac{L}{l} \right)^{\tau_q},
\eea
where $\tau_q$ and $A_q$ are constants.   The constants $M$ and $\sigma$ are related to each other as follows.  By setting $q=1$ in Eq.~(\ref{eq:K41:Pi_l^q}), we obtain $M = -\sigma^2/2$.  Hence,
\bea
\frac{\la \Pi_l^q \ra}{\la \Pi_l \ra^q} = \exp \left(\frac{1}{2} \sigma^2 q (q-1) \right).
\label{eq:K41:Pi_l_log_normal_0}
\eea
By definition, substitution of $q=2$ in Eq.~(\ref{eq:K41:Pi_l_log_normal_0}) yields the variance of $\Pi_l$ as
\bea
\frac{\la \Pi_l^2 \ra}{ \la \Pi_l \ra^2} = \exp(\sigma^2) \equiv  A \left( \frac{L}{l} \right)^{\mu},
\label{eq:K41:sigma}
\eea
or
\bea
\sigma^2 = \log A + \mu \log(L/l),
\eea
$A_2 =A$, and $\tau_2 = \mu$.    Substitution of  Eq.~(\ref{eq:K41:sigma})  in Eq.~(\ref{eq:K41:Pi_l_log_normal_0}) yields
\bea
\la \Pi_l^q \ra = A_q \la \Pi_l \ra^q   \left( \frac{l}{L} \right)^{-\frac{\mu}{2} q(q-1)},
\label{eq:K41:Pi_l_log_normal}
\eea
and hence,
\bea
\tau_q = -\frac{\mu}{2} q(q-1).
\eea
Therefore, using Eq.~(\ref{eq:K41:Pi_l_log_normal}) we derive
\bea
\la (\Delta_l u)^q \ra & = &  \la \Pi_l^{(q/3)} \ra  l^{q/3}  \approx A_{q/3}  \la \Pi \ra^{q/3} l^{\zeta_q} L^{-\tau_{q/3}},
\eea
where
\bea
\zeta_q =  \frac{q}{3} + \tau_{q/3} = \frac{q}{3} - \frac{\mu}{18} q (q-3).  
\eea
Note that $\zeta_6 = 2-\mu$, a relation that is used to estimate $\mu$.  The experiments~ \cite{Benzi:PRE1993,Frisch:book} reveal that $\mu \approx 0.22$, albeit with significant errors.

The predictions of lognormal model are illustrated in Fig.~\ref{fig:K41:zeta_q} for $\mu=0.2$.  The model predictions match with the experimental data quite well till $q \approx 12$ (see Fig.~\ref{fig:K41:zeta_q}(a)), but they deviate significantly for larger $q$'s.  In the next subsection we present another model that corrects for the above deficiency.

\subsection{She Leveque model}

The intermittency correction ($q/3-\zeta_q$) for the lognormal model is larger than those observed in experiments, especially for large $q$'s.  This is because for large $q$, $\la \Pi_l^q \ra$ is suppressed very strongly  due to  lognormal nature of $\Pi_l$.  In particular, for large $q$, Eq.~(\ref{eq:K41:Pi_l_log_normal}) yields
\bea
\la \Pi_l^q \ra \sim \left(\frac{l}{L} \right)^{-\frac{\mu}{2} q^2}.
\eea
\citet{She:PRL1994} corrected this discrepancy by a very interesting postulate.  They argued that  asymptotically (for large $q$), singular nature of the vortex filaments is expected to follow 
\bea
\Pi_l^{(\infty)} = \frac{\la \Pi_l^{(q+1)} \ra}{ \la \Pi_l^q \ra} \sim   \left(\frac{l}{L} \right)^{-2/3}.
\eea
In other words, for large $q$,
\bea
\la \Pi_l^q \ra \sim \left(\frac{l}{L} \right)^{-2q/3},
\eea
and hence,
\bea
\zeta_q \approx  \frac{q}{3}+  \tau_{q/3} \approx \frac{q}{3}-  \frac{2q}{9} \approx \frac{q}{9}.
\eea
The above prediction appears to match with the experimental observations for $q > 12$ where lognormal distribution tends to fail. Note that  \cite{She:PRL1994}'s ansatz allows for presence of larger fluctuations in $\Pi_l$; it is also termed as {\em log-Poisson model}~ \cite{Dubrulle:PRL1994,She:PRL1995}.

To compute $\zeta_q$ for all $q$'s,  \citet{She:PRL1994} postulated that
\bea
\la \Pi_l^{(q+1)} \ra = \la \Pi_l^{(q)} \ra^\beta  \left(\Pi_l^{(\infty)}\right)^{(1-\beta)}.
\eea
The above equation has a solution:
\bea
\tau_q= -\frac{2}{3} q + 2\left[ 1- \left(\frac{2}{3}\right)^q \right]
\eea
that leads to
\bea
\zeta_q = \frac{q}{9} + 2\left[ 1- \left(\frac{2}{3}\right)^{(q/3)} \right]
\eea
The above $\zeta_q$ fits with the experimental and numerical results quite well, as illustrated in Fig.~\ref{fig:K41:zeta_q}~ \citet{She:PRL1994,Frisch:book}.   

Now let us compare the results of the four models and K41.  She-Leveque's model predicts that $\zeta_2 \approx 0.696 \approx (2/3)+0.03$, thus it deviates from K41 prediction by 0.03.  As a result, the energy spectrum is 
\bea
E_u(k) \sim \Pi^{2/3} k^{-5/3-0.03} L^{0.03}.
\eea
Thus, the energy spectrum and structure function depend on the system size (though weakly), contrary to the assumptions of universal theory of turbulence.  Note however that the constancy of energy flux still holds; it is related to the energy conservation and Kolmogorov's four-fifth law.  The deviation from the scale-symmetric exponent $\zeta_q = q/3$ is due to the multifractal nature of turbulence~ \cite{Frisch:PRSA1991}.  See~\citet{Frisch:book} for more details on symmetry arguments.

As shown in Fig.~\ref{fig:K41:zeta_q}, the predictions of  the four models are quite close to each other for $q=2$ to 5, and those of multifractal, lognormal and She-Leveque's models match till $q \approx 10$~ \cite{Frisch:book}.    For larger $q$'s, only She-Leveque's model provides  good fit to the experimental results.  As described earlier in this section, the deviation from $\zeta_q=q/3$ is due to the fact that the turbulent structures are multifractal (not fractal).  Singular dissipation at small scales contribute more significantly to the structure functions for large $q$'s.  

It is interesting to contrast the distributions of kinetic energy spectrum and  flux.  The former has a power law behaviour in $l$ or wavenumber $k$, while latter exhibits lognormal or log-Poisson distribution.  This difference is essentially due to fact that the energy spectrum is a distribution across various scales. On the contrary, Eq.~(\ref{eq:K41:log_normal_distr}) describes the lognormal distribution of energy flux at a given scale $l$.     The latter distribution occurs in a branching process after many iterations. Note that a sum of $\Pi_l$ for all the eddies at scale $l$ is a constant (statistically), and it equals the total energy flux $\Pi$.  

Branching process is observed in many natural phenomena, for example, in fragmentation, population development, gene propagation, natural growth, etc.  Following this idea, the mass distribution in the universe is conjectured to be log-normal~\cite{Coles:MNRAS1991}; this distribution can be easily explained using multiscale fragmentation process in the universe.  Though cosmic anisotropy is a complex problem, the above simple idea may provide  interesting connections between the observed anisotropy and fragmentation caused by the big bang.    Similarly, branching process in a tree accounts for the lognormal distribution of length of terminal twigs~ \cite{Koyama:PRSB2017}.

\section{Structure function for 2D turbulence}
\label{sec:Struct_fn_2d}

In this section, we briefly describe the structure functions of 2D turbulence~ \cite{Gotoh:PRE1998,Lindborg:JFM1999, Bernard:PRE1999}.  The situation for 2D turbulence is more complex than for 3D counterpart due to forcing at the intermediate wavenumbers.  We inspect the structure functions for $k<k_f$ and $k>k_f$ regimes that exhibit different scaling.  Here we focus only on the inertial range, hence ignore the effects of viscous force and Ekman friction.

For the inertial range in the $k<k_f$ regime, the equation for the velocity correlation function is same as Eq.~(\ref{eq:K41:d_uiuj_dt}).  For the correlation function $\mathcal{F}_u(l)$, it is assumed that
\bea
\mathcal{F}_u(l) \approx -\epsilon_{u,\alpha} \approx -\epsilon_u.
\eea
In addition, the dissipation is assumed to negligible in this range\footnote{The  assumption of constancy of $\mathcal{F}_u(l)$  is questionable because $k_f \gg 1$.  This is unlike 3D turbulence for which $k_f \rightarrow 0$.  Also note that in 2D turbulence, the dissipation is significant at very small $k$.}.  Hence,
\bea
- \epsilon_u & = & - T_u(l) = - \frac{1}{4} \nabla_l \cdot \la |{\bf u'-u}|^2 ({\bf u'-u}) \ra.  
\label{eq:2D:steady_Kolm}
\eea  
We denote
\bea
{\bf Q(l)} & = &  \la |{\bf u'-u}|^2 ({\bf u'-u}) \ra, \\
{S}_3(l) & = &  \la |{\bf u'-u}|^2 \{  ({\bf u'-u})\cdot \hat{ l} \} \ra .
\eea 
Due to the isotropic nature of ${\bf Q(l)} $, we deduce that
\bea
{\bf Q(l)} = {S}_3(l) \hat{ l},
\eea
whose substitution in Eq.~(\ref{eq:2D:steady_Kolm}) yields
\bea
\frac{1}{4} \frac{1}{l} \frac{d}{dl} \left[ l {S}_3(l) \right] = \epsilon_u.
\label{eq:2D:S3_epsilon}
\eea
Note that  the divergence operator for a  2D radially-symmetric vector  field ${\bf Q(l)}= Q_l \hat{ l}$ is $(1/l) d (l Q_l)/dl$.  Hence, the solution of Eq.~(\ref{eq:2D:S3_epsilon}) is~  \cite{Lindborg:JFM1999, Bernard:PRE1999}
\bea
{S}_3(l) = 2  \epsilon_u l.
\label{eq:2D:S3_bar}
\eea
Contrast the above relation with Eq.~(\ref{eq:RS:S3_bar_4by3_rule}) for 3D hydrodynamic turbulence.  The opposite sign of Eq.~(\ref{eq:2D:S3_bar}) indicates inverse energy cascade for 2D hydrodynamic turbulence.

The above relation was first derived by  \citet{Lindborg:JFM1999} and  \citet{Bernard:PRE1999}. Further, using the symmetry properties of the tensors,  \citet{Lindborg:JFM1999} showed that
\bea
\la [{\bf (u'-u)} \cdot \hat{l}]^3 \ra   = \frac{3}{4} {S}_3(l).
\eea
Therefore,
\bea
\la [{\bf (u'-u)} \cdot \hat{l}]^3 \ra & = &   \frac{3}{2} \epsilon_u l, 
\label{eq:2D:long_S3}\\
\la {\bf (u'_\perp-u_\perp)}^2  \{ {\bf (u'-u)} \cdot \hat{l}\}  \ra & = &  \frac{1}{2} \epsilon_u l, 
\label{eq:2D:transverse_S3}
\eea
where ${\bf u}_\perp$ is the velocity component perpendicular to ${\bf l}$.  The relation of Eq.~(\ref{eq:2D:long_S3})  has been verified numerically by  \citet{Boffetta:PRE2000}.  

In the $k > k_f$ regime, the enstrophy flux dominates the energy flux. Hence, the structure function for the vorticity is reported for this regime~ \cite{Gotoh:PRE1998,Lindborg:JFM1999,Bernard:PRE1999}.  Following similar arguments as that for passive scalar, one can derive the evolution equation for the vorticity correlation function $\la \omega \omega' \ra$ as
\bea
\frac{\partial}{\partial t} \frac{1}{2}  \la \omega \omega' \ra & = &\frac{1}{4} \nabla_l \cdot \la |\omega-\omega'|^2 ({\bf u'-u}) \ra  +   \la F_{\omega} \omega'  \ra + \nu \nabla'^2 \la  \omega \omega' \ra \nonumber \\
& = & T_\omega({\bf l}) + \mathcal{F}_\omega({\bf l}) - D_\omega({\bf l}).
\label{eq:2D:omega_evol}
\eea     
Hence, in the inertial range,
\bea
\la  |\omega-\omega'|^2 ({\bf u'-u})\cdot \hat{ l} \ra = -2  \epsilon_\omega l,
\label{eq:2D:vorticity_struct_fn}
\eea
where $ \epsilon_\omega $ is the enstrophy dissipation rate.   Further,  \citet{Lindborg:JFM1999} generalised Eq.~(\ref{eq:2D:omega_evol}) to
\bea
\frac{\partial}{\partial t} \frac{1}{2} \la \omega \omega' \ra & = &\frac{1}{4} \nabla_l^2 \nabla_l \cdot \la |{\bf u - u'} |^2 ({\bf u'-u}) \ra + \mathcal{F}_\omega({\bf l}) - D_\omega({\bf l}).
\label{eq:2D:omega_evol_2}
\eea    
From the above equation, it is straightforward to derive that  \cite{Lindborg:JFM1999} 
\bea
\la  |{\bf u'-u}|^2 ({\bf u'-u})\cdot \hat{ l} \ra = -\frac{1}{8} \epsilon_\omega l.
\label{eq:2D:vorticity_struct_fn_Lindborg}
\eea
\citet{Gotoh:PRE1998} started from Eq.~(\ref{eq:2D:vorticity_struct_fn}) and derived that $E_u(k) \propto  k^{-(3+\delta)}$ for $k<k_{d2D}$, and $E_u(k) \propto  k^{-(3+\delta)/2} \exp(-\alpha_2 k/k_{d2D})$ for $k>k_{d2D}$, where $\alpha_2, \delta$ are constants.   \cite{Gotoh:PRE1998} also verified the above scaling using numerical simulations.

Regarding the higher order structure function,  \cite{Boffetta:PRE2000} and  \citet{Tabeling:PR2002} argue that in the $k<k_f$ regime, $\zeta_q \approx q/3$, thus signalling an absence of intermittency  in 2D turbulence.  These authors also claim that in the inertial range, the probability distribution function of  velocity increments  is close to gaussian, which is consistent with the non-intermittent nature of the flow.      \cite{Boffetta:ARFM2012} however show that longitudinal velocity increment deviates  from gaussian distribution. These issues may require further investigation.  Using numerical simulations,  \citet{Babiano:PRE1995} and  \citet{Babiano:PRE1997} computed the multiscaling exponents for the $k<k_f$ regime of 2D turbulence.   \citet{Babiano:PRE1997} related the intermittency exponents to the homogeneity and stationarity of the  transfers in the inverse energy cascade.

\subsection{Field-theoretic description of energy flux and  intermittency}
\label{subsec:K41:field-theory-hydro}

K41 is an analytical theory of turbulence.    In addition, there are several field-theoretic calculations of turbulence, which are first-principle treatment of the problem.  Field theory is a complex and vast field, and it is covered in several books, and hundreds of papers. Hence, it is impossible to describe it in a short amount of space. Here, we provide a brief summary of the main results of this topic.
One of first field-theoretic calculations of turbulent flow was by  \cite{Kraichnan:JFM1959} who employed direct interaction approximation (DIA) to compute the {\em effective viscosity} and energy flux.  Later works in this field include  \cite{Wyld:AP1961,Orszag:CP1973,Yakhot:JSC1986, McComb:book:Turbulence,McComb:book:HIT,DeDominicis:PRA1979,Zhou:PR2010,Adzhemyan:book:RG,Zakharov:book:WaveTurb,Verma:PR2004}, etc. These works provide renormalized viscosity as well as energy flux; here we sketch the energy flux aspects of the above works.

It is very hard to compute the fluctuations in the energy flux or dissipation rate using field theory.  The computations, if successful, would have yielded the intermittency exponents.  In spite of many valiant attempts, there is no fully consistent calculation that achieves this objective.   \citet{Belinicher:JSP1998} developed a field-theoretic procedure to compute the scaling exponents $\zeta_q$ of the $q$th-order structure function.  In a series of papers,  \citet{Lvov:PRE1995I, Lvov:PRE1995II, Lvov:PRE1996} employed exact resumption of all the Feynman diagrams of hydrodynamic turbulence and derived scaling relations among the intermittency exponents.  Their perturbative theory is divergence-free both in infrared and ultraviolet regime.  These results are summarized as {\em fusion rules}~\cite{Lvov:PRL1996}.    \citet{Fairhall:PRL1997} showed that the predictions of the fusion rules are in good agreement with the experimental results of atmospheric turbulence.    Another important and related issue is Euler singularity and dissipative anomaly, which are discussed in  \citet{Onsagar:Nouvo1949_SH,Frisch:book}, and  \citet{Eyink:RMP2006}.    Using mode coupling and renormalization group method,  \citet{Das:EPL1994} computed the second-order correlation of the energy flux, $\la \Pi_u({\bf r}) \Pi_u({\bf r+l}) \ra$.     \citet{Gurarie:PRE1996} and  \citet{Apolinario:PRE2019} have employed instantons to derive the exponential tails of the probability distribution function of the velocity gradients of  randomly forced Burgers equation.  Similar methods are being applied to  hydrodynamic turbulence.    The above calculations are quite complex; the reader is referred to the original paper for details.


\section{Conclusion}
In this short review, we describe basics of Kolmogorov's theory of turbulence, energy transfers in turbulence, and field-theoretic description of hydrodynamic turbulence.  We briefly cover intermittency as well. 

\appendix

\section{Fourier Series vs. Fourier Transform for Turbulent Flows\label{sec:Fourier-Series}}

In statistical theory turbulence we typically assume the flow field
to be homogeneous. Therefore, Fourier transform is not applicable
to these flows in strict sense. However, we can define these quantities
by taking limits carefully. This issue has been discussed by Batchelor
\cite{Batchelor:book:Turbulence} and McComb \cite{McComb:book:Turbulence,McComb:book:HIT} We briefly discuss
them here because they form the basis of the whole paper.

A periodic function $\mathbf{u}(\mathbf{x})$ in box $L^{d}$ can
be expanded using Fourier series as following: 

\begin{eqnarray}
\mathbf{u}\left(\mathbf{x}\right) & = & \sum\hat{\mathbf{u}}\left(\mathbf{k}\right)\exp\left(i\mathbf{k\cdot x}\right),\\
\mathbf{\hat{u}(k}) & = & \frac{1}{L^{d}}\int d\mathbf{xu}\left(\mathbf{x}\right)\exp\left(-i\mathbf{k\cdot x}\right),\end{eqnarray}
where $d$ is the space dimensionality. When we take the limit $L\rightarrow\infty$,
we obtain Fourier transform. Using $\mathbf{u}(\mathbf{k})=\mathbf{\hat{\mathbf{u}}}(\mathbf{k})L^{d}$,
 it can be easily shown that\begin{eqnarray}
\mathbf{u}\left(\mathbf{x}\right) & = & \int\frac{d\mathbf{k}}{(2\pi)^{d}}\mathbf{u}\left(\mathbf{k}\right)\exp\left(i\mathbf{k\cdot x}\right),\\
\mathbf{u}(\mathbf{k}) & = & \int d\mathbf{xu}\left(\mathbf{x}\right)\exp\left(-i\mathbf{k\cdot x}\right),\end{eqnarray}
with integrals performed over the whole space. Note however that Fourier
transform (integral converges) makes sense when $u(x)$ vanishes as
$|x|\rightarrow\infty$, which is not the case for homogeneous flows.
However, correlations defined below are sensible quantities. Using
the above equations, we find that\begin{eqnarray}
\left\langle u_{i}(\mathbf{k})u_{j}(\mathbf{k'})\right\rangle  & = & \int d\mathbf{x}d\mathbf{x'}\left\langle u_{i}(\mathbf{x})u_{j}(\mathbf{x'})\right\rangle \exp-i(\mathbf{k}\dot{\cdot\mathbf{x}+\mathbf{k'}\cdot\mathbf{x'})}\nonumber \\
 & = & \int d\mathbf{r}C_{ij}(\mathbf{r})\exp-i\mathbf{k}\cdot\mathbf{r}\int d\mathbf{x}\exp-i(\mathbf{k}\dot{+\mathbf{k')}\cdot\mathbf{x}}\nonumber \\
 & = & C_{ij}(\mathbf{k})(2\pi)^{d}\delta(\mathbf{k+k'})\label{eq:ui_uj}\end{eqnarray}
We have used the fact that $\delta(\mathbf{k})\approx L^{d}/(2\pi)^{d}$.
The above equation holds the key. In experiments we measure correlation
function $C(\mathbf{r})$ which is finite and decays with increasing
$r$, hence spectra $C(\mathbf{k})$ is well defined. Now energy spectrum
as well as total energy can be written in terms of $C(\mathbf{k})$
as the following:

\begin{eqnarray}
\left\langle u^{2}\right\rangle =\frac{1}{L^{d}}\int d\mathbf{x}u^{2}=\sum_{\mathbf{k}}\left|\mathbf{\hat{u}(k)}\right|^{2} & = & \frac{1}{L^{d}}\int\frac{d\mathbf{k}}{(2\pi)^{d}}\left\langle \left|\mathbf{u(k)}\right|^{2}\right\rangle 
\nonumber \\
 & = & (d-1)\int\frac{d\mathbf{k}}{(2\pi)^{d}}C(\mathbf{k)}\end{eqnarray}
We have used the fact that $\delta(\mathbf{k})\approx L^{d}/(2\pi)^{d}$.
Note that $\left\langle \left|\mathbf{u(k)}\right|^{2}\right\rangle =(d-1)C(\mathbf{k})L^{d}$
{[}see Eq. (\ref{eq:ui_uj}){]} is not well defined in the limit $L\rightarrow\infty$.

In conclusion, the measurable quantity in homogeneous turbulence is
the correlation function, which is finite and decays for large $r$.
Therefore, energy spectra etc. are well defined objects in terms of
Fourier transforms of correlation functions.

We choose a finite box, typically $(2\pi)^{d}$, in spectral simulations
for fluid flows. For these problems we express the equations \emph{(incompressible}
MHD) in terms of Fourier series. We write them below for reference.
\begin{eqnarray}
\left(\frac{\partial}{\partial t}+\nu k^{2}\right)\hat{u}_{i}(\mathbf{k},t) & = & -ik_{i}\hat{p}_{tot}\left(\mathbf{k},t\right)-ik_{j}\sum[\hat{u}_{j}(\mathbf{q},t)\hat{u}_{i}(\mathbf{p},t)]
\end{eqnarray}

The energy spectrum can be computed using $\hat{u}_{i}(\mathbf{k},t)$:\be
\int E(k)dk=\sum\left|\mathbf{\hat{u}}(\mathbf{k})\right|^{2}/2=\int d\mathbf{n}\left|\mathbf{\hat{u}}(\mathbf{k})\right|^{2}/2=\int d\mathbf{k}\left|\mathbf{\hat{u}}(\mathbf{k})\right|^{2}/2\ee
where $\mathbf{n}$ is the lattice vector in $d$-dimensional space.
The above equation implies that\be
E(k)=\frac{\left|\mathbf{\hat{u}}(\mathbf{k})\right|^{2}}{2}S_{d}k^{d-1}.\ee

A natural question is why the results of numerical simulations or
experiments done in a finite volume should match with those obtained
for infinite volume. The answer is straight forward. When we go from
size $2\pi$ to $L$, the wavenumbers should be scaled by $(2\pi)/L$.
The velocity and frequency should be should be scaled by $(2\pi)/L$
and $\left[(2\pi)/L\right]^{2}$ to keep dimensionless $\nu$ fixed.
The evolution of the two systems will be identical apart from the
above factors. Hence, numerical simulations in a box of size $2\pi$
can capture the behaviour of a system with $L\rightarrow\infty$,
for which Fourier transform in defined.

\section{ Perturbative Calculation of Navier Stokes Equation}

The Navier Stokes can be written as~\cite{Leslie:book}
\begin{equation}
u_i(\hat{k}) = G(\hat{k}) 
-\frac{i}{2} P^+_{ijm}({\bf k}) 
\int d\hat{p} [ u_j (\hat{p}) u_m (\hat{k}-\hat{p}) ] 
\end{equation}
where the Greens function $G$ 
can be written as 
\begin{equation}
G^{-1}(k,\omega) = -i\omega - \Sigma^{uu}  
\end{equation}

We solve the above equation perturbatively keeping the
terms upto the first nonvanishing order.  The corresponding
Feynmann diagram is~\cite{Wyld:AP1961,Yakhot:JSC1986}

\unitlength=1mm
\begin{fmffile}{fmfub1}
\begin{equation}
\parbox{20mm}
{\begin{fmfgraph*}(20,15) 
  \fmfleft{i} \fmfright{o} \fmf{plain,label=$u_i$}{i,o}
\end{fmfgraph*}} = 
\parbox{20mm}
{\begin{fmfgraph*}(20,15) 
  \fmfleft{i} \fmfright{o1,o2} 
  \fmfv{d.sh=circle,d.f=full,d.si=0.1w}{v}
  \fmf{photon,label=$G$}{i,v} 
  \fmf{plain,label=$u_m$,l.d=5}{v,o1}
  \fmf{plain,label=$u_j$,l.d=5}{v,o2}
\end{fmfgraph*}}
\end{equation}

\end{fmffile}


The solid line represents fields $u$, and the wiggly line  (photon) denotes
$G$.  The filled
circle denotes $-(i/2) P^+_{ijm}$ vertex.  These diagrams appear in
renormalization calculations as well as in energy flux calculations.

\subsection{Viscosity Renormalization}

The expansion of $u$ in terms of Feynman diagrams are given below:
\unitlength=1mm
\begin{fmffile}{fmfub2}
\begin{eqnarray}
I^u & = & 
 \parbox{20mm}
{\begin{fmfgraph*}(20,15) 
  \fmfleft{i} \fmfright{o1,o2} 
  \fmfv{d.sh=circle,l.a=180,l.d=.1w,d.f=full,d.si=0.1w}{v}
  \fmf{phantom}{i,v}
  \fmf{plain,label=$<$,l.d=5}{v,o2}
  \fmf{plain,label=$<$,l.d=5}{v,o1}
\end{fmfgraph*}}  +
 \parbox{20mm}
{\begin{fmfgraph*}(20,15) 
  \fmfleft{i} \fmfright{o1,o2} 
  \fmfv{d.sh=circle,l=2,l.a=180,l.d=.1w,d.f=full,d.si=0.1w}{v}
  \fmf{phantom}{i,v}
  \fmf{plain,label=$>$,l.d=5}{v,o2}
  \fmf{plain,label=$<$,l.d=5}{v,o1}
\end{fmfgraph*}}   +
 \parbox{20mm}
{\begin{fmfgraph*}(20,15) 
  \fmfleft{i} \fmfright{o1,o2} 
  \fmfv{d.sh=circle,l.a=180,l.d=.1w,d.f=full,d.si=0.1w}{v}
  \fmf{phantom}{i,v}
  \fmf{plain,label=$>$,l.d=5}{v,o2}
  \fmf{plain,label=$>$,l.d=5}{v,o1}
\end{fmfgraph*}} 
\end{eqnarray}
\end{fmffile}

Factor of 2 appears in $I^u$ because of $<>$ symmetry in the corresponding
term. To zeroth order, the terms with $<>$ are zero because of quasi-gaussian
nature of $>$ modes.  To the next order in perturbation, the third term
of $I^u$ is
\unitlength=1mm
\begin{fmffile}{fmfub3}
\begin{eqnarray}
 \parbox{20mm}
{\begin{fmfgraph*}(20,15) 
  \fmfleft{i} \fmfright{o1,o2} 
  \fmfv{d.sh=circle,d.f=full,d.si=0.1w}{v}
  \fmf{phantom}{i,v}
  \fmf{plain,label=$>$,l.d=5}{v,o2}
  \fmf{plain,label=$<$,l.d=5}{v,o1}
\end{fmfgraph*}}
& = &
\parbox{20mm}
{\begin{fmfgraph*}(20,15) 
  \fmfv{d.sh=circle,l=2,l.a=180,l.d=0.1w,d.f=full,d.si=0.1w}{v1}
  \fmfv{d.sh=circle,d.f=full,d.si=0.1w}{v2}
  \fmfleft{i}  
  \fmf{phantom}{i,v1}
  \fmf{photon,tension=0.1,left}{v1,v2}
  \fmf{plain,right,label=$<>$}{v1,v2} 
  \fmfright{o}
  \fmf{plain,label=$>$,l.d=5}{v2,o} 
\end{fmfgraph*}}  +
\parbox{20mm}
{\begin{fmfgraph*}(20,15) 
  \fmfv{d.sh=circle,d.f=full,d.si=0.1w}{v1}
  \fmfv{d.sh=circle,d.f=full,d.si=0.1w}{v2}
  \fmfleft{i}  
  \fmf{phantom}{i,v1}
  \fmf{photon,tension=0.1,left}{v1,v2}
  \fmf{plain,right,label=$<>$}{v1,v2} 
  \fmfright{o}
  \fmf{plain,label=$<$,l.d=5}{v2,o} 
\end{fmfgraph*}}  +
\parbox{20mm}
{\begin{fmfgraph*}(20,15) 
  \fmfv{d.sh=circle,d.f=full,d.si=0.1w}{v1}
  \fmfv{d.sh=circle,d.f=full,d.si=0.1w}{v2}
  \fmfleft{i}  
  \fmf{phantom}{i,v1}
  \fmf{photon,tension=0.1,left}{v1,v2}
  \fmf{plain,right,label=$<<$}{v1,v2} 
  \fmfright{o}
  \fmf{plain,label=$>$,l.d=5}{v2,o} 
\end{fmfgraph*}}  +
\parbox{20mm}
{\begin{fmfgraph*}(20,15) 
  \fmfv{d.sh=circle,l=2,l.a=180,l.d=.1w,d.f=full,d.si=0.1w}{v1}
  \fmfv{d.sh=circle,d.f=full,d.si=0.1w}{v2}
  \fmfleft{i}  
  \fmf{phantom}{i,v1}
  \fmf{photon,tension=0.1,left}{v1,v2}
  \fmf{plain,right,label=$<<$}{v1,v2} 
  \fmfright{o}
  \fmf{plain,label=$<$,l.d=5}{v2,o} 
\end{fmfgraph*}} -  \nonumber \\
& & 
\parbox{20mm}
{\begin{fmfgraph*}(20,15) 
  \fmfv{d.sh=circle,label=2,l.a=180,l.d=.1w,d.f=full,d.si=0.1w}{v1}
  \fmfv{d.sh=circle,d.f=full,d.si=0.1w}{v2}
  \fmfleft{i}  
  \fmf{phantom}{i,v1}
  \fmf{photon,tension=0.1,left}{v1,v2}
  \fmf{dots,right,label=$<>$}{v1,v2} 
  \fmfright{o}
  \fmf{dashes,label=$>$,l.d=5}{v2,o} 
\end{fmfgraph*}}  -
\parbox{20mm}
{\begin{fmfgraph*}(20,15) 
  \fmfv{d.sh=circle,d.f=full,d.si=0.1w}{v1}
  \fmfv{d.sh=circle,d.f=full,d.si=0.1w}{v2}
  \fmfleft{i}  
  \fmf{phantom}{i,v1}
  \fmf{photon,tension=0.1,left}{v1,v2}
  \fmf{dots,right,label=$<>$}{v1,v2} 
  \fmfright{o}
  \fmf{dashes,label=$<$,l.d=5}{v2,o} 
\end{fmfgraph*}}  -
\parbox{20mm}
{\begin{fmfgraph*}(20,15) 
  \fmfv{d.sh=circle,d.f=full,d.si=0.1w}{v1}
  \fmfv{d.sh=circle,d.f=full,d.si=0.1w}{v2}
  \fmfleft{i}  
  \fmf{phantom}{i,v1}
  \fmf{photon,tension=0.1,left}{v1,v2}
  \fmf{dots,right,label=$<<$}{v1,v2} 
  \fmfright{o}
  \fmf{dashes,label=$>$,l.d=5}{v2,o} 
\end{fmfgraph*}}  -
\parbox{20mm}
{\begin{fmfgraph*}(20,15) 
  \fmfv{d.sh=circle,label=2,l.a=180,l.d=.1w,d.f=full,d.si=0.1w}{v1}
  \fmfv{d.sh=circle,d.f=full,d.si=0.1w}{v2}
  \fmfleft{i}  
  \fmf{phantom}{i,v1}
  \fmf{photon,tension=0.1,left}{v1,v2}
  \fmf{dots,right,label=$<<$}{v1,v2} 
  \fmfright{o}
  \fmf{dashes,label=$<$,l.d=5}{v2,o} 
\end{fmfgraph*}} + \nonumber \\
& & 
\parbox{20mm}
{\begin{fmfgraph*}(20,15) 
  \fmfv{d.sh=circle,d.f=full,d.si=0.1w}{v1}
  \fmfv{d.sh=circle,d.f=empty,d.si=0.1w}{v2}
  \fmfleft{i}  
  \fmf{phantom}{i,v1}
  \fmf{photon,width=1,tension=0.1,left}{v1,v2}
  \fmf{plain,right,label=$<>$}{v1,v2} 
  \fmfright{o}
  \fmf{dashes,label=$>$,l.d=5}{v2,o} 
\end{fmfgraph*}} +
\parbox{20mm}
{\begin{fmfgraph*}(20,15) 
  \fmfv{d.sh=circle,d.f=full,d.si=0.1w}{v1}
  \fmfv{d.sh=circle,d.f=empty,d.si=0.1w}{v2}
  \fmfleft{i}  
  \fmf{phantom}{i,v1}
  \fmf{photon,width=1,tension=0.1,left}{v1,v2}
  \fmf{plain,right,label=$<>$}{v1,v2} 
  \fmfright{o}
  \fmf{dashes,label=$<$,l.d=5}{v2,o} 
\end{fmfgraph*}} +
\parbox{20mm}
{\begin{fmfgraph*}(20,15) 
  \fmfv{d.sh=circle,d.f=full,d.si=0.1w}{v1}
  \fmfv{d.sh=circle,d.f=empty,d.si=0.1w}{v2}
  \fmfleft{i}  
  \fmf{phantom}{i,v1}
  \fmf{photon,width=1,tension=0.1,left}{v1,v2}
  \fmf{plain,right,label=$<<$}{v1,v2} 
  \fmfright{o}
  \fmf{dashes,label=$>$,l.d=5}{v2,o} 
\end{fmfgraph*}} + 
\parbox{20mm}
{\begin{fmfgraph*}(20,15) 
  \fmfv{d.sh=circle,d.f=full,d.si=0.1w}{v1}
  \fmfv{d.sh=circle,d.f=empty,d.si=0.1w}{v2}
  \fmfleft{i}  
  \fmf{phantom}{i,v1}
  \fmf{photon,width=1,tension=0.1,left}{v1,v2}
  \fmf{plain,right,label=$<<$}{v1,v2} 
  \fmfright{o}
  \fmf{dashes,label=$<$,l.d=5}{v2,o} 
\end{fmfgraph*}} +  \nonumber \\
& & 
\parbox{20mm}
{\begin{fmfgraph*}(20,15) 
  \fmfv{d.sh=circle,d.f=full,d.si=0.1w}{v1}
  \fmfv{d.sh=circle,d.f=empty,d.si=0.1w}{v2}
  \fmfleft{i}  
  \fmf{phantom}{i,v1}
  \fmf{photon,width=1,tension=0.1,left}{v1,v2}
  \fmf{dots,right,label=$<>$}{v1,v2} 
  \fmfright{o}
  \fmf{plain,label=$>$,l.d=5}{v2,o} 
\end{fmfgraph*}} +
\parbox{20mm}
{\begin{fmfgraph*}(20,15) 
  \fmfv{d.sh=circle,d.f=full,d.si=0.1w}{v1}
  \fmfv{d.sh=circle,d.f=empty,d.si=0.1w}{v2}
  \fmfleft{i}  
  \fmf{phantom}{i,v1}
  \fmf{photon,width=1,tension=0.1,left}{v1,v2}
  \fmf{dots,right,label=$<>$}{v1,v2} 
  \fmfright{o}
  \fmf{plain,label=$<$,l.d=5}{v2,o} 
\end{fmfgraph*}} +
\parbox{20mm}
{\begin{fmfgraph*}(20,15) 
  \fmfv{d.sh=circle,d.f=full,d.si=0.1w}{v1}
  \fmfv{d.sh=circle,d.f=empty,d.si=0.1w}{v2}
  \fmfleft{i}  
  \fmf{phantom}{i,v1}
  \fmf{photon,width=1,tension=0.1,left}{v1,v2}
  \fmf{dots,right,label=$<<$}{v1,v2} 
  \fmfright{o}
  \fmf{plain,label=$>$,l.d=5}{v2,o} 
\end{fmfgraph*}} +
\parbox{20mm}
{\begin{fmfgraph*}(20,15) 
  \fmfv{d.sh=circle,d.f=full,d.si=0.1w}{v1}
  \fmfv{d.sh=circle,d.f=empty,d.si=0.1w}{v2}
  \fmfleft{i}  
  \fmf{phantom}{i,v1}
  \fmf{photon,width=1,tension=0.1,left}{v1,v2}
  \fmf{dots,right,label=$<<$}{v1,v2} 
  \fmfright{o}
  \fmf{plain,label=$<$,l.d=5}{v2,o} 
\end{fmfgraph*}} \label{eq:feyn_I3ub} \nonumber \\
& & + \mbox{higer oder diagrams}
\label{eq:feyn_I3u}
\end{eqnarray} 
\end{fmffile}   


In the above diagrams solid lines denote $<u({\bf k}) u({\bf k'})>$. 
  As mentioned earlier, the wiggly  line denotes 
Green's functions.  All the diagrams except 4,8,12, and 16th can be
shown to be trivially zero using
Eqs.~(\ref{eqn:avgbegin},\ref{eqn:avgend}).  We assume that 4,8,12,
and 16th diagrams are also zero, as usually done in turbulence RG
calculations \cite{Yakhot:JSC1986,McComb:book:Turbulence,Zhou:PRA1988,Zhou:PRA1988,Zhou:PR2010}.
Hence, the term is zero. Now we are left with $>>$ terms 
(3rd term of $I^u$), which is

\unitlength=1mm
\begin{fmffile}{fmfub4}
\begin{eqnarray}
I_3^u & = &
\parbox{20mm}
{\begin{fmfgraph*}(20,15) 
  \fmfleft{i} \fmfright{o1,o2} 
  \fmfv{d.sh=circle,d.f=full,d.si=0.1w}{v}
  \fmf{phantom}{i,v}
  \fmf{plain,label=$>$,l.d=5}{v,o2}
  \fmf{plain,label=$>$,l.d=5}{v,o1}
\end{fmfgraph*}}   =  
- \delta \Sigma(k)   
\parbox{20mm}
{\begin{fmfgraph*}(20,15) 
  \fmfleft{i} \fmfright{o}
  \fmf{plain,label=$<$,l.s=left}{i,o} 
\end{fmfgraph*}}
\end{eqnarray}
where
\begin{eqnarray}
-(d-1) \delta \Sigma & = &
\parbox{20mm}
{\begin{fmfgraph*}(20,15) 
  \fmfv{l=4,l.a=180,l.d=.1w,d.sh=circle,d.f=full,d.si=0.1w}{v1}
  \fmfv{d.sh=circle,d.f=full,d.si=0.1w}{v2}
  \fmfleft{i}  
  \fmf{phantom}{i,v1}
  \fmf{photon,tension=0.3,right}{v1,v2}
  \fmf{plain,tension=0.3,left,label=$>>$}{v1,v2} 
  \fmfright{o}   \fmf{phantom}{v2,o}
\end{fmfgraph*}}
\end{eqnarray}
\end{fmffile}

In the above equation we have omitted all the 
vanishing diagrams (similar
to those appearing in Eq.~[\ref{eq:feyn_I3u}]).  
These terms contribute to $\Sigma$s.

The algebraic expressions for the above diagrams are given in Section 
\ref{sec:Renormalization-Group}.
For isotropic flows, the algebraic factors $S(k,p,q)$ resulting from tensor 
contractions are given below.
\begin{eqnarray}
S(k,p,q)   & = & P^{+}_{bjm}(k) P^{+}_{mab}(p) P_{ja}(q) 
            =  kp \left( (d-3)z+2 z^3+(d-1)xy \right) 
\end{eqnarray}

In the next subsection we will derive the terms for
mode-to-mode energy transfer function.

\subsection{Mode-to-Mode Energy Transfer in fluid Turbulence}

In Section 3, we studied the ``mode-to-mode'' energy transfer
$S^{uu}({\bf k'|p|q})$  from mode {\bf p} to
mode {\bf k'}, with mode {\bf q} acting as a mediator.
The perturbative calculation of $S$ involves the following terms
\unitlength=1mm
\begin{fmffile}{fmfub5}
\begin{eqnarray}
\left\langle S^{uu}(k'|p|q) \right\rangle & = &
\parbox{30mm}
{\begin{fmfgraph*}(30,20) 
  \fmfv{d.sh=circle,d.f=full,d.si=0.01w}{v1}
  \fmfv{d.sh=circle,d.f=full,d.si=0.1w}{v2}
  \fmfleft{i}  
  \fmf{phantom}{i,v1}
  \fmf{plain,tension=0.1,right,label=$q$}{v1,v2}
  \fmf{plain,tension=0.1,label=$p$}{v1,v2}
  \fmf{photon,tension=0.1,left,label=$k$}{v1,v2} 
  \fmfright{o}   \fmf{phantom}{v2,o}
\end{fmfgraph*}} +
\parbox{30mm}
{\begin{fmfgraph*}(30,20) 
  \fmfv{d.sh=circle,d.f=full,d.si=0.01w}{v1}
  \fmfv{d.sh=circle,d.f=full,d.si=0.1w}{v2}
  \fmfleft{i}  
  \fmf{phantom}{i,v1}
  \fmf{plain,tension=0.1,right,label=$q$}{v1,v2}
  \fmf{photon,tension=0.1,label=$p$}{v1,v2}
  \fmf{plain,tension=0.1,left,label=$k$}{v1,v2} 
  \fmfright{o}   \fmf{phantom}{v2,o}
\end{fmfgraph*}} +
\parbox{30mm}
{\begin{fmfgraph*}(30,20) 
  \fmfv{d.sh=circle,d.f=full,d.si=0.01w}{v1}
  \fmfv{d.sh=circle,d.f=full,d.si=0.1w}{v2}
  \fmfleft{i}  
  \fmf{phantom}{i,v1}
  \fmf{photon,tension=0.1,right,label=$q$}{v1,v2}
  \fmf{plain,tension=0.1,label=$p$}{v1,v2}
  \fmf{plain,tension=0.1,left,label=$k$}{v1,v2} 
  \fmfright{o}   \fmf{phantom}{v2,o}
\end{fmfgraph*}} 
\end{eqnarray}
\end{fmffile}
\vspace{1cm}

In all the diagrams, the left vertex denotes
$k_i$, while the filled circle  of the right vertex
represent $(-i/2) P^+_{ijm}$.
For isotropic nonhelical flows, the algebraic factors are given
below. The factors for the diagrams are
$T_1, T_2, T_3$ in sequential order.
\begin{eqnarray}
T_1(k,p,q) & = & k_i P^{+}_{jab}(k) P_{ja}(p) P_{ib}(q)
            =  kp\left( (d-3)z + (d-2)xy +2 z^3+ 2 x y z^2
			  + x^2 z \right) \\
T_3(k,p,q) & = & -k_i P^{+}_{jab}(p) P_{ja}(k) P_{ib}(q)
             =  -kp\left( (d-3)z + (d-2)xy +2 z^3+ 2 x y z^2
			  + y^2 z \right)  \\
T_3(k,p,q) & = & -k_i P^{+}_{iab}(q) P_{ja}(k) P_{jb}(p)
            =  -k q \left(x z - 2 x y^2 z - y z^2 \right) \\
\end{eqnarray}

These terms are similar to those given in Leslie \cite{Leslie:book}.

\section{Energy transfer in scalar turbulence and MHD turbulence}

The equations for passive scalar turbulence are
\begin{eqnarray}
\frac{\partial\mathbf{u}}{\partial t}+(\mathbf{u}\cdot\nabla)\mathbf{u} 
& = & 
- \nabla p +\nu\nabla^{2}\mathbf{u},\label{eq:conv1}  \\
\frac{\partial\theta}{\partial t}+(\mathbf{u}\cdot\nabla)\theta 
& = & \kappa\nabla^{2}\theta,\label{eq:conv2} \\
\nabla\cdot\mathbf{u} & = & 0,
\label{eq:conv3}
\end{eqnarray}
where $\theta$ is the scalar density.  Note that the scalar is
convected by velocity field, but the scalar does not affect the
flow.  For details refer to Lesieur \cite{Leslie:book} and
Stani\u{s}i\'{c} \cite{Stanisic:book}.

For energy transfer in scalar field, we can follow the same procedure
as in fluid turbulence.  If we take only a single triad $({\bf
k',p,q})$ with ${\bf k'+p+q=0}$, energy is conserved for $\kappa=0$.
Following fluid turbulence, we can show that the energy equation for
scalar turbulence can be written in terms of 'mode-to-mode energy
transfer' $S^{\theta \theta}({\bf k'|p|q})$ from mode $\theta({\bf
p})$ to $\theta({\bf k'})$ with $\theta({\bf q}),{\bf u}({\bf q})$ as
a mediator, which is
\begin{eqnarray}
S^{\theta \theta}({\bf k'|p|q}) &  = & -\Im \left({\bf \left[k'.u(q)\right]}
 \left[\theta(k') \theta(p) \right] \right)  \label{eq:Spsikpsip_def} 
\end{eqnarray}
where $\Im$ stands for the imaginary part of the argument. 
 The energy equation for scalar field is
\begin{eqnarray}
\left(\frac{\partial}{\partial t}  + 2 \kappa k^2 \right) 
			C^{\theta}({\bf k}) 
& = &	       [S^{\theta \theta}({\bf k'|p|q})
		+S^{\theta \theta}({\bf k'|q|p})]
			\label{eqn:Cpsi_t} 
\end{eqnarray}

Note that there is no cross-transfer between $u$ and $\theta$ energy.
It is also important to note that both $C^u$ and $C^{\theta}$ are
conserved in every triad interaction, i.e.,
\begin{eqnarray}
S^{uu}({\bf k'|p|q}) + S^{uu}({\bf k'|q|p}) + 
S^{uu}({\bf p|k'|q}) + S^{uu}({\bf p|q|k'}) + 
S^{uu}({\bf q|k'|p}) + S^{uu}({\bf q|p|k'}) & = & 0  \\
S^{\theta \theta}({\bf k'|p|q}) + S^{\theta \theta}({\bf k'|q|p}) + 
S^{\theta \theta}({\bf p|k'|q}) + S^{\theta \theta}({\bf p|q|k'}) + 
S^{\theta \theta}({\bf q|k'|p}) + S^{\theta \theta}({\bf q|p|k'}) & = & 0
\end{eqnarray}
These are the statements of  ``detailed conservation of energy''
in triad interaction  (when $\nu=\kappa=0$) \cite{Leslie:book}.

The energy flux $\Pi^{\theta}$ from a wavenumber sphere of
radius $k_0$ is \cite{Dar:PD2001}
\begin{eqnarray}
\Pi^{\theta}(k_0) & = & \int_{k'>k_0} \frac{d {\bf k'}}{(2 \pi)^d} 
		       \int_{p<k_0} \frac{d {\bf p}}{(2 \pi)^d}  
			\la S^{\theta \theta}({\bf k'|p|q}) \ra
			\label{eqn:psi_flux}	
\end{eqnarray}

MHD turbulence has both velocity and magnetic field.  Detailed calculations show that there are six energy fluxes in MHD turbulence~\cite{Dar:PD2001,Verma:PR2004}. For field-theoretic treatment of MHD turbulence, refer to the reference~\cite{Verma:PP1999,Goldreich:ApJ1995,Verma:PR2004,Verma:Pramana2003Nonhelical,Verma:Pramana2003Helical}.  We refer to the reader to these references for the details.

\section{Energy Transfers in Rayleigh B\'{e}nard
Convection}

One of the most studied model of convection is Rayleigh B\'{e}nard Convection (RBC). In this model the fluid confined between two
parallel plates is heated from below. Constant temperature is 
maintained across these plates. Nondimensionalized equations for RBC under Boussinesq approximation 
are~\cite{Chandrasekhar:book:Instability,Verma:book:BDF}
\begin{eqnarray}
\frac{\partial\mathbf{u}}{\partial t}+(\mathbf{u}\cdot\nabla)\mathbf{u} & = & -\nabla \sigma + RP\theta\hat{{\mathbf{z}}}+P\nabla^{2}\mathbf{u},\\
\frac{\partial\theta}{\partial t}+(\mathbf{u}\cdot\nabla)\theta & = & u_{3}+\nabla^{2}\theta,\end{eqnarray}
where the nondimensinal parameters $R$ and $P=\nu/\kappa$ are Rayleigh
and Prandtl numbers respectively. The fluid is assumed to
incompressible except for the buoyancy term.

We can apply 'mode-to-mode energy transfer' model to RBC. 
The above hydrodynamical equations in Fourier space are
\begin{eqnarray}
\frac{\partial u_{i}(\mathbf{k})}{\partial t} &=& -i k_{i}\sigma{\bf{k}}
-ik_{j}\sum_{{\mathbf{p + q} = {\mathbf k}}} u_{j}({\bf q})u_{i}({\bf p})
 + RP\theta({\mathbf{k}})\delta_{i3} - P k^2 u_{i}({\mathbf{k}}) \\
\frac{\partial\theta({\mathbf{k}})}{\partial t} &=& u_{3}(k)-ik_{j}
\sum_{\mathbf p + q = k} u_{j}({\mathbf q})\theta({\mathbf p})- k^{2}
\theta({\mathbf k}), \\ 
k_{i}u_{i}(\mathbf{k}) &=& 0,
\end{eqnarray}
where $\mathbf{k}=\mathbf{p+q}$. We can derive interesting results by
focussing on a single triad $(\mathbf{k}',\mathbf{p},\mathbf{q})$ such
that $\mathbf{k'+p+q}=0$. Clearly $\mathbf{k'}=-\mathbf{k}$.  We can
easily derive the following energy equations:
\begin{eqnarray}
\frac{\partial}{\partial t}\frac{|\mathbf{u}({\mathbf{k}})|^{2}}{2} & = & 
S^{uu}({\mathbf{k}'|\mathbf{p}|\mathbf{q}})
+S^{uu}(\mathbf{k}'|\mathbf{q}|\mathbf{p}) 
+ RP\Re\left[\theta({\mathbf{k}})u_{3}({\mathbf{k}'})\right]
-2 P k^{2}\frac{|\mathbf{u}(\mathbf{k})|^{2}}{2},\label{eq:u-S}\\
\frac{\partial}{\partial t}\frac{|\theta({\mathbf{k}})|^{2}}{2} & = & 
S^{\theta\theta}({\mathbf{k}'|\mathbf{p}|\mathbf{q}})
+S^{\theta\theta}({\mathbf{k}'|\mathbf{q}|\mathbf{p}})
+\Re\left[\theta({\mathbf{k}})u_{3}({\mathbf{k}'})\right]
-2 k^{2}\frac{|\theta(\mathbf{k})|^{2}}{2},\
\label{eq:theta-S}\end{eqnarray}
where \begin{eqnarray} S^{uu}(\mathbf{k'|p|q}) & = &
-\Im\left(\left[\mathbf{k'}\cdot\mathbf{u}(\mathbf{q})\right]\left[\mathbf{u}(\mathbf{k}')\cdot\mathbf{u}(\mathbf{q})\right]\right),\label{eq:Suu}\\
S^{\theta\theta}(\mathbf{k'|p|q}) & = &
-\Im\left(\left[\mathbf{k'}\cdot\mathbf{u}(\mathbf{q})\right]
\left[\theta(\mathbf{k}')\cdot\theta(\mathbf{q})\right]\right).
\label{eq:Stheta_theta}
\end{eqnarray}
Here $\Re$ and $\Im$ represents the real and imaginary part of the
argument. The quantity $S^{\theta\theta}(\mathbf{k'|p|q})$ represents
the energy transfer from mode $\theta(\mathbf{p})$ (the field variable
with the second argument) to mode $\theta(\mathbf{k'})$ (the field
variable with the first argument) with the help of the mode
$\theta({\mathbf q})$ (the field variable with the third argument)
acting as a mediator~\cite{Dar:PD2001,Verma:PR2004}. 
 The above energy equations can be interpreted as follows: The
field variables with wavenumber $\mathbf{k}'$
{[}$\mathbf{u}(\mathbf{k'})$, $\theta(\mathbf{k'})${]} receives energy
from the modes $\mathbf{p}$ and $\mathbf{q}$ through mode-to-mode
energy transfer terms, and it also receives energy due to
interaction term $\theta({\mathbf{k}})u_{3}({\mathbf{k}'})$.

Following the same line of arguments as those for passive scalars, we deduce that  the sums of the energy transfer
rates along $u$-$u$ and $\theta-\theta$ channels are zero, i.e.,
\begin{eqnarray}
S^{XX}({\mathbf{k'|p|q}})+S^{XX}({\mathbf{k'|q|p}})+S^{XX}({\mathbf p|k'|q}) \\
+S^{XX}({\mathbf p|q|k})+S^{XX}({\mathbf q|k'|p})
+S^{XX}({\mathbf q|p|k'}) & = & 0,
\end{eqnarray}
where $XX$ could be $uu$ or $\theta\theta$.  Using these identities we can easily show that without viscous and thermal
diffusion
\begin{eqnarray}
\frac{\partial}{\partial t} |\mathbf{u}(\mathbf{k})|^{2}+
|\mathbf{u}(\mathbf{p})|^{2}+|\mathbf{u}(\mathbf{q})|^{2} 
& = & 2RP\Re[\theta({\mathbf{k}})u_{3}^{*}({\mathbf{k}}) 
+ \theta({\bf{p}})u_{3}^{*}({\bf{p}}) 
+\theta({\bf{q}})u_{3}^{*}({\bf{q}})],\\
\frac{\partial}{\partial t}\left[|\theta(\bf{k})|^{2}+|\theta(\bf{p})|^{2}
+|\theta(\bf{q})|^{2}\right] 
& = & 2\Re[\theta({\bf{k}})u_{3}^{*}({\bf{k}}) +
\theta({\bf{p}})u_{3}^{*}({\bf{p}})+\theta({\bf{q}})u_{3}^{*}({\bf{q}})],
\end{eqnarray}
The interpretation of the above equations is that the triads
$[\mathbf{u}(\mathbf{k}'),\mathbf{u}(\mathbf{p}),\mathbf{u}(\mathbf{q})]$
and $[\theta(\mathbf{k}'),\theta(\mathbf{p}),\theta(\mathbf{q})]$
exchange energy between each other via
$\theta({\mathbf{k}})u_{3}^{*}({\mathbf{k}})$ interaction terms. The
mode-to-mode interactions conserve energy within a triad. The viscous
and diffusive terms dissipate kinetic energy and $\theta$-energy
respectively.

For more details on the energy fluxes and spectra, refer to \cite{Verma:NJP2017,Verma:book:BDF}. Such studies are also useful for studying instabilities and patterns~\cite{Pal:EPL2009,Mishra:EPL2010}.  Also note that this formalism also works for stably stratified turbulence.

\bibliographystyle{apsrev}



\end{document}